\documentclass[twocolumn,tighten]{aastex63}

\usepackage{hyperref}

\newcommand{\msun}{\ensuremath{\mathrm{M}_\sun}}
\newcommand{\unitspace}{\,}
\usepackage{xstring}
\newcommand{\unit}[2]{%
	\IfEqCase{#2}{%
		{\msun}{\ensuremath{{#1}\unitspace\msun}}%
	}[\ensuremath{{#1}\unitspace\mathrm{#2}}]%
}
\newcommand{\eqref}[1]{(\ref{#1})}
\newcommand{\nue}{\ensuremath{\nu_{e}}}
\newcommand{\nuebar}{\ensuremath{\bar{\nu}_{e}}}
\newcommand{\nux}{\ensuremath{\nu_{x}}}
\newcommand{\nui}{\ensuremath{\nu_{i}}}
\newcommand{\Enutot}{\ensuremath{E_{\nu}^\mathrm{tot}}}
\newcommand{\Enuitot}{\ensuremath{E_{\nui}^\mathrm{tot}}}
\newcommand{\Lnutot}{\ensuremath{L_\mathrm{tot}}}
\newcommand{\Emean}{\ensuremath{\langle E \rangle}}
\newcommand{\Emeanx}{\ensuremath{\langle E_{\nux} \rangle}}
\newcommand{\alphax}{\ensuremath{\overline{\alpha}_{\nux}}}
\newcommand{\alphaea}{\ensuremath{\overline{\alpha}_{\nuebar}}}
\newcommand{\lambdaE}{\ensuremath{\lambda_{E}^{\scriptsize\textsc{PV}}}}
\newcommand{\lambdaa}{\ensuremath{\lambda_{\overline{\alpha}}^{\scriptsize\textsc{PV}}}}
\newcommand{\MBH}{\ensuremath{M_\mathrm{NS,b}^\mathrm{lim}}}
\newcommand{\alphaBH}{\ensuremath{\alpha_\mathrm{BH}}}
\newcommand{\RCC}{\ensuremath{R_\mathrm{CC}}}
\newcommand{\RLM}{\ensuremath{R_\mathrm{LM}}}
\newcommand{\diff}{\mathrm{d}}
\newcommand{\E}{\mathrm{e}}

\received{---}
\revised{---}
\accepted{---}

\shorttitle{}
\shortauthors{}

\begin{document}

\title{Stellar Collapse Diversity and the Diffuse Supernova Neutrino Background}



\author[0000-0003-1120-2559]{Daniel Kresse}
\affiliation{Max-Planck-Institut f\"ur Astrophysik, Karl-Schwarzschild-Stra\ss e~1, 85748 Garching, Germany}
\affiliation{Physik-Department, Technische Universit\"at M\"unchen, James-Franck-Stra\ss e 1, 85748 Garching, Germany}

\author[0000-0002-5711-7969]{Thomas Ertl}
\affiliation{Max-Planck-Institut f\"ur Astrophysik, Karl-Schwarzschild-Stra\ss e~1, 85748 Garching, Germany}

\author[0000-0002-0831-3330]{Hans-Thomas Janka}
\affiliation{Max-Planck-Institut f\"ur Astrophysik, Karl-Schwarzschild-Stra\ss e~1, 85748 Garching, Germany}

\begin{abstract}

The diffuse cosmic supernova neutrino background (DSNB) is observational target of the gadolinium-loaded Super-Kamiokande (SK) detector and the forthcoming JUNO and Hyper-Kamiokande detectors. Current predictions are hampered by our still incomplete understanding of the supernova (SN) explosion mechanism and of the neutron star (NS) equation of state and maximum mass. In our comprehensive study we revisit this problem on grounds of the landscapes of successful and failed SN explosions obtained by Sukhbold et al.\ and Ertl et al.\ with parametrized one-dimensional neutrino engines for large sets of single-star and helium-star progenitors, with the latter serving as proxy of binary evolution effects. Besides considering engines of different strengths, leading to different fractions of failed SNe with black-hole (BH) formation, we also vary the NS mass limit, the spectral shape of the neutrino emission, and include contributions from poorly understood alternative NS-formation channels such as accretion-induced or merger-induced collapse events. Since the neutrino signals of our large model sets are approximate, we calibrate the associated degrees of freedom by using state-of-the-art simulations of proto-neutron star cooling. Our predictions are higher than other recent ones because of a large fraction of failed SNe with long delay to BH formation. Our best-guess model predicts a DSNB \nuebar-flux of $28.8^{+24.6}_{-10.9}$\,cm$^{-2}$s$^{-1}$ with $6.0^{+5.1}_{-2.1}$\,cm$^{-2}$s$^{-1}$ in the favorable measurement interval of [10,30]\,MeV, and $1.3^{+1.1}_{-0.4}$\,cm$^{-2}$s$^{-1}$ with \nuebar\ energies $>$\,17.3\,MeV, which is roughly a factor of two below the current SK limit. The uncertainty range is dominated by the still insufficiently constrained cosmic rate of stellar core-collapse events.

\end{abstract}


\section{Introduction} \label{sec:intro}

When a massive star (above $\sim$9\,\msun) ends its life with the collapse of the inner core to a neutron star (NS) or a black hole (BH), a tremendous amount of gravitational binding energy (several $\unit{10^{53}}{erg}$) is released, predominantly in the form of neutrinos and antineutrinos \citep[see, e.g.,][]{2012ARNPS..62..407J, 2017hsn..book.1575J, 2013RvMP...85..245B}. In 1987, when the blue supergiant Sanduleak -69$^\circ$ 202 \citep{1987ApJ...321L..41W} in the Large Magellanic Cloud exploded as supernova (SN) 1987A, such an associated neutrino burst was detected for the first (and so far only) time as a $\sim$10\,s long signal, however, with the sparse yield of only two dozen counts \citep{1987PhRvL..58.1490H, 1987PhRvL..58.1494B, 1988PhLB..205..209A}. Nowadays, the size of the neutrino observatories all over the world has grown significantly such that a galactic SN would lead to a high-statistics signal \citep[e.g.,][]{2007ApJ...669..519I, 2011A&A...535A.109A}, which the scientific community is eagerly waiting for.

While such a nearby SN is a rare event \citep{2006Natur.439...45D, 2007ApJ...669..519I, 2015ApJ...802...47A}, a vast number of massive stars already ended their lives in the cosmic history, generously radiating neutrinos. The integral flux from all those past core collapses at cosmological distances, which is steadily flooding Earth, constitutes the so-called diffuse supernova neutrino background (DSNB). It makes for a ``guaranteed'' (isotropic and stationary) signal of MeV neutrinos, comprising rich information on the entire population of stellar core collapses \citep[for dedicated reviews, see][]{2004NJPh....6..170A, 2010ARNPS..60..439B, 2016APh....79...49L, 2019arXiv191011878V}. Intriguingly, the Super-Kamiokande (SK) experiment set upper flux limits on the DSNB \citep{2003PhRvL..90f1101M, 2012PhRvD..85e2007B, 2015APh....60...41Z} which are already close to theoretical predictions. This indicates the excellent discovery prospect within the next decade in the gadolinium-loaded SK detector and the forthcoming JUNO experiment \citep[see, e.g.,][]{2004PhRvL..93q1101B, 2006PhRvC..74a5803Y, 2009PhRvD..79h3013H, 2016JPhG...43c0401A, 2017JCAP...11..031P, 2018JCAP...05..066M}, as well as, in the longer term, with the Hyper-Kamiokande detector \citep{2011arXiv1109.3262A}, with DUNE \citep{2015arXiv151206148D}, or with the proposed \textsc{Theia} detector \citep{2020EPJC...80..416A, 2020arXiv200714705S}.

To exploit the full potential of future observations, comprehensive theoretical models will be needed for comparison. First predictions of the DSNB date back to the 1980s and 1990s \citep[e.g.,][]{1982SvA....26..132B, 1984Natur.310..191K, 1997APh.....7..137H} and have been refined ever since. Its link to the cosmic history of star formation has been studied in detail \citep[e.g.,][]{2004ApJ...607...20A, 2005JCAP...04..017S, 2006ApJ...651..142H, 2014ApJ...790..115M, 2020RNAAS...4....4A, 2020arXiv200702951R}; and also the dependence on the SN source spectra, which will be in the focus of this paper, has been subject of intense research. For instance, \citet{2007PhRvD..75g3022L} took an analytical approach based on the work by \citet{2003ApJ...590..971K}, while \citet{2006APh....26..190L} and \citet{2007PhRvD..76h3007Y} employed constraints from the measured neutrinos from SN 1987A for their DSNB predictions. The impact of the SN shock revival time has been investigated \citep{2013PhRvD..88h3012N, 2015ApJ...804...75N}, as well as the effect of neutrino flavor conversions \citep{2003PhLB..559..113A, 2011PhLB..702..209C, 2012JCAP...07..012L}.

Particularly the contribution from BH-forming, failed explosions to the DSNB has caught much attention in recent years. It might significantly enhance the high-energy tail of the flux spectrum, which is most relevant for the detection \citep[e.g.,][]{2009PhRvL.102w1101L}. Several studies varied the (still unknown) fraction of failed SNe \citep{2009PhRvL.102w1101L, 2010PhRvD..81h3001L, 2012PhRvD..85d3011K, 2017JCAP...11..031P, 2018MNRAS.475.1363H, 2018JCAP...05..066M}; in this regard, \citet{2015ApJ...804...75N} and \citet{2015PhLB..751..413Y} further considered the cosmic evolution of stellar metallicities; and also the dependence on the high-density equation of state (EoS), which is closely related to the mass limit up to which a NS can be stabilized against its own gravity, has been explored tentatively \citep{2009PhRvL.102w1101L, 2012PhRvD..85d3011K, 2014ApJ...790..115M, 2015ApJ...804...75N, 2016ApJ...827...85H, 2018ApJ...869...31H, 2018MNRAS.475.1363H}.

Detailed neutrino signals from successful and failed SNe are the premise for reliable DSNB predictions. While most previous works employed rather approximate neutrino source spectra or spectra representative of some typical cases, numerical modeling of stellar core collapse has reached a high level of sophistication nowadays. An increasing number of three-dimensional (3D) simulations with detailed microphysics has become available \citep[e.g.,][]{2014ApJ...786...83T, 2014PhRvD..90d5032T, 2015ApJ...807L..31L, 2015ApJ...801L..24M, 2017MNRAS.472..491M, 2018ApJ...865...81O, 2018ApJ...855L...3O, 2018ApJ...852...28S, 2019MNRAS.485.3153B, 2019ApJ...873...45G, 2019MNRAS.482..351V, 2020ApJ...891...27M}. Nonetheless, high computational costs are still causing limitations. Up to now, only about twenty selected progenitors have been considered in 3D SN models, none of them evolved longer than roughly one second.

At the same time, it was shown that the outcome of a core-collapse event (successful explosion or BH formation) as well as the neutrino emission strongly depend on the progenitor structure, with large variations between different stars \citep{2011ApJ...730...70O, 2012ApJ...757...69U, 2014MNRAS.445L..99H, 2015PASJ...67..107N, 2015ApJ...801...90P, 2016ApJ...818..124E, 2016MNRAS.460..742M, 2016ApJ...821...38S, 2019ApJ...870....1E}. This has been neglected (or oversimplified) in most previous DSNB studies, which typically employed only a few exemplary models. Particularly the signals from BH-forming, failed SNe are strongly dependent on the progenitor-specific mass-accretion rate \citep{2009A&A...499....1F, 2011ApJ...730...70O}. Comprehensive sets of neutrino signals over the entire range of pre-SN stars are therefore required to adequately account for the diversity of stellar core collapse. In light of this, \citet{2018MNRAS.475.1363H} employed a set of 101 axisymmetric (2D) SN simulations and seven models of BH formation from spherically symmetric (1D) simulations, however with the need to extrapolate the neutrino signals at times later than $\sim$1\,s. Due to the limited number of their failed explosions, they (linearly) interpolated the spectral parameters of the time-integrated neutrino emission (total energetics, mean energy, and shape parameter) of their few BH simulations as a function of the ``progenitor compactness'' \citep{2011ApJ...730...70O} to account for a larger scope of failed SNe.

In this paper, we take a different angle of approach. Referring to the studies by \citet{2012ApJ...757...69U}, \citet{2016ApJ...821...38S}, and \citet{2016ApJ...818..124E, 2020ApJ...890...51E}, we use spherically symmetric simulations over a wide range of pre-SN stars exploded by means of a ``calibrated central neutrino engine''. In this way, our analysis of the DSNB is based on detailed information about the ``landscape'' of successful and failed explosions with individual neutrino signals for every progenitor, including cases of long-lasting mass accretion with relatively late BH formation. Using our large sets of (approximately calculated) long-time neutrino signals, which we cross-check by comparing and normalizing them to the outcome of more sophisticated simulations (see appendices), we aim at providing refined predictions of the DSNB. In a systematic parameter study, we further investigate the impact of three critical source properties on the DSNB flux spectrum: (1) We vary the fractions of successful and failed SNe through different calibrations of the neutrino engine used for the explosion modeling of our large progenitor set. (2) We consider different values for the critical mass at which the neutrino signals stop due to BH formation and follow the continued mass accretion of failed explosions. (3) We consider different spectral shapes of the neutrino emission based on the study by \citet{2003ApJ...590..971K}.

As in previous DSNB studies \citep[e.g.,][]{2014ApJ...790..115M, 2018MNRAS.475.1363H}, we also include the contribution from electron-capture SNe (ECSNe) of degenerate oxygen-neon-magnesium (ONeMg) cores \citep{1980PASJ...32..303M, 1984ApJ...277..791N, 1987ApJ...322..206N}, for which we employ the neutrino signals from \citet{2010PhRvL.104y1101H}. Moreover, we explore other possible channels for the formation of low-mass NSs, such as accretion-induced collapse \citep[AIC;][]{1990ApJ...353..159B, 1991ApJ...367L..19N, 2004ApJ...601.1058I, 2010MNRAS.402.1437H, 2016A&A...593A..72J, 2018RAA....18...36W, 2019MNRAS.484..698R} and merger-induced collapse \citep[MIC;][]{1985A&A...150L..21S, 2008MNRAS.386..553I, 2016MNRAS.463.3461S, 2019MNRAS.484..698R, 2018ApJ...869..140K} of white dwarfs (WDs), or ultra\-stripped SNe from close binaries \citep{1994Natur.371..227N, 2002MNRAS.331.1027D, 2013ApJ...778L..23T, 2015MNRAS.451.2123T, 2015MNRAS.454.3073S, 2018MNRAS.479.3675M}. Using simplified assumptions, we estimate the flux from such a combined ``low-mass component'' and comment on its relevance.

While stellar explosion models typically employ single-star progenitors thus far, recent observations suggest that most of the massive stars are in binary systems \citep[see, e.g.,][]{2009AJ....137.3358M, 2012Sci...337..444S}. In view of this we also investigate, for the first time, how the inclusion of binary models affects predictions of the DSNB using the helium-star progenitors from \citet{2019ApJ...878...49W} and the explosion models of \citet{2020ApJ...890...51E}.

The paper is organized as follows. In Section~\ref{sec:simulation_setup}, we describe the setup of our simulations and discuss the overall properties of the neutrino signals used in our study. Section~\ref{sec:DSNB_formulation} is dedicated to our approach of formulating the DSNB. We present our fiducial predictions in Section~\ref{sec:fiducial_model}. In Section~\ref{sec:parameter_study}, we discuss the results of our detailed parameter study: We investigate the sensitivity of the DSNB flux spectrum of electron antineutrinos to the fraction of failed explosions, the BH mass threshold, and the spectral shape of the neutrino emission. We further explore an additional contribution from low-mass NS-forming events (such as AIC, MIC, and ultra\-stripped SNe) and study the impact of including binary progenitors. In Section~\ref{sec:nue}, we briefly comment the DSNB flux spectrum of electron neutrinos. In Section~\ref{sec:uncertainties}, the effects of neutrino flavor conversions are discussed along with remaining uncertainties, followed by a comparison of our models with the SK-flux limits and with previous works (Section~\ref{sec:comparison}). In Section~\ref{sec:summary_uncertainties}, we categorize and rank	the DSNB parameter variations and uncertainties considered in our work. We conclude in Section~\ref{sec:conclusions}. Supplementary material can be found in appendices.\\

\section{Simulation Setup and Neutrino Signals}\label{sec:simulation_setup}

In spherical symmetry, self-consistent SN explosions turned out to be possible only for a few low-mass stars \citep{2006A&A...450..345K, 2008A&A...485..199J, 2012PTEP.2012aA309J, 2010A&A...517A..80F, 2015ApJ...801L..24M, 2017ApJ...850...43R}. To still explore the outcome of stellar core collapse in 1D over a wide range of progenitor masses, we adopt the parametric approach of \citet{2016ApJ...818..124E}, where a ``calibrated neutrino engine'' is placed in the center of all pre-SN models. By these means, we obtain neutrino signals for a large set of individual stars, in satisfactory agreement with more sophisticated simulations and including cases of long-term accretion with late BH formation, as we will elaborate in this section. For more details on our computational setup, the reader is also referred to \citet{2012ApJ...757...69U}, \citet{2016ApJ...821...38S}, and \citet{2016ApJ...818..124E, 2020ApJ...890...51E}.

\subsection{Pre-SN Models}\label{subsec:pre-SN_models}

In this work, we use a combined set of 200 solar-metallicity progenitor models from \citet[``WH07'' and ``WH15'']{2007PhR...442..269W, 2015ApJ...810...34W} and \citet[``SW14'']{2014ApJ...783...10S}, which was already applied in \citet{2016ApJ...821...38S} and can be downloaded from the Garching Core-collapse Supernova Archive.\footnote{\url{https://wwwmpa.mpa-garching.mpg.de/ccsnarchive/data/SEWBJ_2015/index.html} (\url{http://doi.org/10.17617/1.b})} All models are non-rotating single stars, evolved with the \textsc{Kepler} code \citep{1978ApJ...225.1021W} up to the onset of iron-core collapse. The resulting grid of progenitors, unevenly distributed over the zero-age main sequence (ZAMS) mass interval of 9--120\,\msun, spans the commonly assumed range of ``conventional'' iron-core collapse SNe (or BH-forming, failed SNe, respectively).

Below that, in the narrow band between $\unit{8.7}{\msun}$ and $\unit{9}{\msun}$, we additionally consider ECSNe of degenerate ONeMg cores as another channel for NS formation \citep{1980PASJ...32..303M, 1984ApJ...277..791N, 1987ApJ...322..206N}; yet it should be stressed that the exact mass range of ECSNe in the local universe is not finally clear according to current knowledge \citep[see, e.g.,][]{2008ApJ...675..614P, 2013ApJ...772..150J, 2015MNRAS.446.2599D, 2016A&A...593A..72J, 2019PhRvL.123z2701K, 2019ApJ...886...22Z, 2020ApJ...889...34L}. We employ a simulation by \citet[``model Sf'']{2010PhRvL.104y1101H} for the neutrino signal of such core-collapse events. The upper-mass end of the ZAMS mass grid is similarly uncertain and depends strongly on the physics of mass loss. However, as will be detailed in Section~\ref{subsec:IMF}, high-mass contributions are suppressed by the steeply declining initial mass function (IMF) and are therefore of subordinate importance for the DSNB. In Sections~\ref{subsec:dsnb_LM} and \ref{subsec:dsnb_binary}, we will further consider progenitors from binary systems. The helium-star models used in this context were published by \citet{2019ApJ...878...49W} and their explosions were investigated by \citet{2020ApJ...890...51E}.

\subsection{SN Simulations}\label{subsec:SN_simulations}

\begin{figure}[t!]
	\includegraphics[width=\columnwidth]{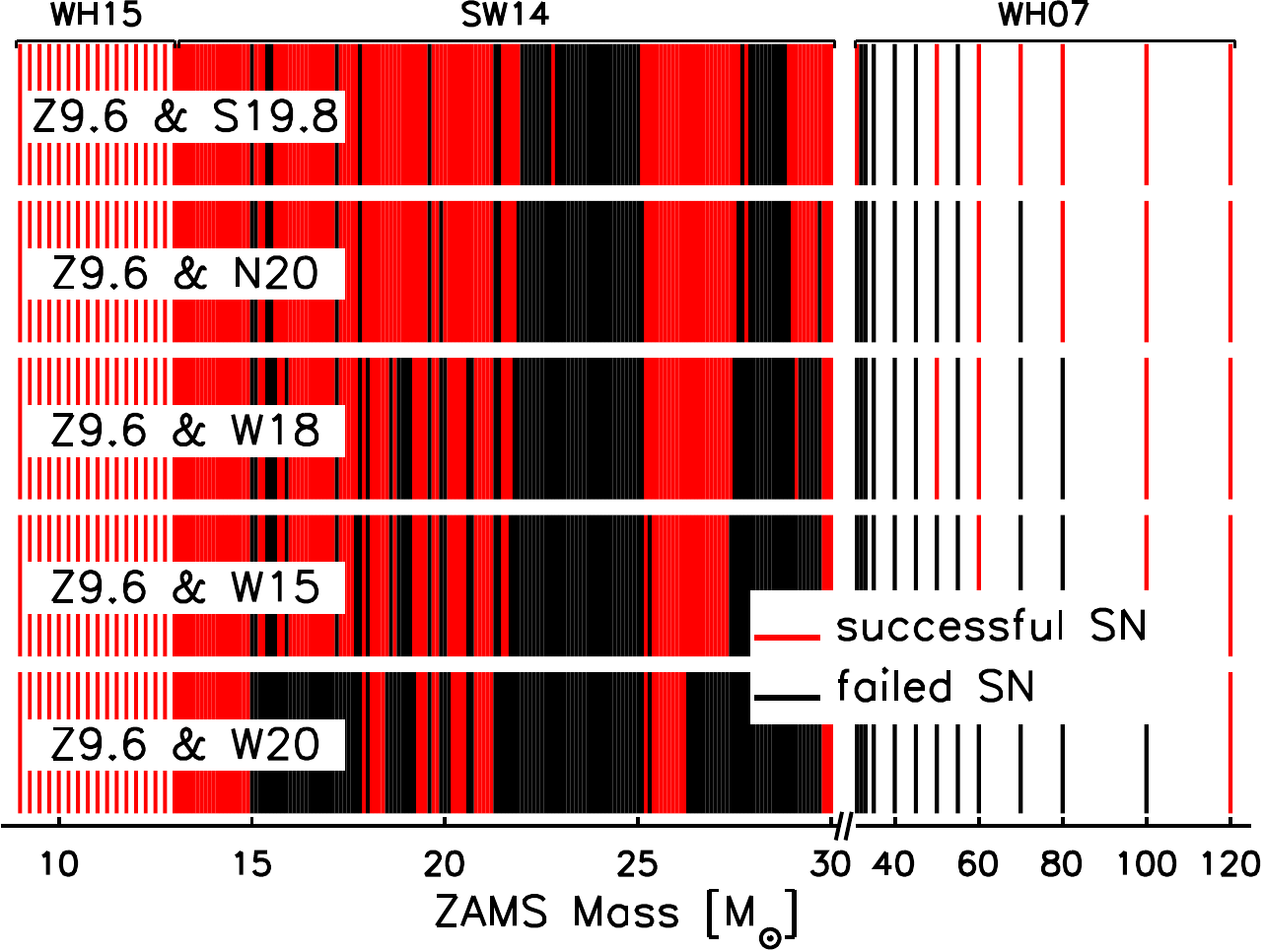}
	\caption{``Landscapes of explodability'' for our five different engine models (see main text for details). The ZAMS mass ranges of our three progenitor-model sets (WH15, SW14, and WH07) are indicated on the top of the figure. Successful SN explosions are marked in red, while black bars indicate the formation of a BH in a failed SN. From top to bottom, the IMF-weighted fraction of successful explosions decreases from 82.2\% (Z9.6\,\&\,S19.8) to 58.3\% (Z9.6\,\&\,W20); see Table~\ref{tab:explodability}. ECSNe are not shown in the plot.\label{fig:explodability}}
\end{figure}

Our stellar collapse and explosion simulations were performed with the \textsc{Prometheus-HotB} code \citep{1996A&A...306..167J, 2003A&A...408..621K, 2006A&A...457..963S, 2016ApJ...818..124E, 2020ApJ...890...51E}. The innermost $\unit{1.1}{\msun}$ of the nascent proto-NS (PNS) were excised and replaced by a contracting inner-grid boundary and an analytic one-zone core-cooling model with tuneable parameters \citep[for the details, see][]{2012ApJ...757...69U}. This ``central neutrino engine'' was calibrated to yield explosions in agreement with the well studied cases of SN 1987A and the Crab SN (SN 1054). More specifically, for pre-SN stars with ZAMS masses above $\unit{12}{\msun}$, which \citet{2016ApJ...821...38S} termed ``87A-like'', a PNS core model was applied and adjusted such that a given progenitor in the range of 15--20\,\msun, namely S19.8, N20, W18, W15, or W20 \citep[as described in][]{2016ApJ...821...38S}, reproduced the observed explosion energy \citep[(1.2--1.5)$\times 10^{51}$\,erg;][]{1989ARA&A..27..629A, 2015A&A...581A..40U}, $^{56}$Ni yield \citep[$\sim$0.07\,\msun;][]{1991A&A...245..490B, 1992ApJ...384L..33S} and the basic neutrino-emission features \citep{1987PhRvL..58.1490H, 1987PhRvL..58.1494B} of SN 1987A. The low-mass end (9--12\,\msun) was connected to the 87A-like cases by an interpolation of the core-model parameters. As a second anchor point, we used the progenitor z9.6 by A. Heger (2012, private communication), which explodes with low energy \citep[$\sim$10$^{50}$\,erg;][]{2012PTEP.2012aA309J, 2015ApJ...801L..24M} and a small $^{56}$Ni yield \citep[$\sim$0.0025\,\msun;][]{2018ApJ...852...40W} in self-consistent simulations, in good agreement with the observational constraints for the Crab SN \citep{2013MNRAS.434..102S, 2013ApJ...771L..12T, 2015ApJ...806..153Y}. For more details on our calibration procedure, the reader is referred to \citet{2016ApJ...821...38S}.

Depending on the engine model, we obtained more or less energetic or failed explosions over the range of considered pre-SN stars, as can be seen in Figure~\ref{fig:explodability}. While the S19.8 and N20 calibrations lead to the largest fraction of successful SNe (red), W20 is a rather weak engine, resulting in the largest fraction of BH-forming cases (black). W18 and W15 reside between these two extremes, as can also be seen in Table~\ref{tab:explodability}, which shows the IMF-weighted fractions of successful and failed explosions for the different neutrino engines. The outcome in the low-mass range (9--12\,\msun) is the same for all five cases, since our interpolation towards z9.6 is independent of the high-mass calibration. The non-monotonic pattern of successful SNe and BH-forming collapses in Figure~\ref{fig:explodability} was described in previous works \citep{2012ApJ...757...69U, 2015ApJ...801...90P, 2016ApJ...818..124E, 2016MNRAS.460..742M, 2016ApJ...821...38S, 2019ApJ...870....1E}. It grounds on the progenitor structure, which is strongly varying with ZAMS mass \citep{2011ApJ...730...70O, 2014MNRAS.445L..99H, 2015PASJ...67..107N, 2018ApJ...860...93S}.

\begin{deluxetable}{lCC}[t!]
	\tablecaption{Fractions of successful and failed SNe.\label{tab:explodability}}
	\tablehead{
		\colhead{Engine Model} & \colhead{Successful SNe} & \colhead{Failed SNe}
	}
	\startdata
	Z9.6\,\&\,S19.8 & 82.2\% & 17.8\%\\
	Z9.6\,\&\,N20 & 77.2\% & 22.8\%\\
	Z9.6\,\&\,W18 & 73.1\% & 26.9\%\\
	Z9.6\,\&\,W15 & 70.9\% & 29.1\%\\
	Z9.6\,\&\,W20 & 58.3\% & 41.7\%
	\enddata
	\tablecomments{NS and BH predictions from our five engine models were IMF weighted according to Equation~\eqref{eq:IMF}.}
\end{deluxetable}

Compared to the simulations of \citet{2016ApJ...818..124E} and \citet{2016ApJ...821...38S}, the neutrino transport outside of the PNS core, which is treated by a gray approximation \citep{2006A&A...457..963S, 2007A&A...467.1227A}, was slightly improved such that we were able to follow cases of long-lasting mass accretion until late collapse to a BH. For numerical reasons, the neutrino-nucleon scattering rate \citep[equation~(D.68) of][]{2006A&A...457..963S} is now split into two separate source terms, one for absorption ($\propto \langle \epsilon_{\nu}^4 \rangle$, with $\epsilon_{\nu}$ denoting the neutrino energy) and one for emission ($\propto T \langle \epsilon_{\nu}^3 \rangle$), to avoid sign fluctuations for large temperatures $T$ (for details, see appendix~B of \citealt{2020MNRAS.496.2039S}). Furthermore, an adaptive grid was implemented to better resolve the steep density gradient at the PNS surface. Our new code was applied without recalibrating the core models, which led to slightly increased explosion energies because of decreased neutrino luminosities compared to the models reported by \citet{2016ApJ...821...38S}. Accordingly, a few scattered progenitors which failed to explode with the old code \citep[cf.][figure~13]{2016ApJ...821...38S} yield successful SNe with our new treatment. (A detailed report of the code changes and consequences for the model results is provided in the appendices of \citealt{2020ApJ...890...51E}.) In the work at hand, we moreover neglect the neutrino emission from the late-time fallback in so-called fallback SNe, in which the fallback matter pushes the NS beyond the BH limit after a successful explosion was initiated. This is justified because such cases turned out to be rare in the considered set of solar-metallicity progenitors \citep{2016ApJ...818..124E, 2016ApJ...821...38S} and additionally reside in the IMF-suppressed high-mass regime. In the context of our paper we therefore consider fallback SNe as successful SN events with the corresponding neutrino emission from NS formation. BH-forming events are only those cases where the BH does not form by fallback but by continuous accretion, and we use the terms ``BH formation'' and ``failed SN'' equivalently.

\subsection{Neutrino Signals}\label{subsec:neutrino_signals}

\begin{figure*}
	\includegraphics[width=\textwidth]{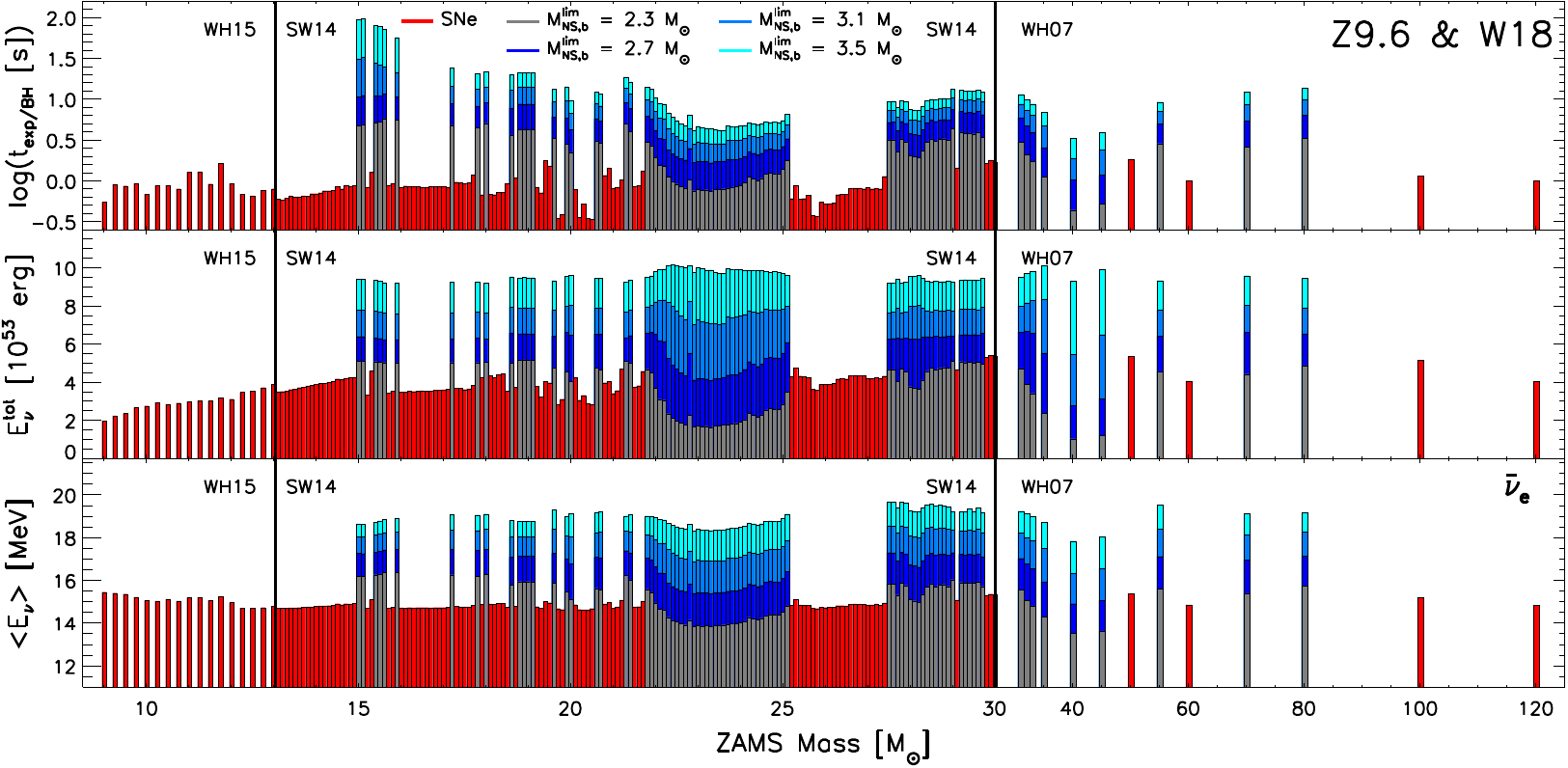}
	\caption{Landscape of SN and BH-formation cases for the combined progenitor sets of WH15, SW14, and WH07, simulated with the neutrino engine model of Z9.6\,\&\,W18. From top to bottom: time of explosion or BH formation, total energy radiated in all species of neutrinos, and mean energy of electron antineutrinos versus ZAMS mass of the progenitors. Note the logarithmic scale in the top panel. Red bars indicate successful SN explosions and fallback SNe, while the outcomes of BH-forming, failed SNe are shown for our different cases of baryonic NS mass limits in gray ($\unit{2.3}{\msun}$), dark blue ($\unit{2.7}{\msun}$), light blue ($\unit{3.1}{\msun}$), and cyan ($\unit{3.5}{\msun}$). The outcome of the ECSN by \citet{2010PhRvL.104y1101H} is not shown in the figure, but discussed in the main text.\\
	\label{fig:neutrino_outcomesystematics}}
\end{figure*}

For each progenitor, we obtain the total energy release in neutrinos through time integration of the time-dependent neutrino luminosities, $L_{\nui}(t)$, and the mean values of the energies of the radiated neutrinos by computing the (luminosity-weighted) time average of the time-dependent mean neutrino energies, $\langle E_{\nui}(t) \rangle$, for all three considered neutrino species $\nui = \nue, \nuebar, \nux$, where $\nux$ denotes a representative heavy-lepton neutrino ($\nu_{\mu}, \bar{\nu}_{\mu}, \nu_{\tau}, \bar{\nu}_{\tau}$). Successful SNe were simulated up to a post-bounce time of $t\,\mathord{=}\,\unit{15}{s}$, when the neutrino luminosities from PNS cooling have already declined to an insignificant level (see Appendix~\ref{appendix:extrapolation}). In the cases of failed explosions, however, the continued accretion of infalling mass onto the PNS releases gravitational binding energy, leading to an ongoing accretion component of the neutrino luminosities. The signals of such cases are truncated only when the PNS is pushed beyond the (still unknown) limit of BH formation, for which we consider four different values of the baryonic mass, \MBH, namely 2.3, 2.7, 3.1, and 3.5\,\msun, which are motivated as follows.

Assuming a NS radius of $\unit{(11\pm1)}{km}$ for maximum-mass NSs,\footnote{\label{fn:R_NS}This range is motivated by recent publications, constraining the NS radius from observations of the binary NS merger event GW170817 \citep{2017ApJ...850L..34B, 2017ApJ...848L..18N, 2018ApJ...857L..23R, 2018PhRvL.121p1101A, 2020NatAs...4..625C}, as well as by the studies of \citet{2010ApJ...722...33S}, \citet{2016ApJ...820...28O}, \citet{2016ARA&A..54..401O}, and \citet{2016PhR...621..127L}. For NSs at the upper mass end, we consider (circumferential) radii of $\unit{10}{km}\,\mathord{\leqslant}\,R_\mathrm{NS}\,\mathord{\leqslant}\,\unit{12}{km}$, while we assume $R_\mathrm{NS}\,\mathord{\geqslant}\,\unit{11}{km}$ in Appendix~\ref{appendix:total_energies} for ``average-mass'' NSs, as suggested by \citet{2017ApJ...850L..34B}, and also compatible with recent results by NICER \citep{2019ApJ...887L..24M}.} and utilizing equation~(36) of \citet{2001ApJ...550..426L}, which we provide as Equation~\eqref{eq:LattimerPrakash} in Appendix~\ref{appendix:total_energies}, a baryonic NS mass of $\unit{2.3}{\msun}$ converts to a gravitating mass of $\unit{1.95^{+0.02}_{-0.03}}{\msun}$. This is marginally below the largest currently measured pulsar masses of $\sim$2\,\msun\ \citep{2010Natur.467.1081D, 2013Sci...340..448A, 2016ARA&A..54..401O, 2020NatAs...4...72C}, setting a lower limit for the maximum NS mass.

From the first gravitational-wave observation of a binary NS merger \citep[GW170817;][]{2017PhRvL.119p1101A} and its electromagnetic counterparts \citep{2017ApJ...848L..12A}, \citet{2017ApJ...850L..19M} placed a tentative upper bound on the maximum gravitational NS mass of $\unit{2.17}{\msun}$ (at 90\% confidence level), which follows from their reasoning that the merger remnant was a relatively short-lived, differentially-rotating hyper-massive NS, disfavoring both the prompt collapse to a BH as well as the formation of a long-lived, supermassive NS. Their mass limit is compatible with other recent publications \citep[e.g.,][]{2017PhRvD..96l3012S, 2018MNRAS.478.1377A, 2018ApJ...852L..25R, 2018PhRvD..97b1501R, 2019EPJA...55..209L, 2020PhRvD.101f3007E}. Consistently, we take our case of a baryonic mass of $\unit{2.7}{\msun}$ (corresponding to $\unit{2.23^{+0.03}_{-0.04}}{\msun}$ gravitational mass), which is close to this bound, as our reference threshold for BH formation.

Nonetheless, \citet{2017ApJ...850L..19M} pointed out several uncertainties related to their analysis. For instance, they neglected the effects of thermal pressure support on the stability of the compact merger remnant, which may change their conclusions.\footnote{In fact, two competing effects play a role and can dominate under different circumstances: destabilization because of an enhanced gravitational potential due to additional thermal energy, or stabilization due to increased support by thermal pressure \citep[see, e.g.,][]{1995A&A...296..145K, 2011ApJ...730...70O, 2013ApJ...774...17S, 2020ApJ...894....4D}.} Thermal effects are also important for the stability of hot PNSs on their way towards BH formation in the cases of failed SNe, possibly increasing the limiting mass compared to the value for cold NSs \citep{2011ApJ...730...70O, 2013ApJ...774...17S, 2020ApJ...894....4D}. For these reasons, we additionally explore two more extreme cases for the baryonic (gravitational) mass limit, namely $\unit{3.1}{\msun}$ ($\unit{2.50^{+0.04}_{-0.05}}{\msun}$) and $\unit{3.5}{\msun}$ ($\unit{2.75^{+0.05}_{-0.05}}{\msun}$). Eventually, further pulsar timing measurements \citep[cf.][]{2010Natur.467.1081D, 2013Sci...340..448A, 2016ARA&A..54..401O, 2020NatAs...4...72C} as well as an increased number of observed binary-NS mergers \citep[see, e.g.,][]{2010CQGra..27q3001A} should be able to shed more light on the maximum mass of NSs.

The most important results of the SN and BH-formation simulations to be used in our DSNB calculations are the values of the time-integrated total energy release in neutrinos of all species and the time-averaged mean energies of the emitted electron antineutrinos. Figure~\ref{fig:neutrino_outcomesystematics} provides an overview of the corresponding values over the entire range of iron-core progenitors as a function of ZAMS mass for the exemplary case of the Z9.6\,\&\,W18 engine model, whose results will serve as a reference point in our later discussion (see Section~\ref{sec:fiducial_model}). The three sets of pre-SN stars, WH15, SW14, and WH07, are separated by black vertical lines. Red bars indicate successful explosions and fallback SNe, whereas the outcomes of failed SNe are marked by gray, dark blue, light blue, or cyan, depending on the different choices of the critical baryonic mass for BH formation.

The upper panel shows the explosion time, $t_\mathrm{exp}$, for successful SNe, defined as the time when the shock passes $\unit{500}{km}$ (and not to be confused with the termination of our successful SN simulations and neutrino-signal calculations at $\unit{15}{s}$, which was mentioned above). In cases of failed explosions, the time of BH formation, $t_\mathrm{BH}$, is shown, which coincides with a sudden termination of the neutrino signal. Depending on the assumed NS mass limit and the progenitor-dependent mass-accretion rate, these times range from below $\unit{1}{s}$ up to $\unit{100}{s}$ in the most extreme cases (note the logarithmic scale).\footnote{\label{fn:compactness}Using general-relativistic simulations in spherical symmetry, \citet{2011ApJ...730...70O} found a functional dependence of the time to BH formation on the progenitor structure, to first order compliant with a simple power-law scaling, $t_\mathrm{BH} \propto (\xi_{2.5})^{-3/2}$, where $\xi_{2.5}$ denotes the progenitor compactness parameter at bounce for an enclosed mass of $\unit{2.5}{\msun}$, as defined by their equation~(10). Less compact progenitors of failed SNe, e.g., in the ZAMS mass range around $\unit{15}{\msun}$ \citep[see figure~4 of][]{2016ApJ...818..124E}, with lower densities in the mass shells surrounding the PNS, need longer accretion times until BH formation, in contrast to the fast-accreting high-compactness progenitors at around $\unit{22-25}{\msun}$ and $\sim$$\unit{40}{\msun}$.} This illustrates the need for a large set of long-time simulations to properly sample the neutrino contribution from the BH-formation events.

The middle panel of Figure~\ref{fig:neutrino_outcomesystematics} displays the total radiated neutrino energies, \Enutot, computed as the time-integrals of the summed-up neutrino luminosities of all species, $\Lnutot(t)\,\mathord{=}\,L_{\nue}(t)\,\mathord{+}\,L_{\nuebar}(t)\,\mathord{+}\,4 L_{\nux}(t)$, from core bounce ($t\,\mathord{=}\,0$) until the end of the simulations ($t\,\mathord{=}\,\unit{15}{s}$) or the termination of the signals ($t\,\mathord{=}\,t_\mathrm{BH}$) for SN or BH-formation cases, respectively. Due to the afore-mentioned numerical improvements in the neutrino transport, these energies are slightly lower than those in \citet{2016ApJ...818..124E} and \citet{2016ApJ...821...38S}. In Appendix~\ref{appendix:total_energies}, we cross-check the values of \Enutot\ by comparing them to the available budget of gravitational binding energy released during the cooling of the PNS, as estimated by using an analytic, radius-dependent approximate fit-formula from \citet{2001ApJ...550..426L}. We find good overall agreement, although our values might overestimate the neutrino energy loss by up to about 10--20\%, depending on the NS radius. In Section~\ref{sec:uncertainties}, we will discuss this and other uncertainties related to our DSNB predictions in more detail. In our work, we neglect contributions to the neutrino loss from fallback of matter after the successful launch of an explosion, since the amount of fallback was shown to be small (typically below $\unit{10^{-2}}{\msun}$) for most progenitors \citep{2016ApJ...818..124E, 2016ApJ...821...38S} and since our values for the release of NS binding energy in neutrinos are on the high side anyway. In addition, fallback SNe with substantial late-time fallback (possibly turning NSs to BHs) are rare, as noted above.

The mean neutrino energies of the time-integrated energy emission are displayed in the bottom panel of Figure~\ref{fig:neutrino_outcomesystematics} for electron antineutrinos, which are most relevant for our study. Values around $\unit{15}{MeV}$ are the rather uniform outcome of successful SNe, in agreement with other publications \citep[e.g.,][]{2016NCimR..39....1M, 2018MNRAS.475.1363H}. The mean energies from failed explosions, on the other hand, vary considerably between the progenitors and depend strongly on the NS mass limit. On the way to BH formation, the temperatures in the PNS's accretion mantle rise gradually, yielding increasingly harder neutrino spectra \citep[see, e.g.,][]{2006PhRvL..97i1101S, 2007ApJ...667..382S, 2008ApJ...688.1176S, 2009A&A...499....1F, Huedepohl:2014, 2016NCimR..39....1M}.

In a few cases, we needed to extrapolate our neutrino signals, either because the simulations could not be carried out to sufficiently late times or due to numerical problems, albeit such problems occurred only after $\sim$$\unit{10}{s}$ (see Appendix~\ref{appendix:extrapolation} and Figure~\ref{fig:extrapolation} there). Furthermore, we should point out that the luminosities of heavy-lepton neutrinos are underestimated compared to \nue\ and \nuebar\ in our simulations. This is a consequence of our approximate treatment of the microphysics (e.g., nucleon-nucleon bremsstrahlung is not included) and of the relatively modest contraction of the inner grid boundary and thus underestimated temperatures in the accretion layer. To cure this shortcoming compared to more sophisticated transport calculations, we use information from such calculations with the \textsc{Prometheus-Vertex} code \citep{2002A&A...396..361R, 2006A&A...447.1049B}, which employs a state-of-the-art treatment of neutrino transport based on a Boltzmann-moment-closure scheme and a mixing-length treatment of PNS convection, to rescale the integrated energy loss in the different neutrino flavors, as detailed in Appendix~\ref{appendix:rescaling}.

The neutrino signal of ECSNe from \citet[``model Sf'']{2010PhRvL.104y1101H} was followed for $\unit{8.9}{s}$ after core bounce and yields a total radiated neutrino energy of $\unit{1.63\times10^{53}}{erg}$, with a time-integrated \nuebar\ mean energy of $\unit{11.6}{MeV}$. These results are not shown in Figure~\ref{fig:neutrino_outcomesystematics}, yet we use them for our DSNB estimates, which we describe in the next section.

\section{Formulation of the DSNB} \label{sec:DSNB_formulation}

The differential number flux, $\diff\Phi(E)/\diff E$, of DSNB neutrinos or antineutrinos, isotropically flooding the Earth in the energy interval $[E, E\,\mathord{+}\,\diff E]$, is computed as the line-of-sight integral of the IMF-weighted neutrino spectrum of past core-collapse events ($\diff N_\mathrm{CC}/\diff E'$; see Sections~\ref{subsec:spectra} and \ref{subsec:IMF}) multiplied by the comoving core-collapse rate density ($R_\mathrm{CC}(z)$; see Section~\ref{subsec:CC_rate}) over the cosmic history \citep[e.g.,][]{2010ARNPS..60..439B}:
\begin{equation}\label{eq:DSNB_general}
\frac{\diff\Phi}{\diff E} = c \int \frac{\diff N_{\mathrm{CC}}}{\diff E'} \frac{\diff E'}{\diff E} R_\mathrm{CC}(z) \left|\frac{\diff t_\mathrm{c}}{\diff z}\right| \diff z~,
\end{equation}
where $c$ is the speed of light,\footnote{Due to their small masses \citep[$\lesssim$\,1eV;][]{2019PhRvL.123v1802A}, neutrinos can be approximated to propagate with the speed of light.} $E'\,\mathord{=}\,(1+z)E$ denotes the energy at the time of emission from sources at redshift~$z$, and the term $|\diff t_\mathrm{c}/\diff z|$ accounts for the assumed cosmological model, which relates $z$ to the cosmic time $t_\mathrm{c}$ (see Section~\ref{subsec:cosmology}).

\subsection{Time-integrated Neutrino Spectra}\label{subsec:spectra}

For each progenitor, we compute the differential number spectrum $\diff\mathcal{N}(t)/\diff E$ (in units of $\mathrm{MeV^{-1}s^{-1}}$) as a function of time $t$ after core-bounce from the time-dependent luminosity, $L(t)\,\mathord{=}\,L_{\nui}(t)$, and mean energy, $\langle E(t) \rangle\,\mathord{=}\,\langle E_{\nui}(t) \rangle$, for all neutrino species ($\nui = \nue, \nuebar, \nux$):
\begin{equation} \label{eq:temp_spec}
\frac{\diff\mathcal{N}(t)}{\diff E} = \frac{L(t)}{\langle E(t) \rangle} \frac{f_\alpha(E)}{\int_{0}^{\infty}\diff E f_\alpha(E)}~,
\end{equation}
where we assume a spectral shape $f_\alpha(E)$ according to \citet{2003ApJ...590..971K},
\begin{equation} \label{eq:spectral_shape}
f_\alpha (E) = \left(\frac{E}{\langle E(t) \rangle}\right)^\alpha \E^{-(\alpha+1)E/\langle E(t)\rangle}~.
\end{equation}
In our models, the shape parameter $\alpha$ of the spectrum is assumed to be constant over time.\footnote{$\alpha\,\mathord{\approx}\,2.3$ corresponds to a Fermi-Dirac distribution with vanishing degeneracy parameter, $\alpha\,\mathord{>}\,2.3$ to a pinched, and $\alpha\,\mathord{<}\,2.3$ to an anti-pinched spectrum; $\alpha\,\mathord{=}\,2.0$ gives the Maxwell-Boltzmann distribution.} Although this is a simplification, sophisticated simulations show that $\alpha$ does not change dramatically with time \citep[e.g.,][]{2012PhRvD..86l5031T, 2016NCimR..39....1M}, justifying this approximation. Instead, we vary $\alpha$ as a free parameter over a range of values ($1\,\mathord{\leqslant}\,\alpha\,\mathord{\leqslant}\,4$), which we motivate in Appendix~\ref{appendix:spectra}.

For each progenitor and neutrino species, we then perform a time-integration over the period of emission, from core bounce at $t\,\mathord{=}\,0$ to a final time of $t\,\mathord{=}\,t_\mathrm{f}$ (with $t_\mathrm{f}\,\mathord{=}\,\unit{15}{s}$ for successful explosions and $t_\mathrm{f}\,\mathord{=}\,t_\mathrm{BH}$ for failed SNe):
\begin{equation} \label{eq:time-integr_spec}
\frac{\diff N}{\diff E} = \frac{\tilde{\xi}}{\xi} \int_{0}^{t_\mathrm{f}}\diff t \:\frac{\diff\mathcal{N}(t)}{\diff E}~.
\end{equation}
Because the luminosities of heavy-lepton neutrinos \nux\ are very approximate in our sets of simulations due to the incomplete microphysics and the relatively moderate core contraction mentioned in Section~\ref{sec:simulation_setup}, we rescale each time-integrated spectrum with a factor $\tilde{\xi}/\xi$ (see Appendix~\ref{appendix:rescaling} for the details). By this procedure we adopt the total radiated neutrino energy (\Enutot) from the simulated core-collapse models, but redistribute them between the different neutrino species with weight factors obtained from SN and BH-formation models with sophisticated neutrino treatment (see Table~\ref{tab:flavor_ratios}). $\tilde{\xi}\,\mathord{=}\,\tilde{\xi}_{\nui}\,\mathord{=}\,(\Enuitot / \Enutot)^\mathrm{new}$ thus constitutes the fraction of the total energy emitted in neutrino species $\nui$. Correspondingly, $\xi\,\mathord{=}\,\xi_{\nui}\,\mathord{=}\,(\Enuitot / \Enutot)^\mathrm{old}$ stands for the relative energy as originally computed in the core-collapse models considered in our study.

In Appendix~\ref{appendix:spectra}, we compare the shapes of our time-integrated spectra with results from sophisticated simulations (with detailed microphysics) by a few exemplary cases to examine the viability of our approximate treatment. We find good agreement with these simulations for values of the instantaneous shape parameter $\alpha$ of $\sim$3 to 3.5 for successful explosions and of $\sim$2 for failed SNe. In Appendix~\ref{appendix:spectral_parameters}, we provide, for a set of representative successful and failed SN models, the total radiated neutrino energies, the mean neutrino energies, and the spectral-shape parameters of the time-integrated neutrino (\nuebar, \nue, and \nux) spectra.

As mentioned in Section~\ref{sec:simulation_setup}, our DSNB flux calculations also include the neutrino signal of the $\unit{8.8}{\msun}$-ECSN simulated by \citet{2010PhRvL.104y1101H}. The corresponding time-integrated spectra are computed according to Equations~\eqref{eq:temp_spec}--\eqref{eq:time-integr_spec}, but with time-dependent shape parameters $\alpha\,\mathord{=}\,\alpha(t)$ as given by the simulation. We use the neutrino data of ``model Sf'', which takes into account the full set of neutrino interactions listed in appendix A of \citet{2006A&A...447.1049B}, including nucleon-nucleon bremsstrahlung, inelastic neutrino-nucleon scattering, and neutrino-pair conversions between different flavors, making rescaling of the spectra unnecessary, i.e., $\tilde{\xi}/\xi\,\mathord{=}\,1$ for all flavors.

\subsection{IMF-weighted Average}\label{subsec:IMF}

The relative number of the pre-SN stars depends on their birth masses. For our DSNB flux predictions, the time-integrated neutrino spectra $\diff N/\diff E$ for each core-collapse case therefore need to be weighted by an IMF (providing the number of stars formed per unit of mass as function of the stellar ZAMS mass $M$). As in \citet{2006ApJ...651..142H}, \citet{2011ApJ...738..154H} and \citet{2014ApJ...790..115M}, we apply the modified Salpeter-A IMF of \citet{2003ApJ...593..258B},
\begin{equation}\label{eq:IMF}
\phi(M) \propto M^{-\zeta}~,
\end{equation}
with $\zeta = 2.35$ for birth masses $M \geqslant \unit{0.5}{\msun}$ and $\zeta = 1.5$ for $\unit{0.1}{\msun} \leqslant M < \unit{0.5}{\msun}$. In our study, we consider masses up to $\unit{125}{\msun}$. However, due to the steep decline of Equation~\eqref{eq:IMF}, the high-mass end is suppressed and thus of minor relevance for the DSNB.

The IMF-weighted neutrino spectrum $\diff N_{\mathrm{CC}}/\diff E$ of all core-collapse events can then be calculated as sum over mass intervals $\Delta M_i$ associated with our discrete set of progenitors stars according to:
\begin{equation}\label{eq:IMF_average}
\frac{\diff N_{\mathrm{CC}}}{\diff E} = \sum_{i} \frac{\int_{\Delta M_i}\diff M\phi(M)}{\int_{\unit{8.7}{\msun}}^{\unit{125}{\msun}}\diff M\phi(M)}\frac{\diff N_i}{\diff E}~,
\end{equation}
where $\Delta M_i$ denotes the mass interval around ZAMS mass $M_i$ with the time-integrated spectrum $\diff N_i/\diff E$ of the corresponding SN, failed-SN, or ECSN simulation.\footnote{We apply $\Delta M_i = [(M_{i-1}+M_i)/2,(M_i+M_{i+1})/2]$ for ZAMS masses $\unit{9.0}{\msun} < M_i < \unit{120}{\msun}$, $\Delta M_i = [\unit{9.0}{\msun},\unit{9.125}{\msun}]$ for the low-mass end ($M_i = \unit{9.0}{\msun}$) and $\Delta M_i = [\unit{110}{\msun},\unit{125}{\msun}]$ for the high-mass end ($M_i = \unit{120}{\msun}$) of our iron-core SN/failed-SN grid. For the $\unit{8.8}{\msun}$-ECSN, we use $\Delta M_i = [\unit{8.7}{\msun},\unit{9.0}{\msun}]$ as relevant range (see Section~\ref{subsec:pre-SN_models}).} Equation~\eqref{eq:IMF_average} is applied separately to the different neutrino species. As in Section~\ref{subsec:spectra}, the indices \nue, \nuebar, and \nux\ are omitted here for the sake of clarity. In the following, we primarily focus on $\nuebar$, since the prospects for a first detection of the DSNB in upcoming detectors are the best for this species \citep[see, e.g.,][]{2004PhRvL..93q1101B, 2006PhRvC..74a5803Y, 2009PhRvD..79h3013H, 2016JPhG...43c0401A}.

\subsection{Cosmic Core-collapse Rate}\label{subsec:CC_rate}

Nuclear burning proceeds fast in massive stars. As a consequence, the progenitors of core-collapse SNe (and failed SNe) have relatively ``short'' ($<$10$^8$\,yr) lives compared to cosmic time scales \citep[cf.][]{1998ARA&A..36..189K}. Therefore, the assumption is well justified that the cosmic core-collapse rate density $R_\mathrm{CC}(z)$ as a function of redshift equals the birth rate density of stars in the relevant ZAMS mass range ($\unit{8.7}{\msun}\,\mathord{\leqslant}\,M\,\mathord{\leqslant}\,\unit{125}{\msun}$), i.e.,
\begin{equation}\label{eq:R_CC}
R_\mathrm{CC}(z) = \psi_\ast(z) \frac{\int_{\unit{8.7}{\msun}}^{\unit{125}{\msun}}\diff M\phi(M)}{\int_{\unit{0.1}{\msun}}^{\unit{125}{\msun}}\diff M M\phi(M)} \simeq \frac{\psi_\ast(z)}{\unit{116}{\msun}}\:.
\end{equation}
Here, $\psi_\ast(z)$ describes the cosmic star-formation history (SFH) in terms of the star-formation rate in units of $\mathrm{\msun Mpc^{-3}yr^{-1}}$, which can be deduced from observations \citep[e.g.,][]{2006ApJ...651..142H, 2008ApJS..175...48R, 2010ApJ...718.1171R} and is thus independent of cosmological assumptions. In our study, we adopt the parametrized description by \citet{2008ApJ...683L...5Y},
\begin{equation}\label{eq:SFH}
\psi_\ast(z) = \dot{\rho}_{0}\left[(1+z)^{\alpha\eta}+\left(\frac{1+z}{B}\right)^{\!\beta\eta}+\left(\frac{1+z}{C}\right)^{\!\gamma\eta}\right]^{\frac{1}{\eta}},
\end{equation}
with the best-fit parameters from \citet{2014ApJ...790..115M}, see table~1 therein. Note that the derivation of a SFH $\psi_\ast(z)$ from observational data requires the use of an IMF, which should be consistent with the one employed in Equation~\eqref{eq:R_CC}. For this reason we use the Salpeter-A IMF \citep{2003ApJ...593..258B} to be consistent with the SFH data sample compiled by \citet{2014ApJ...790..115M}, which is based on the data sets by \citet{2006ApJ...651..142H} and \cite{2011ApJ...738..154H}.\footnote{We point out that \citet{2014ApJ...790..115M} used an equality relation (instead of a proportionality) for Equation~\eqref{eq:IMF}, which leads to a discontinuous behavior at $M\,\mathord{=}\,\unit{0.5}{\msun}$. This seems to be in conflict with the (continuous) IMF employed in the compilations of star-formation-rate data by \citet{2006ApJ...651..142H} and \citet{2011ApJ...738..154H}, which served as a basis for the study of \citet{2014ApJ...790..115M}. For this reason, we construct a continuous IMF by properly choosing the normalization coefficients in the two mass intervals described by Equation~\eqref{eq:IMF}.}

Even though the cosmic core-collapse rate is not yet known to good accuracy (its impact on the DSNB flux is discussed, e.g., by \citealt{2010PhRvD..81h3001L}), our work is focused on variations of the neutrino source properties. To still account for the large uncertainty of \RCC, we additionally employ the $\pm1\sigma$ upper and lower limits to the SFH of \citet{2014ApJ...790..115M}, such that we obtain $\RCC(0)$\,= $\unit{8.93^{+8.24}_{-3.01} \times 10^{-5}}{Mpc^{-3}yr^{-1}}$ for the local universe. In Section~\ref{sec:fiducial_model}, we further test parametrizations of the SFH by \citet{2014MadauDickinson} and the \citet{2018Sci...362.1031F}. The cosmic metallicity evolution and its impact on the DSNB will be discussed briefly in Section~\ref{subsec:remaining_uncertainties}.

\begin{figure*}[ht!]
	\includegraphics[width=\textwidth]{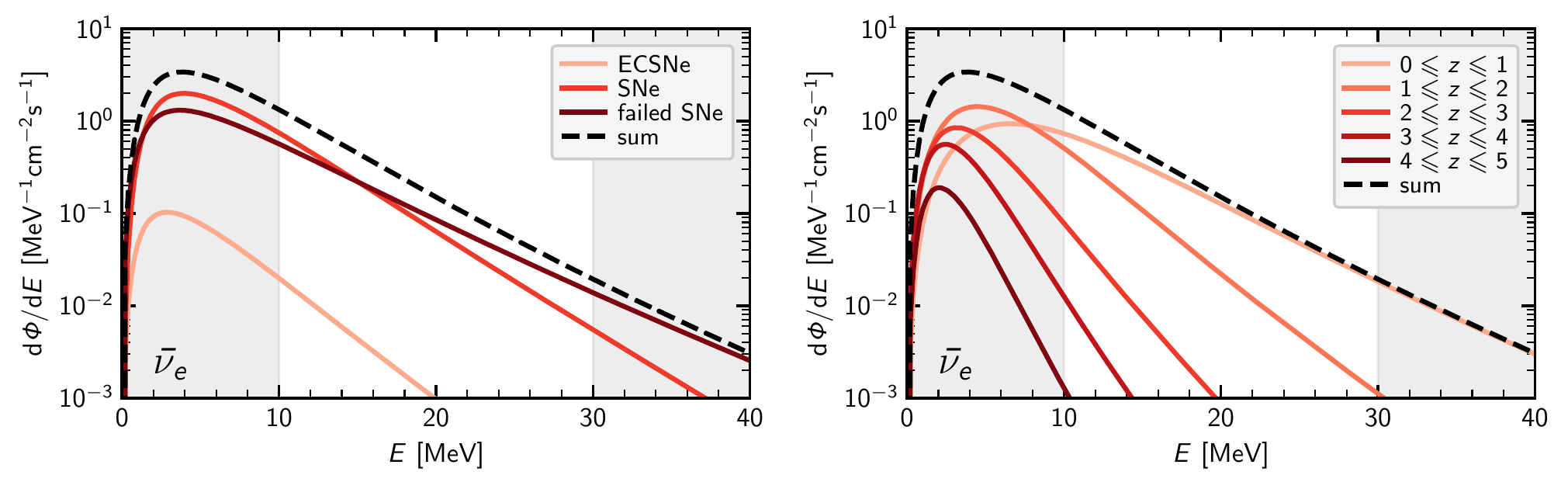}
	\caption{Components of the DSNB flux spectrum, $\diff\Phi/\diff E$, of electron antineutrinos arriving on Earth with energy $E$ for the case of our fiducial model (Z9.6\,\&\,W18; $\MBH\,\mathord{=}\,\unit{2.7}{\msun}$; best-fit $\alpha$). In the left panel, solid lines correspond to the contributions from ECSNe (light), successful iron-core SNe (medium), and failed SNe (dark) to the total DSNB flux (dashed line). The right panel shows the flux originating from different redshift intervals (light to dark for increasing redshift). To guide the eye, the approximate detection window of $\unit{(10-30)}{MeV}$ is bracketed by shaded vertical bands.\label{fig:dsnb_contributions}}
\end{figure*}

For our DSNB calculations, we consider contributions up to a maximum redshift of $z_\mathrm{max}\,\mathord{=}\,5$. This limit is justified because, as pointed out in numerous previous works \citep{2004ApJ...607...20A, 2012PhRvD..85d3011K, 2014ApJ...790..115M, 2015ApJ...804...75N, 2016APh....79...49L}, only sources at lower redshifts ($z\,\mathord{\lesssim}\,1\,\mathord{-}\,2$) noticeably add to the high-energy part of the DSNB, which is most relevant for the detection (cf. Figure~\ref{fig:dsnb_contributions}). Neutrinos from higher $z$ are almost entirely shifted to energies below $\unit{10}{MeV}$, where background sources dominate the flux and thus prevent a clear identification of the DSNB signal \citep[see, e.g.,][]{2016APh....79...49L}.

\subsection{Cosmological Model}\label{subsec:cosmology}

Throughout this work we assume standard $\Lambda$CDM cosmology with the present-day mass-energy density parameters $\Omega_\mathrm{m}\,\mathord{=}\,0.3$ and $\Omega_\mathrm{\Lambda}\,\mathord{=}\,0.7$ of matter and a cosmological constant, respectively, and the Hubble constant $H_0\,\mathord{=}\,\unit{70}{km\unitspace s^{-1}\unitspace Mpc^{-1}}$. The expansion history of the Universe is then given by $\diff z/\diff t_\mathrm{c}$\,= $-H_0(1\,\mathord{+}\,z)\sqrt{\Omega_\mathrm{m}(1\,\mathord{+}\,z)^3\,\mathord{+}\,\Omega_\Lambda}$. Using this together with Equation~\eqref{eq:DSNB_general}, we can write the DSNB flux spectrum (in units of $\mathrm{MeV^{-1}cm^{-2}s^{-1}}$) as
\begin{equation}\label{eq:DSNB}
\frac{\diff\Phi}{\diff E} = \frac{c}{H_0}\int_{0}^{z_\mathrm{max}} \frac{\diff N_{\mathrm{CC}}}{\diff E'} \frac{R_\mathrm{CC}(z)\:\diff z}{\sqrt{\Omega_\mathrm{m}(1+z)^3+\Omega_\mathrm{\Lambda}}}~.
\end{equation}
We do not vary the cosmological assumptions within our work, following most publications on the DSNB topic. For recent studies of the impact of different cosmological models on the DSNB, the reader is referred to \citet{2018JPhG...45e5201B} or \citet{2019PDU....2600397Y}. Having described our computational model with all of its required inputs, we now proceed to the discussion of our results.

\section{Fiducial DSNB Model} \label{sec:fiducial_model}

In this section, we present our fiducial DSNB pre\-dictions and show how the single components (ECSNe, SNe, and failed SNe at various redshifts) contribute to the total flux. The following set of inputs makes up our fiducial model:

\begin{itemize}
\item As in \citet{2016ApJ...818..124E} and \citet{2016ApJ...821...38S}, we take the intermediate engine model Z9.6\,\&\,W18 (with 26.9\% failed SNe) as our reference case.\footnote{The resulting nucleosynthesis yields show a reasonable agreement with the solar element abundances (when type Ia SNe are included); and the NS mass distribution roughly fits observational data \citep{2016ARA&A..54..401O}, as does the distribution of BH masses \citep{2014bsee.confE..37W} if one assumes that only the star's helium core collapses while its hydrogen envelope gets unbound \citep[cf.][]{1980Ap&SS..69..115N, 2013ApJ...769..109L, 2014ApJ...785...28K}. For more details, the reader is referred to \citet{2016ApJ...821...38S}. The rather high fraction of failed explosions (26.9\%; see Table~\ref{tab:explodability}) is not unrealistic given the large discrepancy between the observed SN rate and the SFH \citep{2011ApJ...738..154H}. And also the recent discovery of a disappearing star \citep{2017MNRAS.468.4968A} supports a non-zero fraction of failed explosions. Our weakest engine model, Z9.6\,\&\,W20, which yields by far the largest fraction of failed SNe (41.7\% see Table~\ref{tab:explodability}), is disfavored since it would lead to a significant underproduction of s-process elements \citep{2013ApJ...769...99B, 2016ApJ...821...38S}.}

\item Guided by \citet{2017ApJ...850L..19M}, we assume a fiducial value of $\MBH\,\mathord{=}\,\unit{2.7}{\msun}$ for the NS (baryonic) mass limit, where a PNS is assumed to collapse to a BH and the neutrino signal is truncated (see Section~\ref{subsec:neutrino_signals}).

\item According to our detailed analysis of the spectral shapes in Appendix~\ref{appendix:spectra}, we take a ``best-fit'' value of $\alpha\,\mathord{=}\,3.5$ for the instantaneous spectral-shape parameter of $\nuebar$ for successful SNe with baryonic NS masses of $M_\mathrm{NS,b}\,\mathord{\leqslant}\,\unit{1.6}{\msun}$, of $\alpha\,\mathord{=}\,3.0$ for SNe with $M_\mathrm{NS,b}\,\mathord{>}\,\unit{1.6}{\msun}$, and of $\alphaBH\,\mathord{=}\,2.0$ for the BH-forming, failed explosions.

\item As our reference for the cosmic core-collapse rate, we take Equations~\eqref{eq:R_CC} and \eqref{eq:SFH} with the best-fit parameters for the SFH according to \citet[table~1]{2014ApJ...790..115M}, which yields $\RCC(0)$\,= $\unit{8.93 \times 10^{-5}}{Mpc^{-3}yr^{-1}}$ for the local universe.
\end{itemize}

\begin{deluxetable*}{cccccc}
	\tablecaption{
	DSNB \nuebar-flux contributions.
	\label{tab:dsnb_contributions}}
	\tablehead{
		\colhead{} & \colhead{$\unit{(0-10)}{MeV}$} & \colhead{$\unit{(10-20)}{MeV}$} & \colhead{$\unit{(20-30)}{MeV}$} & \colhead{$\unit{(30-40)}{MeV}$} & \colhead{$\unit{(0-40)}{MeV}$}
	}
	\startdata
	Total DSNB Flux (\nuebar)               & $\unit{22.7}{cm^{-2}s^{-1}}$  & $\unit{5.4}{cm^{-2}s^{-1}}$   & $\unit{0.6}{cm^{-2}s^{-1}}$& $\unit{0.1}{cm^{-2}s^{-1}}$   & $\unit{28.8}{cm^{-2}s^{-1}}$  \\
	\hline
	ECSNe           & $2.6\%$       & $1.2\%$       & $0.5\%$       & $0.2\%$       & $2.3\%$       \\
	Iron-Core SNe           & $57.1\%$      & $51.8\%$      & $37.5\%$      & $23.9\%$      & $55.6\%$      \\
	Failed SNe              & $40.3\%$      & $47.0\%$      & $62.0\%$      & $75.8\%$      & $42.1\%$      \\
	\hline
	$0 \leqslant z \leqslant 1$     & $28.3\%$      & $67.4\%$      & $88.7\%$      & $95.8\%$      & $37.2\%$      \\
	$1 \leqslant z \leqslant 2$     & $40.7\%$      & $29.3\%$      & $11.0\%$      & $4.2\%$       & $37.8\%$      \\
	$2 \leqslant z \leqslant 3$     & $19.0\%$      & $3.1\%$       & $0.3\%$       & $<$\,0.1\%    & $15.6\%$      \\
	$3 \leqslant z \leqslant 4$     & $10.0\%$      & $0.4\%$       & $<$\,0.1\%    & $<$\,0.1\%    & $7.9\%$       \\
	$4 \leqslant z \leqslant 5$     & $2.8\%$       & $<$\,0.1\%    & $<$\,0.1\%    & $<$\,0.1\%    & $2.2\%$
	\enddata
	\tablecomments{Top row: Total DSNB flux of \nuebar\ for our fiducial model (Z9.6\,\&\,W18; $\MBH\,\mathord{=}\,\unit{2.7}{\msun}$; best-fit $\alpha$), integrated over different energy intervals. Second to fourth row: Relative contributions from the various source types (ECSNe/iron-core SNe/failed SNe with BH formation). Rows 5--9: Relative contributions from different redshift intervals (see also Figure~\ref{fig:dsnb_contributions}).}
\end{deluxetable*}

In Figure~\ref{fig:dsnb_contributions}, we first illustrate how the various sources contribute to the total DSNB flux spectrum, $\diff\Phi/\diff E$, of electron antineutrinos, using our fiducial model. The left panel shows the individual fluxes arising from ECSNe, ``conventional'' iron-core SNe, and BH-forming, failed SNe, respectively (light to dark solid lines). Integrated over all energies, ECSNe contribute only 2.3\% ($\unit{0.7}{cm^{-2}s^{-1}}$) to the total flux ($\unit{28.8}{cm^{-2}s^{-1}}$), whose spectrum is shown by a black dashed line. This value is much lower than the $\sim$10\% suggested by \cite{2014ApJ...790..115M} as they assumed a considerably wider ZAMS mass range, $\unit{(8-10)}{\msun}$, compared to $\unit{(8.7-9)}{\msun}$ applied in our work \citep[see][]{2013ApJ...772..150J, 2015MNRAS.446.2599D}. Above $\unit{15}{MeV}$, the contribution of ECSNe accounts for even less than 1\% due to its more rapidly declining spectrum (remember the low mean energy of 11.6\,MeV, as mentioned in Section~\ref{sec:simulation_setup}). However, since the exact mass window of ECSNe is still unclear \citep[see, e.g.,][]{2008ApJ...675..614P, 2013ApJ...772..150J, 2015MNRAS.446.2599D, 2016A&A...593A..72J, 2019PhRvL.123z2701K, 2019ApJ...886...22Z, 2020ApJ...889...34L} and other sources such as ultra\-stripped SNe, AIC, and MIC events might contribute to the DSNB flux with source spectra similar to those of ECSNe, we will consider an enhanced ``low-mass'' component in Section~\ref{subsec:dsnb_LM}.

``Conventional'' iron-core SNe and failed SNe possess comparable integrated fluxes ($\unit{16.0}{cm^{-2}s^{-1}}$ and $12.1$ cm$^{-2}$s$^{-1}$) in case of our fiducial model as shown in Figure~\ref{fig:dsnb_contributions}, yet with distinctly different spectral shapes. Below $\sim$15\,MeV, the contribution from successful explosions is higher, whereas failed explosions dominate the flux at high energies due to their generally harder spectra (see bottom panel of Figure~\ref{fig:neutrino_outcomesystematics}). This was pointed out by previous works \citep[e.g.,][]{2009PhRvL.102w1101L, 2012PhRvD..85d3011K, 2013PhRvD..88h3012N, 2017JCAP...11..031P} and can also be seen in Table~\ref{tab:dsnb_contributions}, where we list the relative flux contributions from the various sources for different ranges of neutrino energies. Between $\unit{20}{MeV}$ and $\unit{30}{MeV}$, failed SNe account for 62\% of the total flux (at still higher energies, even 76\%). Naturally, these numbers (here given for our reference model set) depend strongly on the fraction of failed explosions and their neutrino emission (see Section~\ref{subsec:dsnb_parameter_study}). Compared to previous studies, we obtain a generally increased DSNB flux, advantageous for its imminent detection. We will comment on this issue more thoroughly below.

In the right panel of Figure~\ref{fig:dsnb_contributions}, we compare the DSNB contributions from different redshift intervals (light to dark for increasing redshift). At high energies ($\unit{\gtrsim20}{MeV}$), the flux mainly originates from sources below $z\sim1$, as it was illustrated in several previous works \citep{2004ApJ...607...20A, 2012PhRvD..85d3011K, 2014ApJ...790..115M, 2015ApJ...804...75N, 2016APh....79...49L}. Only at lower energies, the contribution from large redshifts gets increasingly important (see Table~\ref{tab:dsnb_contributions}). In both panels of Figure~\ref{fig:dsnb_contributions}, shaded bands bracket the approximate energy window of $\sim$(10--30)\,MeV which is most relevant for the DSNB detection in upcoming neutrino observatories. Beyond that, background sources (such as reactor and solar neutrinos at low energies and atmospheric neutrinos at high energies) dominate the flux and make the DSNB measurement unfeasible \citep[see, e.g., review by][]{2016APh....79...49L}.

\begin{figure}
	\includegraphics[width=\columnwidth]{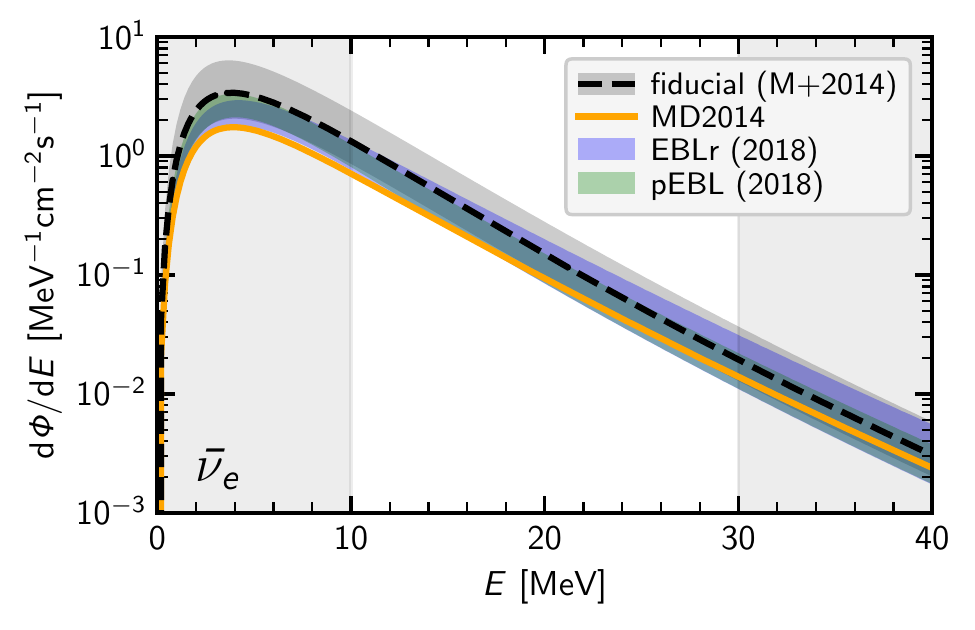}
	\caption{Dependence of the DSNB \nuebar-flux spectrum on the assumed parametrization of the cosmic SFH. In our fiducial model (black dashed line, cf. Figure~\ref{fig:dsnb_contributions}), the SFH of \citet{2014ApJ...790..115M} is employed; the gray shaded band corresponds to their $\pm1\sigma$ upper and lower limits. The orange line indicates the DSNB spectrum for the SFH of \citet{2014MadauDickinson}, whereas the DSNB spectrum for the SFH from the \citet{2018Sci...362.1031F} is given by blue (empirical EBL reconstruction) and green (physical EBL model) shaded bands ($1\sigma$ confidence regions). Note that a conventional Salpeter IMF \citep{1955ApJ...121..161S} and the one by \citet{2003PASP..115..763C} are used (instead of the Salpeter-A IMF from \citealt{2003ApJ...593..258B}) for the conversion of the SFH to the cosmic core-collapse rate (Equation~\eqref{eq:R_CC}), when the SFHs from \citet{2014MadauDickinson} or the \citet{2018Sci...362.1031F} are used, respectively (see main text). As in Figure~\ref{fig:dsnb_contributions}, vertical bands frame the approximate detection window.\label{fig:dsnb_SFR}}
\end{figure}

As already pointed out in Section~\ref{subsec:CC_rate}, the cosmic SFH constitutes one of the major uncertainties in predicting the DSNB. Before we proceed to the main part of our parameter study, we thus test how the DSNB flux spectrum depends on the assumed parametrization of the SFH. In Figure~\ref{fig:dsnb_SFR}, we show our fiducial DSNB model (black dashed line) together with the uncertainty corresponding to the $\pm1\sigma$ confidence interval of the SFH according to \citet[``M+2014''; gray shaded band]{2014ApJ...790..115M}. For comparison, we also employ the more conservative SFH from \citet[][``MD2014''; equation~(15)]{2014MadauDickinson} (orange line) as well as the recent results of the \citet{2018Sci...362.1031F} on the evolution of the extragalactic background light (EBL): an empirical EBL reconstruction (EBLr) and a physical EBL (pEBL) model (blue and green shaded bands, respectively; see their figure~3). We already noted in Section~\ref{subsec:CC_rate} that, to remain consistent with the IMF employed for determining the SFH $\psi_\ast(z)$, the same IMF should be taken also for the conversion of $\psi_\ast(z)$ to the cosmic core-collapse rate density $\RCC(z)$. For this reason, we adopt a conventional Salpeter IMF \citep{1955ApJ...121..161S} or the one by \citet{2003PASP..115..763C} in the cases of using the SFHs from \citet{2014MadauDickinson} or from the \citet{2018Sci...362.1031F}, respectively. Equation~\eqref{eq:R_CC} becomes $\RCC(z)\,\mathord{=}\,\psi_\ast(z)/\unit{151}{\msun}$ in the former case and $\RCC(z)\,\mathord{=}\,\psi_\ast(z)/\unit{95}{\msun}$ in the latter case. Notice the wide spread of the resulting DSNB flux spectra in Figure~\ref{fig:dsnb_SFR}, especially at low energies \citep[cf.][]{2020arXiv200702951R}. Throughout our work, we will assume an uncertainty of the cosmic core-collapse rate corresponding to the $\pm1\sigma$ band of \citet{2014ApJ...790..115M}.

Since our overall findings apply similarly for all neutrino species, we constrain our discussion to electron antineutrinos for now. In Section~\ref{sec:nue}, we will briefly discuss the DSNB flux spectrum of electron neutrinos and, in Section~\ref{subsec:flavor_conversions}, we will comment on the influence of heavy-lepton neutrinos in the context of neutrino flavor oscillation effects.

\section{DSNB Parameter Study} \label{sec:parameter_study}

In this section, we present the results of our detailed DSNB parameter study. Using large grids of long-time neutrino signals (see Section~\ref{sec:simulation_setup}), we probe the sensitivity of the DSNB to three critical source properties (in Section~\ref{subsec:dsnb_parameter_study}): the fraction of failed explosions (by means of our different engine models), the threshold mass for BH formation, and the spectral shape of the neutrino emission from failed explosions. Moreover, the possible enhancement of the DSNB by an additional generic ``low-mass'' component is explored (Section~\ref{subsec:dsnb_LM}) as well as the effect of including binary progenitor models (Section~\ref{subsec:dsnb_binary}).

\subsection{DSNB Parameter Dependence}\label{subsec:dsnb_parameter_study}

\begin{figure*}[ht!]
	\includegraphics[width=\textwidth]{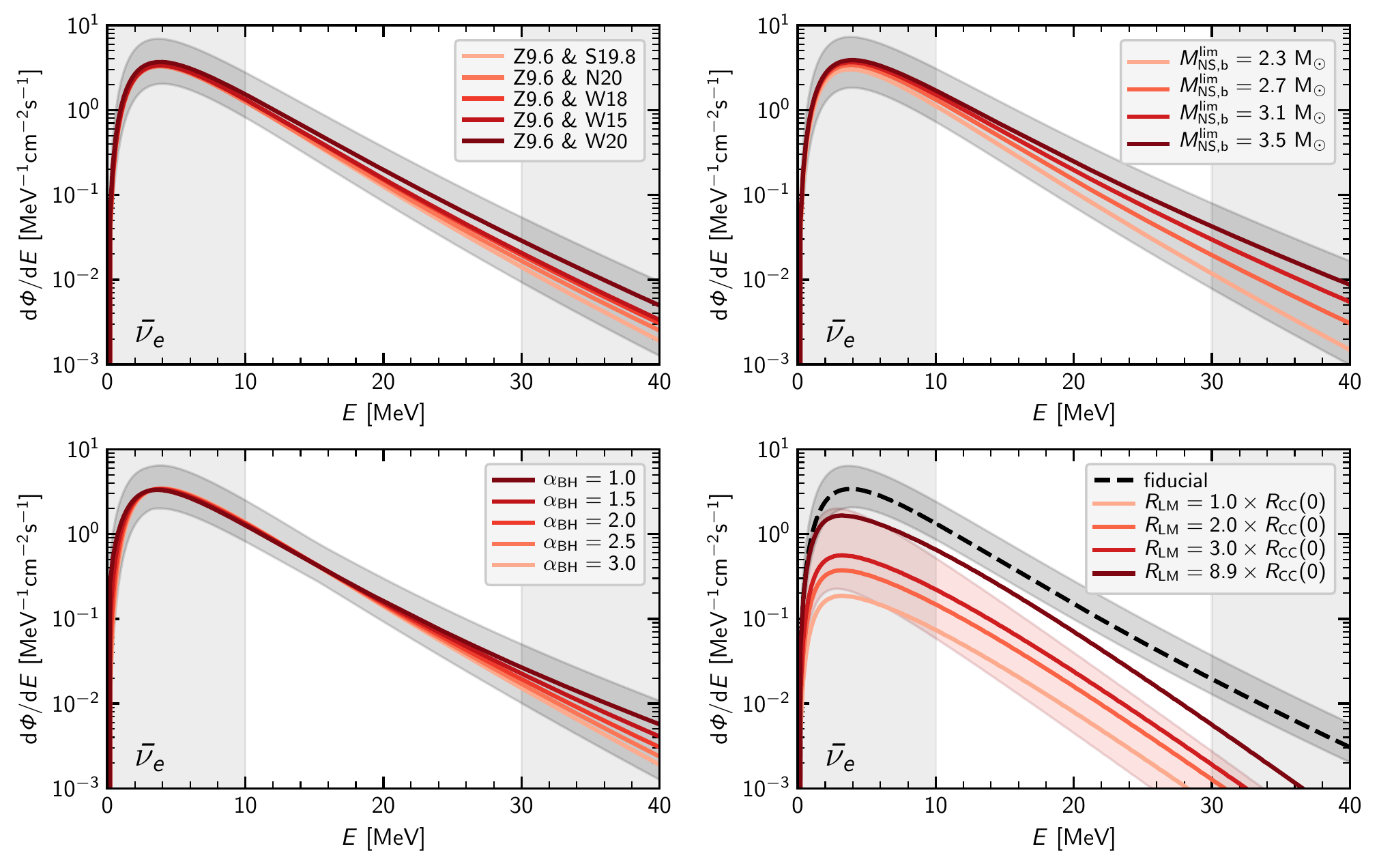}
	\caption{Parameter dependence of the DSNB flux spectrum, $\diff\Phi/\diff E$, for the case of electron antineutrinos. In the different panels the engine models (upper left panel), the NS mass limit for BH formation (upper right panel), and the instantaneous spectral-shape parameter, \alphaBH, of the time-dependent neutrino emission from BH-formation events (lower left panel) are varied, while keeping all other parameters at their reference values (Z9.6\,\&\,W18; $\MBH\,\mathord{=}\,\unit{2.7}{\msun}$; best-fit $\alpha$, i.e. $\alpha$\,=\,3.5 for SNe with $M_\mathrm{NS,b}\,\mathord{\leqslant}\,\unit{1.6}{\msun}$, $\alpha$\,=\,3.0 for those with $M_\mathrm{NS,b}\,\mathord{>}\,\unit{1.6}{\msun}$, and $\alphaBH$\,=\,2.0 for failed SNe; see Section~\ref{sec:fiducial_model}). In the lower right panel, the additional contribution from low-mass (LM) NS-forming events is shown for different constant rate densities \RLM. For comparison, the pale red band marks the LM flux for an evolving rate instead (see main text for details). Our fiducial model with $\RLM\,\mathord{=}\,0$ is plotted as dashed line. In each panel, a gray shaded band indicates the uncertainty arising from the cosmic core-collapse rate \citep[corresponding to the $\pm1\sigma$ upper and lower limits to the SFH of][]{2014ApJ...790..115M}. As in Figure~\ref{fig:dsnb_contributions}, vertical bands frame the approximate detection window.\\
	\label{fig:dsnb_parameter_study}
	}
\end{figure*}

First, we study the impact of our engine model (as described in Section~\ref{subsec:SN_simulations}) on the DSNB flux spectrum. In the upper left panel of Figure~\ref{fig:dsnb_parameter_study}, we show $\diff\Phi/\diff E$ for the various choices of central neutrino engines for our simulations. Sets with a higher percentage of failed explosions (see Figure~\ref{fig:explodability} and Table~\ref{tab:explodability}) yield an enhanced DSNB flux, especially in the high-energy regime. This overall picture is in line with the studies by \citet{2009PhRvL.102w1101L}, \citet{2010PhRvD..81h3001L} and \citet{2012PhRvD..85d3011K}, who varied the fraction of BH-forming collapses while applying generic neutrino spectra and thus neglecting progenitor dependences. More recently, \citet{2017JCAP...11..031P} and \citet{2018JCAP...05..066M} examined the fraction of failed SNe by assuming different ZAMS mass distributions, while \cite{2018MNRAS.475.1363H}, for the first time, employed a larger sample of simulations including seven BH-formation cases, thus taking into account progenitor-dependent variations in the neutrino emission from failed explosions (by linearly interpolating the total energetics, mean energy, and shape parameter of their time-integrated neutrino spectra as a function of the compactness parameter of \citealt{2011ApJ...730...70O}; see footnote~\ref{fn:compactness}). They explored relative fractions of BH-formation cases between 0\% and 45\% by taking different threshold values for the compactness above which they assumed their progenitors to form BHs.

Using our large sets of long-time simulations without predefined outcome (also resulting in BH formation of less compact progenitors with low mass-accretion rates), we can confirm the common result of the previous studies: the larger the fraction of failed explosions, the stronger the enhancement of the DSNB at high energies. To better quantify this behavior, we follow \citet{2007PhRvD..75g3022L} and fit the high-energy tail ($\unit{20}{MeV}\,\mathord{\leqslant}\,E\,\mathord{\leqslant}\,\unit{30}{MeV}$) of our DSNB flux spectra with an exponential function:
\begin{equation}\label{eq:exponential_dsnb_fit}
\frac{\diff\Phi}{\diff E} \simeq \phi_0 \:\E^{-E/E_0}\:.
\end{equation}
Our model set Z9.6\,\&\,S19.8 with the lowest fraction of failed explosions (17.8\%) features the steepest decline (i.e. $E_0\,\mathord{=}\,\unit{4.5}{MeV}$), while Z9.6\,\&\,W20 with 41.7\% BH-formation cases yields a flatter spectrum with $E_0$\,=\,$\unit{5.1}{MeV}$. The ``normalization'' $\phi_0$, on the other hand, is hardly affected by the choice of our engine model. Instead, it is determined by the uncertainty arising from the cosmic core-collapse rate, which shifts the entire flux spectrum vertically without changing the slope by more than $\sim$1\%.\footnote{\label{fn:E0_RCC}The fact that $E_0$ is not entirely unaffected by changes of \RCC\ is due to different functional dependences of the $\pm1\sigma$ upper/lower limits to the cosmic SFH on the redshift~$z$ (see table~1 of \citealt{2014ApJ...790..115M}).} The gray shaded bands in Figure~\ref{fig:dsnb_parameter_study} indicate this severe normalization uncertainty (the $+1\sigma$ upper limit to the SFH of \citet{2014ApJ...790..115M} is taken for our highest-flux, the $-1\sigma$ lower limit for our lowest-flux model). The aspect that the failed-SN fraction is likely to exhibit a dependence on metallicity (and thus redshift) was pointed out by \citet{2015ApJ...804...75N} and \citet{2015PhLB..751..413Y}. We will come back to this point in Section~\ref{subsec:remaining_uncertainties}.

The impact of the NS mass limit on the DSNB has been discussed in the literature to some extent \citep{2009PhRvL.102w1101L, 2012PhRvD..85d3011K, 2015ApJ...804...75N, 2016ApJ...827...85H, 2018ApJ...869...31H}. Commonly, the spectra from exemplary simulations of BH formation with two different EoSs were compared: the stiff Shen EoS \citep[][with incompressibility $K\,\mathord{=}\,\unit{281}{MeV}$]{1998NuPhA.637..435S} and a softer EoS by \citet[``LS180'' or ``LS220'', with $K\,\mathord{=}\,\unit{180}{MeV}$ or $K\,\mathord{=}\,\unit{220}{MeV}$]{1991NuPhA.535..331L}. Generally, a stiff EoS supports the transiently existing PNS of a failed SN against gravity up to a higher limiting mass than a soft EoS does. The final collapse to a BH therefore sets in after a longer period of mass accretion and neutrino emission with the consequence of higher spectral temperatures and an enhanced contribution to the DSNB flux.

Having a large compilation of long-time simulations at hand, we take a different (more rigorous) approach in our work: As described in Section~\ref{sec:simulation_setup}, we directly vary the maximum baryonic NS mass, \MBH, without applying a certain EoS. Our neutrino signals from failed explosions are then truncated when the mass accretion from the collapsing progenitor star pushes the PNS mass beyond this critical threshold for BH formation. In the upper right panel of Figure~\ref{fig:dsnb_parameter_study}, we show the DSNB flux spectra for our different choices of \MBH. Raising the NS mass limit from $\unit{2.3}{\msun}$ to $\unit{3.5}{\msun}$ drastically enhances the flux at higher energies, thus lifting the value of the slope parameter, $E_0$/MeV (see Equation~\eqref{eq:exponential_dsnb_fit}), from 4.4 to 5.6. This strong effect becomes immediately clear from Figure~\ref{fig:neutrino_outcomesystematics}: A higher NS mass limit leads to enhanced time-integrated neutrino luminosities and generally hotter spectra, in line with the studies by \citet{2009PhRvL.102w1101L}, \citet{2012PhRvD..85d3011K}, \citet{2015ApJ...804...75N}, and \citet{2016ApJ...827...85H, 2018ApJ...869...31H}.

We should mention that our study does not consider the possibility of a progenitor-dependent threshold mass for BH formation. \citet{2011ApJ...730...70O} pointed out that thermal pressure support may be stronger for stars with high core compactness, lifting the maximum PNS mass to somewhat larger values. This might slightly reduce differences in the neutrino emission between individual progenitors. The results of Figure~\ref{fig:dsnb_parameter_study} should, however, remain essentially unchanged, because thermal stabilization of the PNS should be most relevant when the mass-accretion rate is high and the PNS becomes very hot. In such cases, however, the critical limit for BH formation is also reached quickly and the neutrino emission is not very extended. The wide range of values for \MBH\ considered in our study should include the true NS mass limit, which depends on the still incompletely known high-density EoS of NS matter.  Once the latter is better constrained by astrophysical observations and nuclear experiments and theory, and thus the maximum mass of cold NS is better constrained, the question of a progenitor-dependent thermal effect on the transient PNS stabilization can be addressed more thoroughly.

\begin{deluxetable*}{lrr}
	\tablecaption{Exponential-fit parameters of Equation~\eqref{eq:exponential_dsnb_fit} for a subset of our DSNB models.\label{tab:exp_fit}}
	\tablehead{
		\colhead{Model} & \colhead{$\phi_0$ [$\mathrm{MeV^{-1}cm^{-2}s^{-1}}$]} & \colhead{$E_0$ [MeV]}
	}
	\startdata
	W18-BH2.7-$\alpha$2.0 (fiducial) & $9.4^{+7.5}_{-3.3}$ ($7.3^{+5.9}_{-2.6}$, $6.7^{+1.7}_{-1.4}$, $4.4$) & $4.82^{+0.04}$ ($4.84^{+0.04}$, $5.1^{+0.2}_{-0.3}$, $5.2$) \\
	W20-BH3.5-$\alpha$1.0 (max.) & $6.6^{+5.3}_{-2.4}$ ($4.8^{+3.8}_{-1.7}$, $4.9^{+1.3}_{-1.0}$, $3.3$) & $6.79^{+0.05}$ ($6.46^{+0.05}$, $7.1^{+0.2}_{-0.3}$, $7.1$) \\
	S19.8-BH2.3-$\alpha$3.0 (min.) & $12.6^{+10.3}_{-4.3}$ ($9.6^{+7.7}_{-3.3}$, $8.7^{+2.1}_{-1.7}$, $5.6$) & $4.09^{+0.03}$ ($4.32^{+0.04}$, $4.4^{+0.2}_{-0.3}$, $4.4$) \\
	\hline
	S19.8-BH2.7-$\alpha$2.0 & $10.6^{+8.6}_{-3.7}$ ($8.5^{+6.9}_{-3.0}$, $7.5^{+1.9}_{-1.6}$, $4.9$) & $4.51^{+0.04}$ ($4.60^{+0.04}$, $4.8^{+0.2}_{-0.3}$, $4.9$) \\
	N20-BH2.7-$\alpha$2.0 & $9.3^{+7.5}_{-3.2}$ ($7.4^{+6.0}_{-2.6}$, $6.6^{+1.7}_{-1.4}$, $4.3$) & $4.71^{+0.04}$ ($4.75^{+0.04}$, $5.0^{+0.2}_{-0.3}$, $5.1$) \\
	W15-BH2.7-$\alpha$2.0 & $9.1^{+7.3}_{-3.2}$ ($7.0^{+5.6}_{-2.4}$, $6.5^{+1.6}_{-1.3}$, $4.2$) & $4.90^{+0.04}$ ($4.90^{+0.05}$, $5.2^{+0.2}_{-0.3}$, $5.3$) \\
	W20-BH2.7-$\alpha$2.0 & $9.6^{+7.6}_{-3.4}$ ($6.7^{+5.3}_{-2.3}$, $6.8^{+1.6}_{-1.3}$, $4.4$) & $5.13^{+0.05}$ ($5.14^{+0.05}$, $5.5^{+0.3}_{-0.3}$, $5.5$) \\
	\hline
	W18-BH2.3-$\alpha$2.0 & $9.9^{+8.0}_{-3.4}$ ($7.7^{+6.2}_{-2.7}$, $7.0^{+1.7}_{-1.4}$, $4.5$) & $4.43^{+0.04}$ ($4.57^{+0.04}$, $4.7^{+0.2}_{-0.3}$, $4.8$) \\
	W18-BH2.7-$\alpha$2.0 & $9.4^{+7.5}_{-3.3}$ ($7.3^{+5.9}_{-2.6}$, $6.7^{+1.7}_{-1.4}$, $4.4$) & $4.82^{+0.04}$ ($4.84^{+0.04}$, $5.1^{+0.2}_{-0.3}$, $5.2$) \\
	W18-BH3.1-$\alpha$2.0 & $9.0^{+7.1}_{-3.1}$ ($6.9^{+5.5}_{-2.4}$, $6.4^{+1.6}_{-1.3}$, $4.3$) & $5.22^{+0.05}$ ($5.14^{+0.05}$, $5.5^{+0.2}_{-0.3}$, $5.6$) \\
	W18-BH3.5-$\alpha$2.0 & $8.7^{+6.9}_{-3.1}$ ($6.6^{+5.3}_{-2.3}$, $6.3^{+1.6}_{-1.3}$, $4.2$) & $5.59^{+0.05}$ ($5.44^{+0.05}$, $5.9^{+0.2}_{-0.3}$, $6.0$) \\
	\hline
	W18-BH2.7-$\alpha$1.0 & $6.3^{+5.1}_{-2.2}$ ($5.5^{+4.5}_{-1.9}$, $4.7^{+1.4}_{-1.1}$, $3.2$) & $5.44^{+0.04}$ ($5.25^{+0.04}$, $5.7^{+0.2}_{-0.3}$, $5.7$) \\
	W18-BH2.7-$\alpha$1.5 & $7.8^{+6.3}_{-2.7}$ ($6.5^{+5.2}_{-2.3}$, $5.7^{+1.5}_{-1.3}$, $3.8$) & $5.09^{+0.04}$ ($5.02^{+0.04}$, $5.4^{+0.2}_{-0.3}$, $5.4$) \\
	W18-BH2.7-$\alpha$2.0 & $9.4^{+7.5}_{-3.3}$ ($7.3^{+5.9}_{-2.6}$, $6.7^{+1.7}_{-1.4}$, $4.4$) & $4.82^{+0.04}$ ($4.84^{+0.04}$, $5.1^{+0.2}_{-0.3}$, $5.2$) \\
	W18-BH2.7-$\alpha$2.5 & $10.9^{+8.7}_{-3.8}$ ($8.1^{+6.5}_{-2.8}$, $7.6^{+1.8}_{-1.5}$, $5.0$) & $4.62^{+0.04}$ ($4.70^{+0.04}$, $4.9^{+0.2}_{-0.3}$, $5.0$) \\
	W18-BH2.7-$\alpha$3.0 & $12.3^{+9.9}_{-4.3}$ ($8.9^{+7.1}_{-3.1}$, $8.5^{+2.0}_{-1.6}$, $5.5$) & $4.46^{+0.04}$ ($4.58^{+0.04}$, $4.8^{+0.2}_{-0.3}$, $4.9$) \\
	\hline
	W18-BH2.7-$\alpha$2.0-He33 & $8.0^{+6.4}_{-2.8}$ ($6.3^{+5.1}_{-2.2}$, $5.7^{+1.4}_{-1.2}$, $3.7$) & $4.76^{+0.04}$ ($4.79^{+0.04}$, $5.1^{+0.2}_{-0.3}$, $5.1$) \\
	W18-BH2.7-$\alpha$2.0-He100 & $5.5^{+4.4}_{-1.9}$ ($4.4^{+3.5}_{-1.5}$, $4.0^{+1.0}_{-0.8}$, $2.5$) & $4.45^{+0.04}$ ($4.59^{+0.04}$, $4.7^{+0.2}_{-0.3}$, $4.8$) \\
	\hline
	S19.8-BH2.3-$\alpha$2.0 & $11.1^{+9.1}_{-3.8}$ ($8.8^{+7.1}_{-3.0}$, $7.8^{+1.9}_{-1.6}$, $5.0$) & $4.23^{+0.03}$ ($4.42^{+0.04}$, $4.5^{+0.2}_{-0.3}$, $4.6$) \\
	W18-BH3.5-$\alpha$1.0 & $5.8^{+4.7}_{-2.1}$ ($4.9^{+4.0}_{-1.7}$, $4.4^{+1.3}_{-1.1}$, $3.0$) & $6.38^{+0.04}$ ($5.96^{+0.04}$, $6.6^{+0.2}_{-0.3}$, $6.7$) \\
	W15-BH3.5-$\alpha$1.0 & $5.7^{+4.6}_{-2.1}$ ($4.7^{+3.8}_{-1.7}$, $4.3^{+1.3}_{-1.0}$, $2.9$) & $6.49^{+0.04}$ ($6.08^{+0.04}$, $6.8^{+0.2}_{-0.3}$, $6.8$) \\
	W20-BH2.7-$\alpha$1.0 & $6.2^{+4.9}_{-2.2}$ ($4.8^{+3.8}_{-1.7}$, $4.5^{+1.2}_{-1.0}$, $3.1$) & $5.94^{+0.05}$ ($5.73^{+0.04}$, $6.2^{+0.2}_{-0.3}$, $6.3$) \\
	W20-BH3.1-$\alpha$1.0 & $6.3^{+5.1}_{-2.3}$ ($4.7^{+3.8}_{-1.7}$, $4.7^{+1.3}_{-1.0}$, $3.2$) & $6.39^{+0.05}$ ($6.12^{+0.05}$, $6.7^{+0.2}_{-0.3}$, $6.7$) \\
	W20-BH3.5-$\alpha$2.0 & $10.2^{+7.9}_{-3.6}$ ($6.7^{+5.3}_{-2.4}$, $7.2^{+1.7}_{-1.3}$, $4.8$) & $5.86^{+0.06}$ ($5.77^{+0.06}$, $6.2^{+0.3}_{-0.4}$, $6.3$)
	\enddata
	\tablecomments{The fits are applied in the energy region $\unit{20}{MeV}\,\mathord{\leqslant}\,E\,\mathord{\leqslant}\,\unit{30}{MeV}$. The listed values correspond to the unoscillated \nuebar\ DSNB flux spectra using the SFH from \citet{2014ApJ...790..115M} with its associated $\pm1\sigma$ uncertainty. In parentheses, the values for the case of a complete flavor swap ($\nuebar \leftrightarrow \nux$) are provided as well as the results for a SFH according to the EBL reconstruction model by the \citet{2018Sci...362.1031F} and for the SFH of \cite{2014MadauDickinson}. The one-sided error intervals of $E_0$ in the cases with the SFH from \cite{2014ApJ...790..115M} are caused by the fact that the functional fits to the SFH scale slightly differently with redshift (see footnote~\ref{fn:E0_RCC}), with the best-fit case by \cite{2014ApJ...790..115M} yielding the largest relative contribution from high-redshift regions and thus smallest value of $E_0$ compared to both the +1$\sigma$ and the $-$1$\sigma$ limits.
	}
\end{deluxetable*}

In our study, the spectral shape of the time-dependent neutrino emission is assumed to obey Equation~\eqref{eq:spectral_shape} with a constant shape parameter $\alpha$. Following our detailed analysis of the spectral shapes in Appendix~\ref{appendix:spectra}, we show in the lower left panel of Figure~\ref{fig:dsnb_parameter_study} how the DSNB flux spectrum changes when different values (between 1.0 and 3.0) of this instantaneous spectral-shape parameter $\alpha\,\mathord{=}\,\alphaBH$ are taken for the emission from failed explosions. For successful SNe, $\alpha$ is not varied but kept constant at the best-fit values of 3.0 and 3.5 (for $M_\mathrm{NS,b}\,\mathord{>}\,\unit{1.6}{\msun}$ and $M_\mathrm{NS,b}\,\mathord{\leqslant}\,\unit{1.6}{\msun}$, respectively). Similar to its influence on the individual failed-SN source spectra, a small value of $\alphaBH$ also broadens the shape of the DSNB such that its high-energy tail gets lifted relative to the peak \citep[cf.][]{2003ApJ...590..971K, 2007PhRvD..75g3022L, 2016APh....79...49L}. For $\alphaBH\,\mathord{=}\,1.0$ (i.e., antipinched failed-SN source spectra), the exponential fit of Equation~\eqref{eq:exponential_dsnb_fit} yields $E_0\,\mathord{=}\,\unit{5.4}{MeV}$ and $\phi_0\,\mathord{=}\,\unit{6.3}{MeV^{-1}cm^{-2}s^{-1}}$ in the range of neutrino energies $\unit{20}{MeV}\,\mathord{\leqslant}\,E\,\mathord{\leqslant}\,\unit{30}{MeV}$. Choosing $\alphaBH\,\mathord{=}\,3.0$, on the other hand, results in a more prominent peak at the cost of a suppressed flux at high energies ($E_0\,\mathord{=}\,\unit{4.5}{MeV}$; $\phi_0\,\mathord{=}\,\unit{12.3}{MeV^{-1}cm^{-2}s^{-1}}$). Because the instantaneous spectral-shape parameter $\alpha$ is only varied for failed SNe while the contribution from successful SNe is unchanged, a slight ``kink'' gets visible in the overall DSNB flux spectrum for the cases of small $\alphaBH$, unveiling its ``two-component'' nature. Notice the crossings of the different curves at $\sim$3\,MeV and $\sim$15\,MeV. Accordingly, we construct the shaded band for the uncertainty of \RCC\ such that the lowest-flux and highest-flux models are considered in each segment.

In Table~\ref{tab:exp_fit}, we provide an overview of the two fit parameters $\phi_0$ and $E_0$ for all models discussed in this section. We use the following naming convention for our DSNB models: ``W18-BH2.7-$\alpha$2.0'' corresponds to our fiducial model with the Z9.6\,\&\,W18 neutrino engine (``W18''), with a baryonic NS mass limit of $\MBH\,\mathord{=}\,\unit{2.7}{\msun}$ (``BH2.7''), and with the best-fit choice for the instantaneous spectral-shape parameter (``$\alpha$2.0''; i.e., $\alphaBH\,\mathord{=}\,2.0$). The two models ``W20-BH3.5-$\alpha$1.0'' and ``S19.8-BH2.3-$\alpha$3.0'', which employ the most extreme parameter combinations, yield the largest or smallest slope parameters $E_0$ of all our models and thus the highest or lowest fluxes at high energies, respectively.

\subsection{Additional Low-mass Component}\label{subsec:dsnb_LM}

As we mentioned in Section~\ref{subsec:pre-SN_models}, the low-mass range of core-collapse SN progenitors is rather uncertain. It is widely believed that in degenerate ONeMg cores electron-capture reactions on $^{20}$Ne and $^{24}$Mg can win against the effects of oxygen deflagration, initiating the collapse to a NS rather than thermonuclear runaway \citep{1980PASJ...32..303M, 1984ApJ...277..791N, 1987ApJ...322..206N}. Nevertheless, the conditions for such an ECSN to occur in Nature are still discussed controversially \citep[see, e.g.,][]{2016A&A...593A..72J, 2019PhRvL.123z2701K, 2019ApJ...886...22Z, 2020ApJ...889...34L}. Moreover, observations suggest that most massive stars are in binary systems \citep[see, e.g.,][]{2009AJ....137.3358M, 2012Sci...337..444S}, and evolution in binaries might lead to a larger population of degenerate ONeMg cores which produce ESCNe \citep{2004ApJ...612.1044P}.

In addition to these uncertain SN progenitors, three other channels are discussed that may lead to preferentially rather low-mass NSs, whose formation might contribute to the DSNB:  Electron-capture initiated collapse may also occur when an ONeMg WD is pushed beyond the Chandrasekhar mass limit due to Roche-lobe overflow from a companion. Such a NS-forming event is referred to as AIC \citep[see, e.g.,][]{1990ApJ...353..159B, 1991ApJ...367L..19N, 2004ApJ...601.1058I, 2010MNRAS.402.1437H, 2016A&A...593A..72J, 2018RAA....18...36W, 2019MNRAS.484..698R}. Similarly, \citet{1985A&A...150L..21S} suggested the MIC of two WDs as another possible scenario to form a single NS \citep[also see][]{2008MNRAS.386..553I, 2016MNRAS.463.3461S, 2019MNRAS.484..698R}. Moreover, close-binary interaction might in some cases lead to the stripping of a star's hydrogen and (most of its) helium envelope onto a companion NS, leaving behind a bare carbon-oxygen core \citep{1994Natur.371..227N, 2002MNRAS.331.1027D}, undergoing subsequent iron-core collapse. The explosion of such ultra\-stripped SNe \citep{2013ApJ...778L..23T, 2015MNRAS.451.2123T, 2015MNRAS.454.3073S, 2018MNRAS.479.3675M} is discussed as the most likely evolutionary pathway leading to the formation of double NS systems \citep{2017ApJ...846..170T, 2020arXiv200703890M}.

Previous works \citep[e.g., by][]{2014ApJ...790..115M, 2018MNRAS.475.1363H} considered the contribution from ECSNe to the DNSB flux and, in a footnote, \citet{2010PhRvD..81h3001L} already mentioned that, to a minor degree, also neutrinos from the AIC of WDs might add to the DSNB.

\begin{deluxetable*}{cccccc}
	\tablecaption{
	DSNB contribution from additional low-mass NS-formation events.
	\label{tab:dsnb_LM_component}}
	\tablehead{
		\colhead{} & \colhead{$\unit{(0-10)}{MeV}$} & \colhead{$\unit{(10-20)}{MeV}$} & \colhead{$\unit{(20-30)}{MeV}$} & \colhead{$\unit{(30-40)}{MeV}$} & \colhead{$\unit{(0-40)}{MeV}$}
	}
	\startdata
	Fiducial DSNB Flux (\nuebar), $R_\mathrm{LM} = 0$ & $\unit{22.7}{cm^{-2}s^{-1}}$ & $\unit{5.4}{cm^{-2}s^{-1}}$ & $\unit{0.6}{cm^{-2}s^{-1}}$ & $\unit{0.1}{cm^{-2}s^{-1}}$ & $\unit{28.8}{cm^{-2}s^{-1}}$ \\
	\hline
	$R_\mathrm{LM} = 1.0 \times R_\mathrm{CC}(0)$, $\chi=0.11$ & 5.7\%  (6.1\%)  & 5.7\%  (4.0\%)  & 4.7\%  (2.5\%)  & 2.6\%  (1.2\%)  & 5.6\%  (5.6\%)  \\
	$R_\mathrm{LM} = 2.0 \times R_\mathrm{CC}(0)$, $\chi=0.23$ & 11.3\% (12.2\%) & 11.3\% (7.9\%)  & 9.3\%  (5.0\%)  & 5.1\%  (2.4\%)  & 11.2\% (11.2\%) \\
	$R_\mathrm{LM} = 3.0 \times R_\mathrm{CC}(0)$, $\chi=0.34$ & 17.0\% (18.4\%) & 17.0\% (11.9\%) & 14.0\% (7.5\%)  & 7.7\%  (3.7\%)  & 16.9\% (16.9\%) \\
	$R_\mathrm{LM} = 8.9 \times R_\mathrm{CC}(0)$, $\chi=1.00$ & 50.0\% (54.2\%) & 50.1\% (35.0\%) & 41.3\% (22.1\%) & 22.8\% (10.8\%) & 49.8\% (49.8\%)
	\enddata
	\tablecomments{First row: DSNB \nuebar-flux for our fiducial model (with $\RLM\,\mathord{=}\,0$, cf. Table~\ref{tab:dsnb_contributions}), integrated over different energy intervals. Rows 2--5: Flux contributions $x$ (Equation~\eqref{eq:x}) from low-mass (LM) NS-formation events (AIC, MIC, ultra\-stripped SNe) relative to the fiducial model for four different choices of the constant (LM$_\mathrm{const}$) rate density \RLM. In parentheses, the values of $x$ for an evolving LM NS-formation rate (LM$_\mathrm{evolv}$) with the same value of $\chi$ (Equation~\eqref{eq:chi}) are given (see main text for details).}
\end{deluxetable*}

In our study we explore the consequences of additional formation channels of (rather) low-mass (LM) NSs on our DSNB predictions in a quantitative and systematic way, subsuming the possible contributions from ultra\-stripped SNe, AIC, and MIC events in addition to the contribution from ECSNe that is included in our standard models. To this end, we employ a generic neutrino spectrum ($\diff N_\mathrm{LM}/\diff E'$) adopted from the ECSN calculations of \citet[``model Sf'']{2010PhRvL.104y1101H} since neutrino signals from sophisticated long-time simulations of AIC, MIC, and ultra\-stripped SNe are still lacking. We expect the neutrino emission properties of all three additional formation channels of LM NSs to be fairly similar to the case of ECSNe. Our approach is therefore meant to serve as an order-of-magnitude estimate, but it cannot capture any details connected to differences in the individual event rates and in the neutrino signals of the three channels of ultra\-stripped SNe, AIC, and MIC events, which we combine to a single, additional LM NS-formation component.

It should be mentioned here that the cosmic rates of such events are highly uncertain, because a large parameter space in the treatment of binary interaction (especially common-envelope physics) makes precise predictions difficult. Using population synthesis methods, \citet{2017A&A...601A..29Z} found that core-collapse events in binary systems are generally delayed compared to those of single stars. More particularly, \citet{2019MNRAS.484..698R} showed that AIC and MIC can proceed in various evolutionary pathways, featuring a variety of delay-times (from below $\unit{10^2}{Myr}$ up to over $\unit{10}{Gyr}$) between star burst and eventual stellar collapse. For simplicity, we thus explore on the one hand different values of a comoving rate density, $\RLM(z)\,\mathord{=}\,\RLM$, which does not change with cosmic time (``LM$_\mathrm{const}$''). On the other hand, we examine how our DSNB results differ in the case of an evolving rate for additional LM NS-formation events (``LM$_\mathrm{evolv}$''). The DSNB flux spectrum (Equation~\eqref{eq:DSNB}) can be rewritten in the generalized form
\begin{equation}\label{eq:DSNB_LM}
\frac{\diff\Phi}{\diff E} = \frac{c}{H_0}\int_{0}^{5}\!\!\diff z\, \frac{\RCC(z)\frac{\diff N_\mathrm{CC}}{\diff E'} + \RLM(z)\frac{\diff N_\mathrm{LM}}{\diff E'}}{\sqrt{\Omega_\mathrm{m}(1+z)^3+\Omega_\mathrm{\Lambda}}}~.
\end{equation}

In the lower right panel of Figure~\ref{fig:dsnb_parameter_study}, we separately plot our fiducial DSNB prediction (dashed line; see Section~\ref{sec:fiducial_model}) and the additional contribution from LM events for four different constant rate densities \RLM\ (solid lines), which we take as multiples of the local stellar core-collapse rate, $\RCC(0)\,\mathord{=}\,\unit{8.93 \times 10^{-5}}{Mpc^{-3}yr^{-1}}$. However, since $\RCC(z)$ varies strongly with redshift (it increases by over an order of magnitude from $z\,\mathord{=}\,0$ to $z\,\mathord{=}\,1$), we also consider the ratio of the comoving rate densities of LM NS-formation events relative to ``conventional'' core-collapse SNe, both integrated over the cosmic history:
\begin{equation}\label{eq:chi}
\chi = \frac{\int_0^5\diff z \,\RLM(z) |\diff t_\mathrm{c}/\diff z|}{\int_0^5\diff z \,\RCC(z) |\diff t_\mathrm{c}/\diff z|}~.
\end{equation}
This serves as a measure of the relative importance of how both types of neutrino sources contribute to the DSNB from the time of the highest considered redshifts ($z_{\mathrm{max}}\,\mathord{=}\,5$) until the present day. In Table~\ref{tab:dsnb_LM_component}, we show the ratios
\begin{equation}\label{eq:x}
x = \frac{\int_{E_1}^{E_2}\diff E \int_0^5\diff z \,\RLM(z)\frac{\diff N_\mathrm{LM}}{\diff E'} |\diff t_\mathrm{c}/\diff z|}{\int_{E_1}^{E_2}\diff E \int_0^5\diff z \,\RCC(z)\frac{\diff N_\mathrm{CC}}{\diff E'} |\diff t_\mathrm{c}/\diff z|}~,
\end{equation}
i.e. the DSNB flux contributions from LM events relative to our fiducial model with $\RLM\,\mathord{=}\,0$, integrated over different energy intervals [$E_1$,$E_2$] for our different choices of $\RLM$. To see an effect of at least 10\% within the detection window (10--30\,MeV), an additional (constant) low-mass rate $\RLM\,\mathord{=}\,\unit{1.55\times10^{-4}}{Mpc^{-3}yr^{-1}}$ is required, which is nearly twice the local stellar core-collapse rate, \RCC(0), and corresponds to $\chi\,\mathord{=}\,0.20$. Such a fraction is well above present estimates for both AIC/MIC events \citep{2009MNRAS.396.1659M, 2019MNRAS.484..698R} and ultra\-stripped SNe \citep{2013ApJ...778L..23T} of at most a few percent of the ``conventional'' core-collapse SN population. However, due to large uncertainties in the physics of binary interaction, the possibility of such a large population of LM NS-formation events may not be ruled out completely.

As a sensitivity check, we additionally consider a comoving rate density, $\RLM(z)$, which linearly increases by a factor of 4 between $z\,\mathord{=}\,0$ and $z\,\mathord{=}\,1$ and stays constant at even larger redshifts, roughly following the observationally inferred rate of type Ia SNe \citep[e.g.,][]{2011MNRAS.417..916G}. In the lower right panel of Figure~\ref{fig:dsnb_parameter_study}, the LM-flux contribution resulting from such an evolving rate is indicated by the pale red band (defined by $0.11\,\mathord{\leqslant}\,\chi\,\mathord{\leqslant}\,1.00$, like for the four cases for constant rates \RLM). The spectra are shifted towards lower energies, as expected due to the relatively increased contribution from events at high redshifts. This can also be seen in Table~\ref{tab:dsnb_LM_component} (values in parentheses). An enhancement of the DSNB flux by 10\% at energies above $\unit{10}{MeV}$ would even require $\chi\,\mathord{=}\,0.26$ for an evolving LM rate, which would mean that, for example, more than roughly half of the WD mergers lead to NS formation instead of a type Ia SN, if merging WDs explain the majority of SNIa events. Although this seems to be disfavored on grounds of current observations and population synthesis models \citep[e.g.,][]{2009MNRAS.396.1659M, 2019MNRAS.484..698R}, it might not be entirely impossible. Nevertheless, within the relevant detection window, the contribution from AIC/MIC events and ultra\-stripped SNe to the DSNB is likely to be hidden by the current uncertainty of the cosmic core-collapse rate (gray shaded band in Figure~\ref{fig:dsnb_parameter_study}). Only when this dominant uncertainty will be reduced significantly, there may be a chance to uncover a contribution to the neutrino background from such LM NS-forming events.

\subsection{Inclusion of Binary Models}\label{subsec:dsnb_binary}

\begin{figure*}[ht!]
	\includegraphics[width=\textwidth]{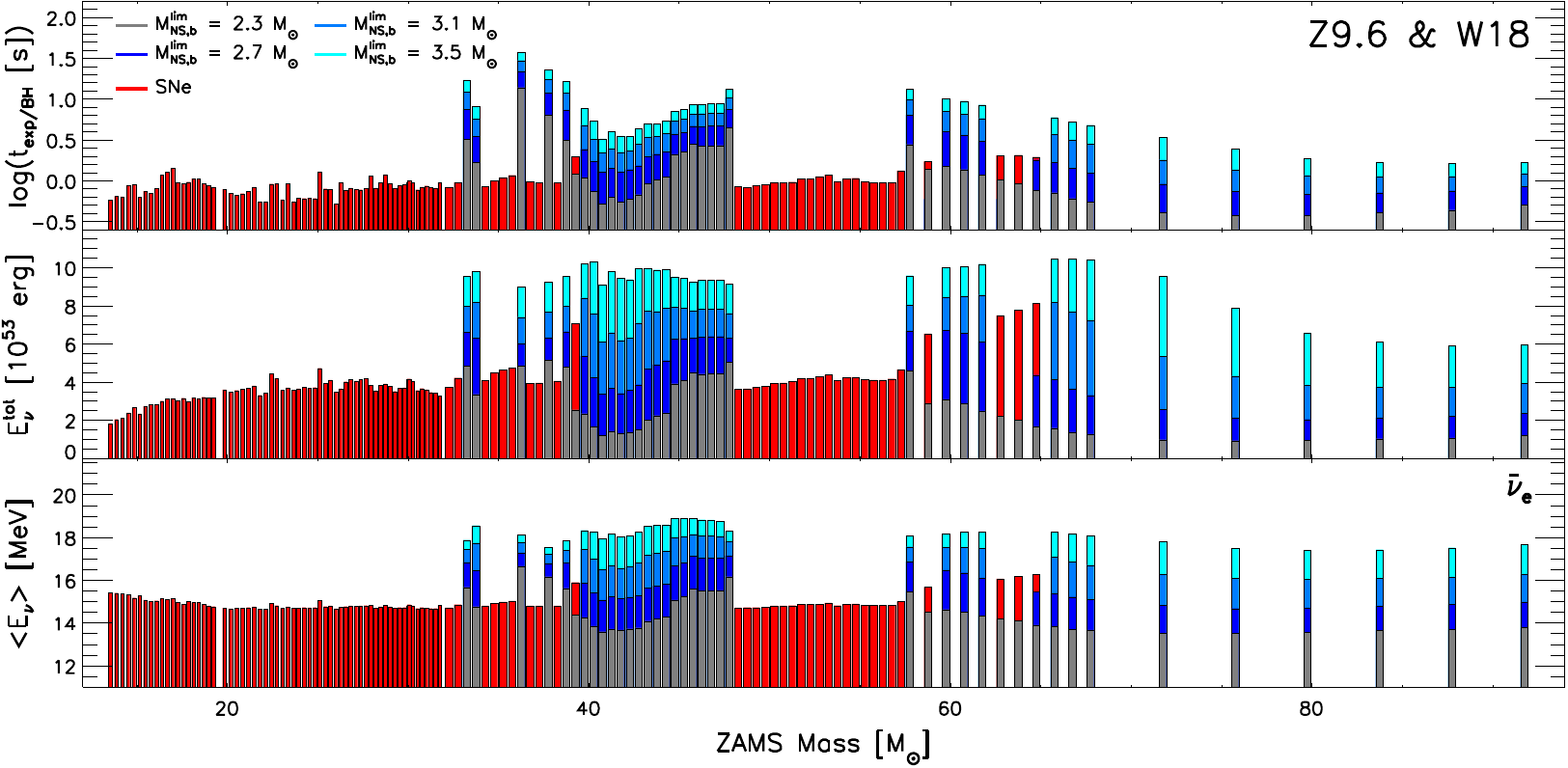}
	\caption{Landscape of NS or BH formation for the set of helium-star progenitors from \citet{2019ApJ...878...49W} as obtained in simulations with the engine model Z9.6\,\&\,W18 (cf. Figure~\ref{fig:neutrino_outcomesystematics} for single-star progenitors). From top to bottom: time of explosion or BH formation, total energy radiated in all species of neutrinos, and mean energy of electron antineutrinos versus ZAMS mass of the progenitors. Note the different mass range for stellar core-collapse progenitors compared to Figure~\ref{fig:neutrino_outcomesystematics}. Red bars indicate successful SN explosions (and fallback SNe), while the outcomes of BH-forming, failed SNe are shown for the different baryonic NS mass limits in gray ($\unit{2.3}{\msun}$), dark blue ($\unit{2.7}{\msun}$), light blue ($\unit{3.1}{\msun}$), and cyan ($\unit{3.5}{\msun}$). Five special progenitors yield successful or failed explosions depending on the NS mass limit (see footnote~\ref{fn:he-stars}).\\
	\label{fig:neutrino_outcomesystematics_He}}
\end{figure*}

A large fraction of massive stars is expected to undergo binary interaction with a companion, possibly shedding their hydrogen envelopes (e.g., via Roche-lobe overflow or common-envelope ejection) and leaving behind bare helium stars \citep{2012Sci...337..444S}. Taking this as a motivation, we explore how the inclusion of binary models affects our DSNB predictions. To this end, we employ a set of 132 helium stars with initial masses in the range of $\unit{2.5-40}{\msun}$, originating from hydrogen burning in non-rotating, solar-metallicity stars \citep{2019ApJ...878...49W}. According to equations~(4) and (5) therein, this range of initial helium-core masses converts to ZAMS masses of $\unit{13.5-91.7}{\msun}$. Stars with masses lower than that are assumed to form WDs, thus not contributing to the DSNB.\footnote{Consistently, the lower integration bounds in Equations~\eqref{eq:IMF_average} and \eqref{eq:R_CC} are raised from $\unit{8.7}{\msun}$ to $\unit{13.5}{\msun}$.} For the details of the pre-SN evolution (which includes wind mass loss), the reader is referred to \citet{2019ApJ...878...49W}.

We used these progenitor models and performed SN simulations with the \textsc{Prometheus-HotB} code as done for the single-star progenitors (see Section~\ref{subsec:SN_simulations}). A detailed and dedicated analysis of the explosions of these helium stars can be found in the recent paper by \cite{2020ApJ...890...51E}. In Figure~\ref{fig:neutrino_outcomesystematics_He}, we show, for engine model Z9.6\,\&\,W18, the landscape of NS- and BH-formation events with basic properties of the neutrino emission of relevance for our DSNB calculations. Compared to Figure~\ref{fig:neutrino_outcomesystematics}, the range of stars experiencing core collapse is shifted towards higher ZAMS masses, starting only at $\unit{13.5}{\msun}$. Moreover, there are no cases of BH formation below a ZAMS mass of $\unit{33}{\msun}$. This can be understood as a consequence of the mass loss by stellar winds during the pre-SN evolution of the helium stars, yielding less compact cores compared to stars which still possess their hydrogen envelope (see figures~1 and 10 in \citealt{2019ApJ...878...49W}).

\begin{figure*}[ht!]
	\includegraphics[width=\textwidth]{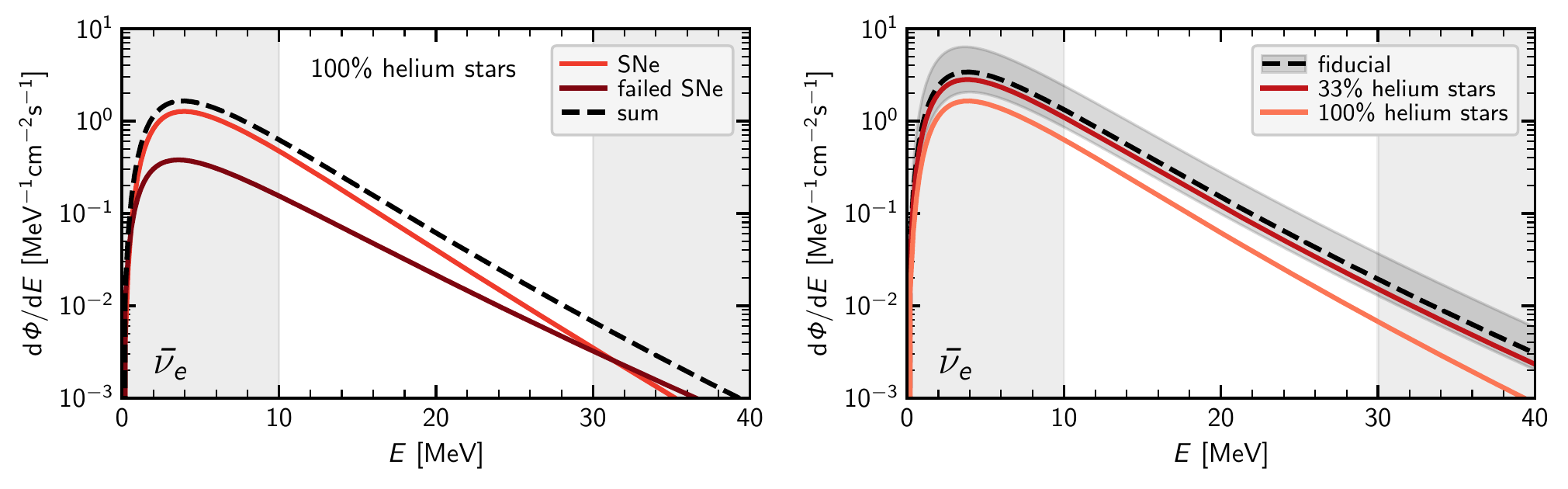}
	\caption{DSNB flux spectrum, $\diff\Phi/\diff E$, of electron antineutrinos from the helium-star progenitors of \citet{2019ApJ...878...49W}, exploded with engine model Z9.6\,\&\,W18, taking $\MBH\,\mathord{=}\,\unit{2.7}{\msun}$, the best-fit choice for the shape parameter~$\alpha$, and the SFH from \citet{2014ApJ...790..115M}. The left panel shows the two components, successful and failed SNe (light and dark red solid lines), contributing to the total DSNB flux (dashed line), assuming that the entire population of progenitors evolves as helium stars (cf. left panel of Figure~\ref{fig:dsnb_contributions} for single-star progenitors). In the right panel, our fiducial model based on single stars only (black dashed line) is compared with the DSNB flux spectra assuming a fraction of 33\% or 100\% of helium stars (dark or light red solid lines, respectively). The gray band around our fiducial single-star DSNB spectrum corresponds to the $\pm1\sigma$ uncertainty of the SFH from \citet{2014ApJ...790..115M}. As in Figures~\ref{fig:dsnb_contributions}, \ref{fig:dsnb_SFR}, and \ref{fig:dsnb_parameter_study}, vertical bands frame the approximate detection window.\\
		\label{fig:dsnb_He}}
\end{figure*}

We note in passing that the values for the neutrino energy loss used in our present study differ in details from the numbers shown in figure~5 of \citet{2020ApJ...890...51E}. First, we do not consider the additional neutrino-energy loss from fallback accretion and consistently treat fallback SNe as NS-formation events, whereas \citet{2020ApJ...890...51E} took fallback into account in their estimates of the compact remnant masses and the associated release of gravitational binding energy through neutrinos.\footnote{\label{fn:he-stars}Note that five such progenitors, which explode at relatively late times ($\sim$2\,s) and consequently reach high PNS masses, are treated either as ``normal'' successful SNe (without fallback) in this work or, if the PNS mass exceeds \MBH\ at any time during the post-bounce evolution, as failed explosions. In the latter case, the neutrino signals are truncated at this time, $t_\mathrm{BH}$.}
Second, in our present study we extrapolate the neutrino emission of non-exploding cases until the accreting PNS in our \textsc{Prometheus-HotB} runs reaches the assumed and parametrically varied baryonic mass limit of stable, cold NSs, \MBH, and therefore should collapse to a BH.  In contrast, \citet{2020ApJ...890...51E} employed for BH cases (with $t_\mathrm{BH}\,\mathrm{>}\,\unit{10}{s}$) the radius-dependent fit formula of \citet{2001ApJ...550..426L} for the gravitational binding energy of a NS with the maximum mass assumed in their work. The energy release (\Enutot) estimated that way is somewhat larger than for our accretion-determined estimates (see Figure~\ref{fig:appendix_E_tot_fSNe} in Appendix~\ref{appendix:total_energies}).

Figure~\ref{fig:dsnb_He} illustrates how the inclusion of binary models impacts our DSNB predictions. In the left panel, we separately show the contributions from successful and failed explosions to the DSNB flux spectrum of electron antineutrinos for our fiducial model parameters (Z9.6\,\&\,W18; $\MBH\,\mathord{=}\,\unit{2.7}{\msun}$; best-fit $\alpha$; SFH from \citealt{2014ApJ...790..115M}), assuming that all (100\%) progenitors evolve as helium stars (``W18-BH2.7-$\alpha$2.0-He100''). Compared to single stars (Figure~\ref{fig:dsnb_contributions} and black dashed line in the right panel of Figure~\ref{fig:dsnb_He}), the overall DSNB flux is reduced by a factor of $\sim$2 owing to the smaller fraction of stars experiencing core collapse. At the same time, the less frequent failed explosions produce a lower high-energy tail of the spectrum compared to our fiducial DSNB spectrum based on single stars (see also Table~\ref{tab:exp_fit}). If we assume that only 33\% of all massive stars strip their hydrogen envelopes (as suggested by \citealt{2012Sci...337..444S}; ``W18-BH2.7-$\alpha$2.0-He33''), the effects of helium stars on the DSNB spectrum are less dramatic and the shifted spectrum lies within the uncertainty band associated with the SFH (gray shaded band; cf. Figure~\ref{fig:dsnb_SFR}).

Applying the other neutrino engines considered in our work to the helium-star models, we obtain similar relative changes of the DSNB spectra as in the case of our fiducial engine model Z9.6\,\&\,W18. We should stress at this point that, if progenitors do not lose their entire hydrogen envelopes, end stages of stellar evolution more similar to those of single-star evolution can be expected \citep{2019ApJ...878...49W}.

\begin{figure*}[ht!]
	\includegraphics[width=\textwidth]{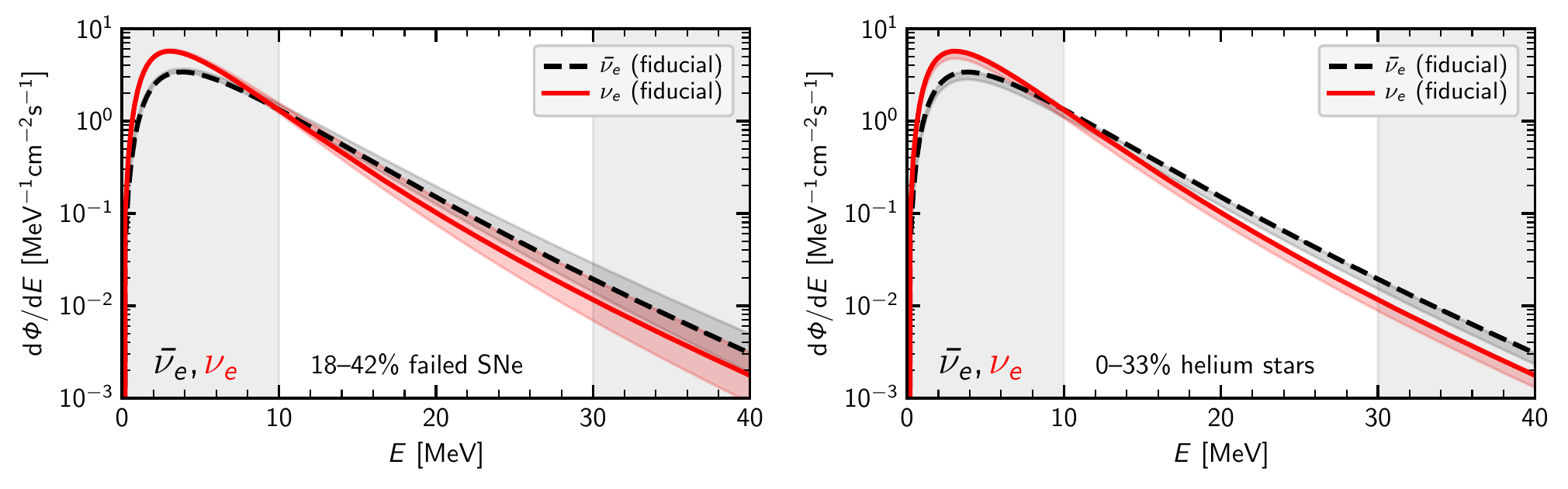}
	\caption{Comparison of the DSNB flux spectra of electron antineutrinos and electron neutrinos. In both panels, the black dashed and the red solid lines correspond to the (unoscillated) DSNB spectra, $\diff\Phi/\diff E$, of our fiducial model (W18-BH2.7-$\alpha$2.0; 26.9\% failed SNe, no helium stars; see Section~\ref{sec:fiducial_model}) for \nuebar\ and \nue, respectively. In the left panel, the shaded bands indicate the spectral DSNB variations for our different neutrino engines, leading to fractions of failed explosions with BH-formation between $\sim$18\% (for the Z9.6\,\&\,S19.8 engine) and up to $\sim$42\% (for the Z9.6\,\&\,W20 engine; cf. upper left panel of Figure~\ref{fig:dsnb_parameter_study}). In the right panel, the shaded bands show the effect of including a 33\% fraction of hydrogen-stripped helium stars (cf. Figure~\ref{fig:dsnb_He}). As in previous figures, vertical bands indicate the approximate \nuebar-detection window. Note, however, that the detection window is different for \nue\ ($\sim$17--40\,MeV; not shown in the figure; see, e.g., \citealt{2004JCAP...12..002C}, \citealt{2019PhRvC..99e5810Z}).\\
	\label{fig:dsnb_nue}}
\end{figure*}

\section{DSNB Spectrum of Electron Neutrinos}\label{sec:nue}

Although the main focus of our study lies on the DSNB's \nuebar\ component, we briefly comment on the flux spectrum of \nue, which is an observational target of DUNE \citep{2015arXiv151206148D}. Combining future DSNB \nue-flux measurements by DUNE with the \nuebar-flux data gathered by the gadolinium-loaded SK (and Hyper-Kamiokande) and by JUNO will yield complementary constraints on the DSNB parameter space \citep[see, e.g.,][]{2018JCAP...05..066M} and will help testing different neutrino oscillation scenarios or non-standard-model physics such as neutrino decays \citep[see, e.g.,][]{2004PhRvD..70a3001F, 2020arXiv200713748D, 2020arXiv201110933T}.

In Figure~\ref{fig:dsnb_nue}, we show our predictions for the DSNB flux spectrum for \nue\ in comparison to the \nuebar\ component for our fiducial model parameters (Z9.6\,\&\,W18 neutrino engine; $\MBH\,\mathord{=}\,\unit{2.7}{\msun}$; $\alphaBH\,\mathord{=}\,2.0$; best-fit SFH from \citet{2014ApJ...790..115M}). The main differences are a more prominent spectral peak (at energies $E\,\mathord{\lesssim}\,\unit{8}{MeV}$) and a faster decline of the spectrum towards high neutrino energies for the case of \nue\ compared to \nuebar. The exponential fit of Equation~\eqref{eq:exponential_dsnb_fit} yields a value of the ``slope parameter'' $E_0$ of 4.49\,MeV for the \nue\ spectrum (compared to $E_0\,\mathord{=}\,\unit{4.82}{MeV}$ for \nuebar). This is a consequence of generally lower mean neutrino energies of \nue\ compared to \nuebar\ (see Table~\ref{tab:spectral_parameters}). Note that for \nue\ a DSNB detection will not be possible below $\sim$17\,MeV due to the overwhelming solar \textit{hep} (and $^8$B) neutrino flux (see, e.g., figure~8 of \citealt{2019PhRvC..99e5810Z}).

To give an impression of the spectral DSNB variability for \nue, we also show the \nue-flux spectra for models with different neutrino engines applied and thus varied fractions of failed SNe with BH formation (left panel), as well as for a model that includes 33\% hydrogen-stripped helium-star progenitors (as suggested by \citealt{2012Sci...337..444S}; right panel). The overall trends (i.e., enhanced high-energy tail of the DSNB spectrum for a larger fraction of failed SNe and reduced DSNB flux for helium stars being included) are similar to the case of \nuebar.

\section{Neutrino Flavor Conversions and Remaining Uncertainties}\label{sec:uncertainties}

\subsection{Neutrino Flavor Conversions}\label{subsec:flavor_conversions}

So far we did not take neutrino flavor oscillations into account but identified the emission of electron antineutrinos (or neutrinos) by the considered astrophysical sources with the measurable DSNB flux of \nuebar\ (or \nue). However, on their way out of a collapsing star, neutrinos (and antineutrinos) undergo collective and matter-induced (MSW) flavor conversions \citep{1978PhRvD..17.2369W, 1985YaFiz..42.1441M, 2010ARNPS..60..569D, 2016NCimR..39....1M}. Hereafter, we discuss how such oscillations can affect our DSNB flux predictions.

\begin{figure*}
	\includegraphics[width=\textwidth]{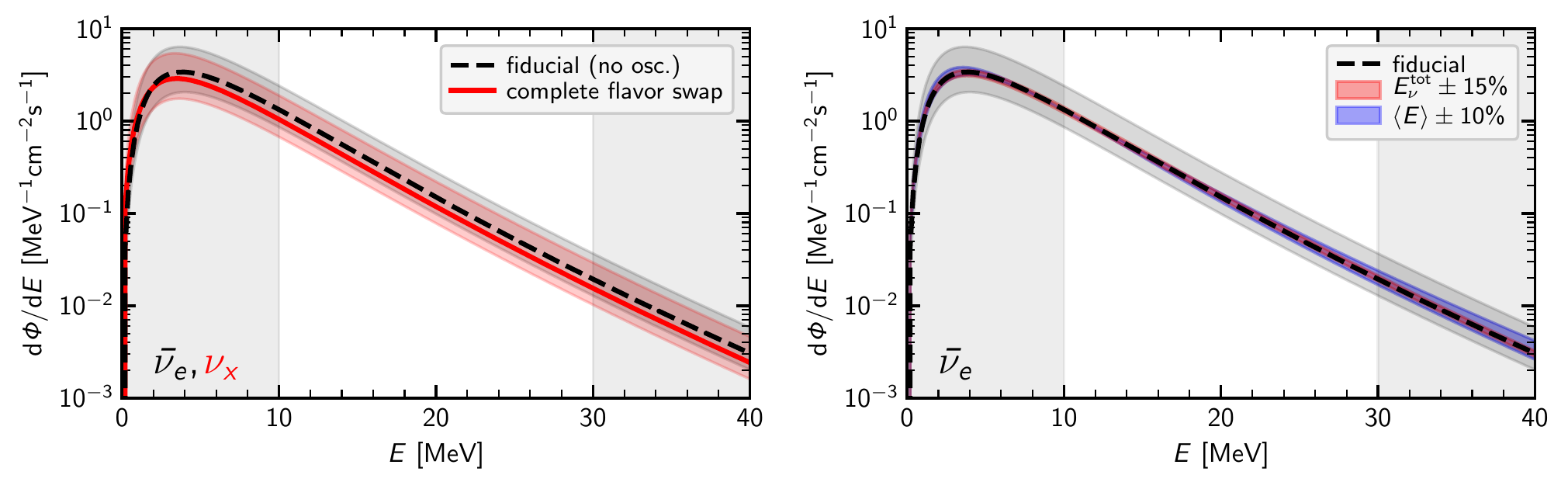}
	\caption{Effects of neutrino flavor conversions on the DSNB flux spectrum and remaining modeling uncertainties for the case of our fiducial model parameters (see Section~\ref{sec:fiducial_model}). The left panel shows the unoscillated DSNB spectrum of electron antineutrinos ($\diff \Phi^0_{\nuebar}/\diff E$; black dashed line) and the predicted DSNB spectrum for one species of heavy-lepton neutrinos ($\diff \Phi^0_{\nux}/\diff E$; red solid line), which would become the measurable \nuebar\ spectrum in the case of a complete flavor swap $\nuebar\,\mathord{\leftrightarrow}\,\nux$ (see Equation~\eqref{eq:flavor_conversion}). The uncertainty arising from the cosmic core-collapse rate \RCC\ \citep[represented by the $\pm1\sigma$ limits to the SFH from][]{2014ApJ...790..115M} is indicated by shaded bands. In the right panel, our fiducial model (black dashed line; unoscillated \nuebar) is compared to DSNB flux spectra where the total radiated neutrino energy, \Enutot, is reduced by 15\% for successful SNe or increased by 15\% for BH-formation cases. The corresponding red band is partly covered by the blue band, which marks the DSNB variation when the time-integrated mean \nuebar\ energies, $\langle E \rangle$, are shifted by $-$10\% for successful SNe or by $+$10\% for failed explosions (see main text for details). Changing \Enutot\ or $\langle E\rangle$ for successful and failed SNe at the same time yields spectra within the uncertainty bands shown. The uncertainty of the fiducial spectrum due to \RCC\ is indicated by the gray band.\\
	\label{fig:dsnb_uncert}}
\end{figure*}

Following \citet{2011PhLB..702..209C} and \citet{2012JCAP...07..012L}, we write the DSNB flux spectrum of electron antineutrinos after including the effect of flavor conversions as
\begin{equation}\label{eq:flavor_conversion}
\frac{\diff \Phi_{\nuebar}}{\diff E} = \bar{p} \, \frac{\diff \Phi^0_{\nuebar}}{\diff E} + (1 - \bar{p}) \, \frac{\diff \Phi^0_{\nux}}{\diff E} \:,
\end{equation}
where $\diff \Phi^0_{\nuebar}/\diff E$ and $\diff \Phi^0_{\nux}/\diff E$ are the unoscillated spectra for electron antineutrinos (\nuebar) and a re\-pre\-sen\-ta\-tive heavy-lepton neutrino (\nux). $\bar{p}\,\mathord{\simeq}\,0.7$ ($\bar{p}\,\mathord{\simeq}\,0$) denotes the survival probability of \nuebar\ in the cases of normal (NH) or inverted (IH) mass hierarchy, respectively.\footnote{\citet{2012JCAP...07..012L} showed that the effects of self-induced (collective) conversions and the MSW resonances can be treated separately, because the latter occur farther away from the central core regions of a SN. The \nuebar\ survival probability is then given by $\bar{p} = \cos^2\theta_{12}\bar{P}_\mathrm{c}$ for NH, and $\bar{p} = \cos^2\theta_{12}(1-\bar{P}_\mathrm{c})$ for IH, with $\bar{P}_\mathrm{c}$ denoting the survival probability when exclusively collective effects play a role \citep[for more details, see also][]{2011PhLB..702..209C}. However, \citet{2012JCAP...07..012L} noted that self-induced conversions affect the DSNB only on a few-percent level and can therefore be neglected. Also the recently discussed fast conversions \citep[see, e.g.,][]{2016NuPhB.908..366C, 2017ApJ...839..132T, 2017PhRvL.118b1101I}, which might lead to partial flavor equilibration, should not modify the DSNB in a more extreme manner than captured by the two discussed extremes of purely MSW-induced conversions (i.e., $\bar{p} \simeq 0$ and $\bar{p} \simeq 0.7$), as pointed out by \citet{2018JCAP...05..066M}.} Recently, \citet{2018JCAP...05..066M} confirmed by numerically solving the neutrino kinetic equations of motion that (matter-induced) neutrino flavor conversions can be well approximated by the simplified analytic description of Equation~\eqref{eq:flavor_conversion} for the small set of \textsc{Prometheus-Vertex} simulations that they used in their study and that we also employ in our work as reference cases to calibrate some degrees of freedom in our modeling approach (see Table~\ref{tab:flavor_ratios} in Appendix~\ref{appendix:rescaling}). We already mentioned earlier that the large sets of core-collapse simulations underlying our DSNB calculations do not provide reliable information of the heavy-lepton neutrino source emission, which is why we use \textsc{Prometheus-Vertex} SN and BH-formation models to rescale the neutrino energy release in the different neutrino species (see Section~\ref{subsec:spectra}). For the same reason, we also adjust the spectral parameters, $\langle E_{\nux} \rangle$ and \alphax, of the time-integrated \nux\ emission (the bar in the symbol \alphax\ indicates that the shape parameter refers to the \textit{time-integrated} spectrum rather than the \textit{instantaneous} spectrum), guided by the sophisticated \textsc{Prometheus-Vertex} models listed in Table~\ref{tab:flavor_ratios}, to get a useful representation of the unoscillated DSNB spectrum of heavy-lepton neutrinos, $\diff \Phi^0_{\nux}/\diff E$ (see Appendix~\ref{appendix:rescaling} for the details).

In the left panel of Figure~\ref{fig:dsnb_uncert}, we show our unoscillated, fiducial DSNB spectrum for \nuebar, $\diff \Phi^0_{\nuebar}/\diff E$ (black dashed line), and the corresponding unoscillated DSNB spectrum for \nux, $\diff \Phi^0_{\nux}/\diff E$ (red solid line), for our fiducial model pa\-ra\-me\-ters (see Section~\ref{sec:fiducial_model}). According to Equation~\eqref{eq:flavor_conversion}, the latter represents the case of IH, where a complete flavor swap ($\nuebar\,\mathord{\leftrightarrow}\,\nux$) takes place. If, instead, the case of NH is realized in Nature, an outcome between the two plotted extremes can be expected. The uncertainty arising from the cosmic core-collapse rate \citep[corresponding to the $\pm1\sigma$ interval of the SFH from][]{2014ApJ...790..115M} is indicated by shaded bands. In Table~\ref{tab:dsnb_flavor_conversions}, we additionally provide the integrated \nuebar-flux for different energy intervals and a complete flavor swap ($\nuebar\,\mathord{\leftrightarrow}\,\nux$) in analogy to what is given in Table~\ref{tab:dsnb_contributions} for the case of no flavor oscillations (\nuebar). The most important difference is a reduced contribution from failed SNe. This can be understood by the small relative fraction of the heavy-lepton neutrino emission, $\tilde{\xi}_{\nux}$, in our two \textsc{Prometheus-Vertex} reference models for BH formation, which we employ for our rescaling (Appendix~\ref{appendix:rescaling}). At the same time, the contribution from successful explosions (including ECSNe) is largely unchanged, which reflects the approximate flavor equipartition in their neutrino emission.

\begin{deluxetable*}{lccccc}
	\tablecaption{
	DSNB-flux components for the case of a complete flavor swap ($\nuebar\,\mathord{\leftrightarrow}\,\nux$).
	\label{tab:dsnb_flavor_conversions}}
	\tablehead{
		\colhead{} & \colhead{$\unit{(0-10)}{MeV}$} & \colhead{$\unit{(10-20)}{MeV}$} & \colhead{$\unit{(20-30)}{MeV}$} & \colhead{$\unit{(30-40)}{MeV}$} & \colhead{$\unit{(0-40)}{MeV}$}
	}
	\startdata
	Total DSNB Flux (\nuebar)               & $\unit{19.3}{cm^{-2}s^{-1}}$  & $\unit{4.3}{cm^{-2}s^{-1}}$   & $\unit{0.5}{cm^{-2}s^{-1}}$& $\unit{0.1}{cm^{-2}s^{-1}}$   & $\unit{24.2}{cm^{-2}s^{-1}}$  \\
	\hline
	ECSNe           & $3.0\%$       & $1.5\%$       & $0.8\%$       & $0.4\%$       & $2.7\%$       \\
	Iron-Core SNe           & $68.4\%$      & $65.4\%$      & $52.3\%$      & $37.9\%$      & $67.5\%$      \\
	Failed SNe              & $28.6\%$      & $33.2\%$      & $47.0\%$      & $61.8\%$      & $29.9\%$
	\enddata
	\tablecomments{
	First row: DSNB \nuebar-flux for the case of a complete flavor swap ($\nuebar\,\mathord{\leftrightarrow}\,\nux$), integrated over different energy intervals. Rows 2--4: Relative contributions from the various source types (ECSNe/iron-core SNe/failed SNe). Our fiducial model paramters (Z9.6\,\&\,W18; $\MBH\,\mathord{=}\,\unit{2.7}{\msun}$; best-fit $\alpha$) are used. Compare with Table~\ref{tab:dsnb_contributions}, where values for the unoscillated \nuebar-flux are provided.}
\end{deluxetable*}

Despite the less relevant contribution from failed SNe, the slope parameter $E_0$/MeV of the exponential fit of Equation~\eqref{eq:exponential_dsnb_fit} is increased marginally from 4.82 to 4.84 in the case of a complete flavor swap (see Table~\ref{tab:exp_fit}) because smaller values of the spectral-shape parameter \alphax\ for heavy-lepton neutrinos (see Table~\ref{tab:flavor_ratios}; $\lambdaa\,\mathord{<}\,1$) partly compensate for the reduced flux of \nux\ in the high-energy region associated with the BH cases. The mean energies of the time-integrated neutrino signals are fairly similar for \nux\ and \nuebar\ (see Table~\ref{tab:flavor_ratios}; $\lambdaE\,\mathord{\sim}\,1$), as suggested by state-of-the-art simulations \citep[e.g.,][]{2009A&A...496..475M, 2014ApJ...788...82M} and a consequence of the inclusion of energy transfers (non-isoenergetic effects) in the neutrino-nucleon scattering reactions \citep[see][]{2003ApJ...590..971K, Huedepohl:2014}. In conflict with this result of modern SN models with state-of-the-art treatment of the neutrino transport, several previous DSNB studies employed spectra with \Emeanx\ being considerably higher than $\langle E_{\nuebar} \rangle$ (particularly for the emission from failed explosions).

In line with the recent studies by \citet{2017JCAP...11..031P} and \citet{2018JCAP...05..066M}, we find that neutrino flavor conversions have a fairly moderate influence on the DSNB (for \nuebar), which is well dominated by other uncertainties. Nonetheless, for our highest-flux models (with a weak central engine and a high maximum NS mass), which possess a large DSNB contribution from BH-forming events, the oscillation effects become more pronounced. We will further comment on this in Section~\ref{subsec:dsnb_SK_limit}.

For the DSNB \nue\ flux the effects of neutrino flavor oscillations can be described in an analogue manner \citep[see, e.g.,][]{2011PhLB..702..209C,2012JCAP...07..012L}. In the most extreme case of NH (and purely MSW-induced flavor conversions), a complete flavor swap ($\nue\,\mathord{\leftrightarrow}\,\nux$) can take place, whereas for IH a measurable DSNB \nue-flux spectrum in between the unoscillated spectra of \nux\ and \nue\ can be expected.

\subsection{Tests of Remaining Uncertainties}\label{subsec:remaining_uncertainties}

As we point out in Appendix~\ref{appendix:total_energies}, the total radiated neutrino energies (\Enutot) of our successful SNe might, on average, be overestimated by a few percent, whereas the neutrino emission from failed explosions could be slightly underestimated in our modeling approach for the neutrino signals. In the right panel of Figure~\ref{fig:dsnb_uncert}, we therefore compare our fiducial DSNB prediction (black dashed line) with a spectrum where \Enutot\ of all exploding progenitors is reduced by 15\% (lower edge of the red band). This choice of the reduction is guided by a comparison of \Enutot\ with the gravitational binding energies BE$_{12}$ of the corresponding NS remnants (Equation~\eqref{eq:LattimerPrakash} with $R_\mathrm{NS}\,\mathord{=}\,\unit{12}{km}$; see Figure~\ref{fig:appendix_E_tot_SNe} and Table~\ref{tab:IMF-weighted_deviations_from_LP2001}), consistent with the cold-NS radius suggested by recent astrophysical observations and constraints from nuclear theory and experiments (see footnote~\ref{fn:R_NS}). Analogously, the upper edge of the red band in Figure~\ref{fig:dsnb_uncert} indicates a model where \Enutot\ of all failed explosions is increased by 15\%. This case is motivated by the circumstance that the maximum neutrino emission in our failed-SN models with late BH formation lies $\sim$10--20\% below the maximally available gravitational binding energy according to Equation~\eqref{eq:LattimerPrakash} of a NS at its mass limit (see Figure~\ref{fig:appendix_E_tot_fSNe} and Table~\ref{tab:BE_fSNe}). Any mix of changes of the NS and BH energy release will lead to intermediate results. Note that the corresponding red uncertainty band is hardly visible on the logarithmic scale.

A somewhat stronger effect can be seen when we vary the mean energies, \Emean, of the time-integrated spectra by $-$10\% for successful SNe or by $+$10\% for failed explosions, respectively (lower and upper edges of the blue shaded band). Particularly at high energies, the spectra fan out noticeably. Such an uncertainty range cannot be ruled out according to present knowledge. Again, changing \Emean\ for both successful and failed SNe at the same time yields a result in between the given limits. In Appendix~\ref{appendix:spectra}, we show that the outcome of our simplified approach is in reasonable overall agreement with results from the sophisticated \textsc{Prometheus-Vertex} simulations; nonetheless, the mean energies of the time-integrated spectra do not match perfectly (they lie $\sim$1\,MeV higher/lower than in the \textsc{Vertex} models to compare with for successful/failed SNe; see Figures~\ref{fig:appendix_spectra} and \ref{fig:appendix_spectra_fSNe}). Besides this fact, we should emphasize that the neutrino emission characteristics depend considerably on the still incompletely known high-density EoS \citep[e.g.,][]{2013ApJ...774...17S, 2019PhRvC.100e5802S} and also depend on the effects of muons, which have been neglected in most previous stellar core-collapse models, but can raise the mean energies of the radiated neutrinos \citep{2017PhRvL.119x2702B}.

Despite these uncertainties associated with the neutrino source, the cosmic core-collapse rate \RCC\ still constitutes the largest uncertainty affecting the DSNB, especially at lower energies (see Figure~\ref{fig:dsnb_SFR}). Accordingly, the gray shaded band in the right panel of Figure~\ref{fig:dsnb_uncert} indicates the $\pm1\sigma$ variation of \RCC\ for the SFH from \citet{2014ApJ...790..115M}. Upcoming wide-field surveys such as LSST \citep{2002SPIE.4836...10T} should be able to pin down the \textit{visible} SN rate (below redshifts of $z\,\mathord{\sim}\,1$) to good accuracy, opening the chance for DSNB measurements to probe particularly the contribution from \textit{faint} and \textit{failed} explosions \citep{2010PhRvD..81h3001L}.

Finally, one should keep in mind that we only employ solar-metallicity progenitor models in our simulations. Obviously, this is a simplification, because the distribution of metals in the Universe is spatially non-uniform (see, e.g., the low metallicities in the Magellanic Clouds) and evolves with cosmic time. Since the fraction of failed explosions depends on metallicity \citep[e.g.,][]{2002RvMP...74.1015W, 2003ApJ...591..288H, 2012ARA&A..50..107L}, \citet{2015ApJ...804...75N} and \citet{2015PhLB..751..413Y} considered a failed-SN fraction that increases with redshift. On the other hand, \citet{2008MNRAS.391.1117P} suggested that the average metallicity does not decline dramatically up to $z\,\mathord{\sim}\,2$. Assuming solar metallicity should therefore be a sufficiently good approximation, in view of the fact that the DSNB flux in the energy window favorable to the DSNB detection is produced almost entirely by sources at moderate redshifts (see Figure~\ref{fig:dsnb_contributions}).

At this point we should also remind the reader that a core-collapse SN is an inherently multi-dimensional phenomenon \citep[see, e.g.,][]{2016PASA...33...48M}. While our simplified 1D approach should be able to capture the overall picture of the progenitor-dependent neutrino emission, an increasing number of fully self-consistent 3D simulations will have to validate our results eventually.

\begin{figure*}
	\includegraphics[width=\textwidth]{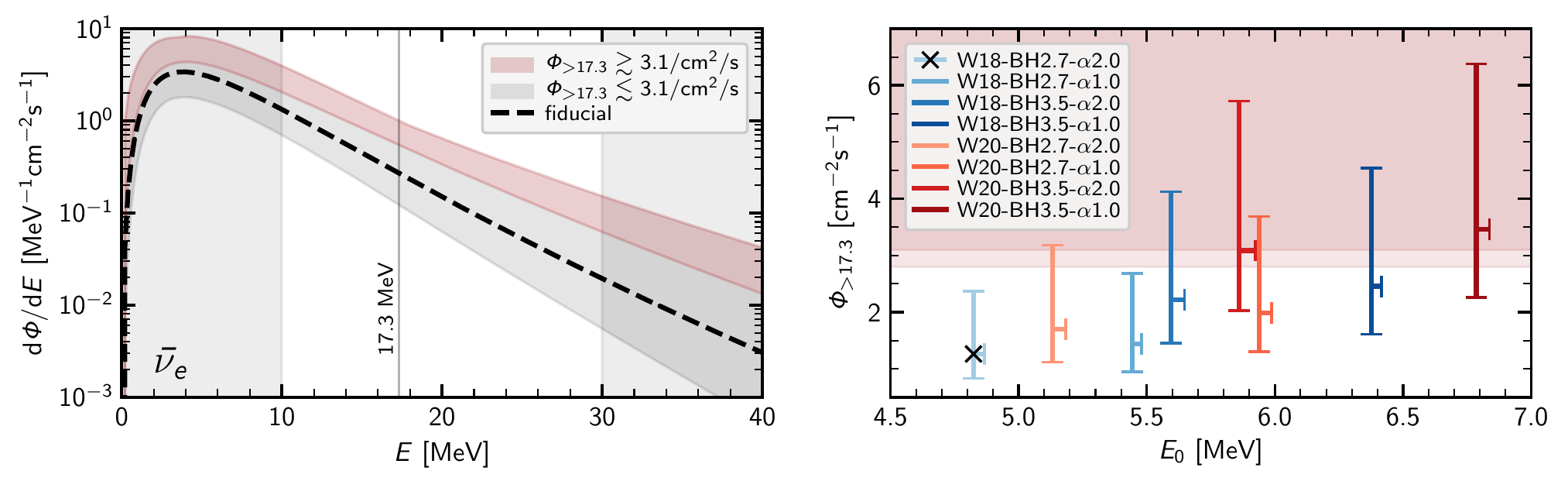}
	\caption{Comparison of our most extreme DSNB predictions with the upper flux limit from SK: $\Phi_{>17.3} \equiv \Phi(E\,\mathord{>}\,\unit{17.3}{MeV}) \lesssim \unit{(2.8-3.1)}{cm^{-2}s^{-1}}$ \citep{2012PhRvD..85e2007B}. The shaded bands in the left panel show the spread between the flux spectra $\diff\Phi/\diff E$ of electron antineutrinos, resulting from various combinations of the source parameters considered in Section~\ref{subsec:dsnb_parameter_study} (see Figure~\ref{fig:dsnb_parameter_study}). Our fiducial model (W18-BH2.7-$\alpha$2.0; Section~\ref{sec:fiducial_model}) is displayed by a dashed line. To guide the eye, we discriminate the approximate ranges for models that yield an integrated flux $\Phi_{>17.3}$ below $\unit{3.1}{cm^{-2}s^{-1}}$ (gray) or exceed this limit (red); see the main text for details. As in the previous figures, vertical bands frame the approximate detection window. In the right panel, $\Phi_{>17.3}$ is shown for a selection of models (including our fiducial case; black cross) that reach close to or beyond the SK limit (pale and dark shaded for $2.8$ and $\unit{3.1}{cm^{-2}s^{-1}}$, respectively) as a function of the fit parameter $E_0$ (Equation~\eqref{eq:exponential_dsnb_fit}). Both vertical and horizontal error bars indicate the uncertainty connected to the cosmic SFH \citep[$\pm1\sigma$ limits of][]{2014ApJ...790..115M}. The one-sided horizontal error intervals are caused by the fact that the functional fits to the SFH scale slightly differently with redshift (see footnote~\ref{fn:E0_RCC}), with the best-fit case by \cite{2014ApJ...790..115M} yielding the largest relative contribution from high-redshift regions and thus smallest value of $E_0$ compared to both the +1$\sigma$ and the $-$1$\sigma$ limits.\\
	\label{fig:dsnb_extremes}}
\end{figure*}

\section{Comparison with the SK-flux limits and previous works} \label{sec:comparison}

\subsection{Comparison with the SK-flux Limits}\label{subsec:dsnb_SK_limit}

After discussing the dependence of the predicted DSNB spectrum on different inputs in Sections~\ref{sec:parameter_study} and \ref{sec:uncertainties}, we compare our results now with the most stringent \nuebar-flux limit set by the SK experiment \citep{2012PhRvD..85e2007B}: $\Phi_{>17.3} \equiv \Phi(E\,\mathord{>}\,\unit{17.3}{MeV}) \lesssim \unit{(2.8-3.1)}{cm^{-2}s^{-1}}$.

\begin{deluxetable*}{lrrr}
	\tablecaption{Total integrated DSNB flux ($\Phi_{\mathrm{tot}}$), flux within the observational window of 10--30\,MeV ($\Phi_{10-30}$), and flux above $\unit{17.3}{MeV}$ ($\Phi_{>17.3}$) for the same subset of our DSNB models as listed in Table~\ref{tab:exp_fit}.
		\label{tab:overview}}
	\tablehead{
		\colhead{Model} & \colhead{$\Phi_\mathrm{tot}$ [$\mathrm{cm^{-2}s^{-1}}$]} & \colhead{$\Phi_{10-30}$ [$\mathrm{cm^{-2}s^{-1}}$]} & \colhead{$\Phi_{>17.3}$ [$\mathrm{cm^{-2}s^{-1}}$]}
	}
	\startdata
	W18-BH2.7-$\alpha$2.0 (fiducial) & $28.8^{+24.6}_{-10.9}$ ($24.2^{+20.7}_{-9.2}$, $20.8^{+6.6}_{-5.3}$, $15.2$) & $6.0^{+5.1}_{-2.1}$ ($4.8^{+4.0}_{-1.7}$, $5.0^{+2.0}_{-1.6}$, $3.4$) & $1.3^{+1.1}_{-0.4}$ ($1.0^{+0.9}_{-0.3}$, $1.2^{+0.6}_{-0.5}$, $0.8$) \\
	W20-BH3.5-$\alpha$1.0 (max.) & $41.7^{+35.7}_{-15.8}$ ($30.4^{+26.0}_{-11.5}$, $30.1^{+9.5}_{-7.7}$, $22.0$) & $10.8^{+8.9}_{-3.8}$ ($7.0^{+5.8}_{-2.5}$, $8.6^{+3.1}_{-2.5}$, $6.0$) & $3.5^{+2.9}_{-1.2}$ ($2.1^{+1.8}_{-0.7}$, $3.0^{+1.2}_{-1.0}$, $2.0$) \\
	S19.8-BH2.3-$\alpha$3.0 (min.) & $24.4^{+20.9}_{-9.2}$ ($22.8^{+19.5}_{-8.6}$, $17.6^{+5.6}_{-4.5}$, $12.9$) & $4.5^{+3.8}_{-1.6}$ ($4.2^{+3.5}_{-1.5}$, $3.8^{+1.6}_{-1.3}$, $2.6$) & $0.7^{+0.7}_{-0.3}$ ($0.8^{+0.7}_{-0.3}$, $0.7^{+0.4}_{-0.3}$, $0.5$) \\
	\hline
	S19.8-BH2.7-$\alpha$2.0 & $27.7^{+23.7}_{-10.5}$ ($24.7^{+21.2}_{-9.4}$, $20.0^{+6.3}_{-5.1}$, $14.6$) & $5.5^{+4.6}_{-1.9}$ ($4.7^{+3.9}_{-1.6}$, $4.6^{+1.8}_{-1.5}$, $3.1$) & $1.0^{+0.9}_{-0.4}$ ($0.9^{+0.8}_{-0.3}$, $1.0^{+0.5}_{-0.4}$, $0.7$) \\
	N20-BH2.7-$\alpha$2.0 & $27.4^{+23.4}_{-10.4}$ ($23.6^{+20.2}_{-8.9}$, $19.8^{+6.2}_{-5.1}$, $14.4$) & $5.6^{+4.7}_{-1.9}$ ($4.5^{+3.8}_{-1.6}$, $4.6^{+1.8}_{-1.5}$, $3.2$) & $1.1^{+1.0}_{-0.4}$ ($0.9^{+0.8}_{-0.3}$, $1.0^{+0.5}_{-0.4}$, $0.7$) \\
	W15-BH2.7-$\alpha$2.0 & $28.7^{+24.6}_{-10.9}$ ($23.7^{+20.3}_{-9.0}$, $20.7^{+6.5}_{-5.3}$, $15.2$) & $6.1^{+5.1}_{-2.1}$ ($4.7^{+3.9}_{-1.6}$, $5.1^{+2.0}_{-1.6}$, $3.5$) & $1.3^{+1.1}_{-0.4}$ ($1.0^{+0.9}_{-0.3}$, $1.2^{+0.6}_{-0.5}$, $0.8$) \\
	W20-BH2.7-$\alpha$2.0 & $32.6^{+27.9}_{-12.3}$ ($24.9^{+21.3}_{-9.4}$, $23.5^{+7.4}_{-6.0}$, $17.2$) & $7.4^{+6.1}_{-2.6}$ ($5.2^{+4.3}_{-1.8}$, $6.1^{+2.3}_{-1.9}$, $4.2$) & $1.7^{+1.5}_{-0.6}$ ($1.2^{+1.0}_{-0.4}$, $1.5^{+0.7}_{-0.6}$, $1.1$) \\
	\hline
	W18-BH2.3-$\alpha$2.0 & $24.8^{+21.2}_{-9.4}$ ($21.7^{+18.6}_{-8.2}$, $17.9^{+5.7}_{-4.6}$, $13.1$) & $4.8^{+4.0}_{-1.7}$ ($4.1^{+3.4}_{-1.4}$, $4.0^{+1.6}_{-1.3}$, $2.8$) & $0.9^{+0.8}_{-0.3}$ ($0.8^{+0.7}_{-0.3}$, $0.8^{+0.4}_{-0.3}$, $0.6$) \\
	W18-BH2.7-$\alpha$2.0 & $28.8^{+24.6}_{-10.9}$ ($24.2^{+20.7}_{-9.2}$, $20.8^{+6.6}_{-5.3}$, $15.2$) & $6.0^{+5.1}_{-2.1}$ ($4.8^{+4.0}_{-1.7}$, $5.0^{+2.0}_{-1.6}$, $3.4$) & $1.3^{+1.1}_{-0.4}$ ($1.0^{+0.9}_{-0.3}$, $1.2^{+0.6}_{-0.5}$, $0.8$) \\
	W18-BH3.1-$\alpha$2.0 & $32.3^{+27.6}_{-12.2}$ ($26.2^{+22.4}_{-9.9}$, $23.3^{+7.3}_{-6.0}$, $17.0$) & $7.3^{+6.1}_{-2.6}$ ($5.4^{+4.5}_{-1.9}$, $6.0^{+2.3}_{-1.9}$, $4.1$) & $1.7^{+1.5}_{-0.6}$ ($1.2^{+1.1}_{-0.4}$, $1.5^{+0.7}_{-0.6}$, $1.1$) \\
	W18-BH3.5-$\alpha$2.0 & $35.4^{+30.3}_{-13.4}$ ($28.1^{+24.0}_{-10.7}$, $25.5^{+8.1}_{-6.5}$, $18.7$) & $8.6^{+7.2}_{-3.0}$ ($6.1^{+5.1}_{-2.1}$, $7.1^{+2.6}_{-2.1}$, $4.9$) & $2.2^{+1.9}_{-0.8}$ ($1.5^{+1.3}_{-0.5}$, $2.0^{+0.9}_{-0.7}$, $1.4$) \\
	\hline
	W18-BH2.7-$\alpha$1.0 & $28.8^{+24.6}_{-10.9}$ ($24.1^{+20.6}_{-9.1}$, $20.8^{+6.6}_{-5.3}$, $15.2$) & $6.0^{+5.0}_{-2.1}$ ($4.8^{+4.0}_{-1.7}$, $5.0^{+1.9}_{-1.5}$, $3.4$) & $1.4^{+1.2}_{-0.5}$ ($1.1^{+0.9}_{-0.4}$, $1.3^{+0.6}_{-0.5}$, $0.9$) \\
	W18-BH2.7-$\alpha$1.5 & $28.8^{+24.6}_{-10.9}$ ($24.2^{+20.7}_{-9.2}$, $20.8^{+6.6}_{-5.3}$, $15.2$) & $6.1^{+5.1}_{-2.1}$ ($4.8^{+4.0}_{-1.7}$, $5.0^{+1.9}_{-1.6}$, $3.4$) & $1.3^{+1.2}_{-0.5}$ ($1.0^{+0.9}_{-0.4}$, $1.2^{+0.6}_{-0.5}$, $0.9$) \\
	W18-BH2.7-$\alpha$2.0 & $28.8^{+24.6}_{-10.9}$ ($24.2^{+20.7}_{-9.2}$, $20.8^{+6.6}_{-5.3}$, $15.2$) & $6.0^{+5.1}_{-2.1}$ ($4.8^{+4.0}_{-1.7}$, $5.0^{+2.0}_{-1.6}$, $3.4$) & $1.3^{+1.1}_{-0.4}$ ($1.0^{+0.9}_{-0.3}$, $1.2^{+0.6}_{-0.5}$, $0.8$) \\
	W18-BH2.7-$\alpha$2.5 & $28.8^{+24.6}_{-10.9}$ ($24.2^{+20.7}_{-9.2}$, $20.8^{+6.6}_{-5.3}$, $15.2$) & $6.0^{+5.1}_{-2.1}$ ($4.7^{+4.0}_{-1.6}$, $5.0^{+2.0}_{-1.6}$, $3.5$) & $1.2^{+1.1}_{-0.4}$ ($1.0^{+0.8}_{-0.3}$, $1.1^{+0.6}_{-0.4}$, $0.8$) \\
	W18-BH2.7-$\alpha$3.0 & $28.8^{+24.6}_{-10.9}$ ($24.2^{+20.7}_{-9.2}$, $20.8^{+6.6}_{-5.3}$, $15.2$) & $6.0^{+5.0}_{-2.1}$ ($4.7^{+4.0}_{-1.6}$, $5.0^{+2.0}_{-1.6}$, $3.4$) & $1.1^{+1.0}_{-0.4}$ ($0.9^{+0.8}_{-0.3}$, $1.1^{+0.6}_{-0.4}$, $0.8$) \\
	\hline
	W18-BH2.7-$\alpha$2.0-He33 & $23.7^{+20.3}_{-9.0}$ ($20.2^{+17.3}_{-7.7}$, $17.2^{+5.4}_{-4.4}$, $12.5$) & $4.9^{+4.1}_{-1.7}$ ($4.0^{+3.3}_{-1.4}$, $4.1^{+1.6}_{-1.3}$, $2.8$) & $1.0^{+0.9}_{-0.3}$ ($0.8^{+0.7}_{-0.3}$, $0.9^{+0.5}_{-0.4}$, $0.7$) \\
	W18-BH2.7-$\alpha$2.0-He100 & $13.6^{+11.6}_{-5.2}$ ($12.4^{+10.6}_{-4.7}$, $10.0^{+3.2}_{-2.6}$, $7.2$) & $2.7^{+2.3}_{-0.9}$ ($2.4^{+2.0}_{-0.8}$, $2.3^{+0.9}_{-0.7}$, $1.6$) & $0.5^{+0.4}_{-0.2}$ ($0.5^{+0.4}_{-0.2}$, $0.5^{+0.3}_{-0.2}$, $0.3$) \\
	\hline
	S19.8-BH2.3-$\alpha$2.0 & $24.4^{+20.9}_{-9.2}$ ($22.8^{+19.5}_{-8.6}$, $17.6^{+5.6}_{-4.5}$, $12.9$) & $4.6^{+3.8}_{-1.6}$ ($4.2^{+3.5}_{-1.5}$, $3.9^{+1.6}_{-1.3}$, $2.6$) & $0.8^{+0.7}_{-0.3}$ ($0.8^{+0.7}_{-0.3}$, $0.8^{+0.4}_{-0.3}$, $0.5$) \\
	W18-BH3.5-$\alpha$1.0 & $35.3^{+30.2}_{-13.4}$ ($28.1^{+24.0}_{-10.6}$, $25.5^{+8.0}_{-6.5}$, $18.6$) & $8.4^{+7.0}_{-3.0}$ ($6.0^{+5.0}_{-2.1}$, $6.8^{+2.5}_{-2.0}$, $4.7$) & $2.5^{+2.1}_{-0.8}$ ($1.6^{+1.4}_{-0.6}$, $2.1^{+0.9}_{-0.7}$, $1.5$) \\
	W15-BH3.5-$\alpha$1.0 & $35.4^{+30.3}_{-13.4}$ ($27.8^{+23.8}_{-10.5}$, $25.6^{+8.0}_{-6.5}$, $18.7$) & $8.6^{+7.2}_{-3.0}$ ($6.1^{+5.1}_{-2.1}$, $7.0^{+2.5}_{-2.1}$, $4.8$) & $2.6^{+2.2}_{-0.9}$ ($1.7^{+1.4}_{-0.6}$, $2.2^{+1.0}_{-0.8}$, $1.5$) \\
	W20-BH2.7-$\alpha$1.0 & $32.5^{+27.8}_{-12.3}$ ($24.8^{+21.2}_{-9.4}$, $23.5^{+7.4}_{-6.0}$, $17.2$) & $7.3^{+6.1}_{-2.6}$ ($5.2^{+4.3}_{-1.8}$, $6.0^{+2.2}_{-1.8}$, $4.1$) & $2.0^{+1.7}_{-0.7}$ ($1.3^{+1.1}_{-0.5}$, $1.7^{+0.8}_{-0.6}$, $1.2$) \\
	W20-BH3.1-$\alpha$1.0 & $37.2^{+31.8}_{-14.1}$ ($27.7^{+23.7}_{-10.5}$, $26.9^{+8.5}_{-6.9}$, $19.6$) & $9.0^{+7.5}_{-3.2}$ ($6.1^{+5.0}_{-2.1}$, $7.3^{+2.6}_{-2.1}$, $5.0$) & $2.7^{+2.3}_{-0.9}$ ($1.7^{+1.5}_{-0.6}$, $2.3^{+1.0}_{-0.8}$, $1.6$) \\
	W20-BH3.5-$\alpha$2.0 & $41.8^{+35.8}_{-15.9}$ ($30.4^{+26.0}_{-11.5}$, $30.2^{+9.5}_{-7.7}$, $22.1$) & $11.1^{+9.2}_{-3.9}$ ($7.2^{+5.9}_{-2.5}$, $9.0^{+3.3}_{-2.7}$, $6.2$) & $3.1^{+2.6}_{-1.1}$ ($1.9^{+1.7}_{-0.7}$, $2.7^{+1.2}_{-1.0}$, $1.9$)
	\enddata
	\tablecomments{The given values correspond to the unoscillated \nuebar\ DSNB flux spectra using the SFH from \citet{2014ApJ...790..115M} with its associated $\pm1\sigma$ uncertainty. In parentheses, the values for the case of a complete flavor swap ($\nuebar \leftrightarrow \nux$) are provided as well as the results for a SFH according to the EBL reconstruction model by the \citet{2018Sci...362.1031F} and for the SFH of \cite{2014MadauDickinson}.}
\end{deluxetable*}

The various parameter combinations considered in our study lead to a wide spread between the DSNB flux spectra, as can be seen in the left panel of Figure~\ref{fig:dsnb_extremes}. At high energies, the spectral tails of our different models fan out over more than an order of magnitude, with our most extreme cases yielding an integrated flux $\Phi_{>17.3}$ that clearly exceeds the SK limit. To guide the eye, we roughly mark the region of such disfavored models (with $\Phi_{>17.3}\,\mathord{\gtrsim}\,\unit{3.1}{cm^{-2}s^{-1}}$) by a red shaded band, while flux spectra with $\Phi_{>17.3}\,\mathord{\lesssim}\,\unit{3.1}{cm^{-2}s^{-1}}$, including our fiducial prediction (dashed line; see Section~\ref{sec:fiducial_model}), lie in the gray band. We take the specific model ``W20-BH3.5-$\alpha$2.0'' (i.e., Z9.6\,\&\,W20 neutrino engine, $\MBH\,\mathord{=}\,\unit{3.5}{\msun}$, $\alphaBH\,\mathord{=}\,2.0$) with the best-fit parameters taken for the SFH from \citet{2014ApJ...790..115M} as a bounding case; it yields an integrated flux $\Phi_{>17.3}\,\mathord{=}\,\unit{3.09}{cm^{-2}s^{-1}}$, just within the uncertainty range of the the SK limit (2.8--3.1\,cm$^{-2}$s$^{-1}$). We should emphasize, however, that this does not define a rigorous border line, since spectra with quite different values of the slope parameter $E_0$ (Equation~\eqref{eq:exponential_dsnb_fit}) can yield similar integrated fluxes in the energy range above $\unit{17.3}{MeV}$.

In the right panel of Figure~\ref{fig:dsnb_extremes}, we therefore plot $\Phi_{>17.3}$ as a function of the fit parameter $E_0$ for a selection of models reaching close to (or beyond) the SK bound, which is marked by the red shaded region (with its uncertainty indicated by two slightly shifted lines). This plot is also intended to facilitate a comparison with other works (see, e.g., table~1 in \citealt{2008JCAP...08..033L}, and figure~19 in \citealt{2012PhRvD..85e2007B}). The tendency of greater integrated fluxes $\Phi_{>17.3}$ for higher values of $E_0$ is obvious, yet there is significant scatter. Especially the large uncertainty connected to the cosmic core-collapse rate ($\pm1\sigma$ interval from \citealt{2014ApJ...790..115M}, indicated by error bars) impedes definite conclusions. Nonetheless, models of our study with the most extreme combinations of parameters such as the different cases of W20-BH3.5, which possess a strong contribution from failed SNe and thus large values of $E_0$ (see Section~\ref{subsec:dsnb_parameter_study} and Table~\ref{tab:exp_fit}), are already disfavored, because their fluxes $\Phi_{>17.3}$ reach beyond the SK limit (unless a minimal \RCC\ is taken). Also a less extreme value of the NS mass limit or a neutrino engine with a lower fraction of BH-formation events can lead to an integrated flux close to the SK bound: models W20-BH3.1-$\alpha$1.0 and W15-BH3.5-$\alpha$1.0 (not shown in Figure~\ref{fig:dsnb_extremes}) yield $\Phi_{>17.3}\,\mathord{=}\,\unit{2.7^{+2.3}_{-0.9}}{cm^{-2}s^{-1}}$ and $\Phi_{>17.3}\,\mathord{=}\,\unit{2.6^{+2.2}_{-0.9}}{cm^{-2}s^{-1}}$, respectively, with a dominant fraction (85\% and 81\%, respectively) of the \nuebar\ above $\unit{17.3}{MeV}$ originating from BH-formation events. In Table~\ref{tab:overview}, we provide the total integrated fluxes ($\Phi_\mathrm{tot}$), the fluxes within the observational window of 10--30\,MeV ($\Phi_{10-30}$), as well as the flux integrals above $\unit{17.3}{MeV}$ ($\Phi_{>17.3}$) for a subset of our DSNB models.

Unlike the experimental DSNB flux limits of \citet{2003PhRvL..90f1101M}, those provided by \citet{2012PhRvD..85e2007B} depend on the DSNB model employed. Nevertheless, for an energy threshold close to $\sim$20\,MeV, the flux limits are rather insensitive to the shape of the DSNB spectrum as pointed out by \citet{2008JCAP...08..033L}. In any case, the (Fermi-Dirac) spectral temperatures ($\unit{3}{MeV}\,\mathord{\leqslant}\,T_{\nu}\,\mathord{\leqslant}\,\unit{8}{MeV}$) that \citet{2012PhRvD..85e2007B} considered for their modeling of a ``typical'' SN source spectrum, lead to DSNB spectra with slope parameters $E_0$ that cover the range of values obtained in our work.\footnote{In an analytic study, \citet{2007PhRvD..75g3022L} showed that the spectral temperatures of the employed SN source spectrum (before integration over redshifts) translates into the slope parameter $E_0$ of the DSNB spectrum up to some tens of percents.} Repeating their analysis of computing upper DSNB flux limits with our DSNB models should therefore lead to comparable bounds. Instead, we simply compare the experimental flux limit of $\unit{(2.8-3.1)}{cm^{-2}s^{-1}}$ to a subset of our model predictions in Figure~\ref{fig:dsnb_extremes}. Naturally, this cannot replace a sophisticated statistical analysis, which is beyond the scope of this work.

Our fiducial model (W18-BH2.7-$\alpha$2.0) yields an integrated flux of $\Phi_{>17.3}\,\mathord{=}\,\unit{1.3^{+1.1}_{-0.4}}{cm^{-2}s^{-1}}$, which is just below the SK bound, possibly not even by a factor of 2. Intriguingly, \citet{2012PhRvD..85e2007B} pointed out that there might already be a hint of a signal in the SK-II and SK-III data, giving hope that the first detection of the DSNB is within close reach now \citep[cf.][]{2004PhRvL..93q1101B, 2006PhRvC..74a5803Y, 2009PhRvD..79h3013H, 2012PhRvD..85d3011K, 2016JPhG...43c0401A, 2017JCAP...11..031P}.

Since the SN neutrino emission is different for heavy-lepton neutrinos compared to electron antineutrinos (see Section~\ref{subsec:flavor_conversions}), a complete (or partial) flavor swap ($\nuebar\,\mathord{\leftrightarrow}\,\nux$) would affect our previous conclusions: In the case of IH (\nuebar\ survival probability $\bar{p}\,\mathord{\simeq}\,0$), the integrated flux above $\unit{17.3}{MeV}$ of our most extreme model (W20-BH3.5-$\alpha$1.0) decreases by 39\% from $\unit{3.5^{+2.9}_{-1.2}}{cm^{-2}s^{-1}}$ to $\unit{2.1^{+1.8}_{-0.7}}{cm^{-2}s^{-1}}$, which is just below the SK bound (however, note the large uncertainties due to the cosmic core-collapse rate). For the case of NH ($\bar{p}\,\mathord{\simeq}\,0.7$), we obtain $\Phi_{>17.3}\,\mathord{=}\,\unit{3.1^{+2.6}_{-1.1}}{cm^{-2}s^{-1}}$, which is still somewhat above the SK limit. Applying neutrino flavor conversions to our fiducial model, the effects are much reduced, as described in Section~\ref{subsec:flavor_conversions} (see left panel of Figure~\ref{fig:dsnb_uncert} and Table~\ref{tab:dsnb_flavor_conversions}): $\Phi_{>17.3}$ decreases by only 7\% (21\%) from $\unit{1.3^{+1.1}_{-0.4}}{cm^{-2}s^{-1}}$ to $\unit{1.2^{+1.0}_{-0.4}(1.0^{+0.9}_{-0.3})}{cm^{-2}s^{-1}}$ for NH (IH), still reaching close to the SK bound. In Table~\ref{tab:overview}, the first values in the parentheses correspond to the case of a complete flavor swap.
	
Taking alternative SFHs such as the ones from \citet{2014MadauDickinson} or the \citet{2018Sci...362.1031F}, which we discussed in Section~\ref{sec:fiducial_model}, leads to lower predictions of the DSNB flux compared to our fiducial model, which employs the SFH from \citet{2014ApJ...790..115M}. This implies weaker constraints by the experimental limit, as can be seen in Table~\ref{tab:overview}, where the second and third values in parentheses show the results for a SFH according to the EBL reconstruction by the \citet{2018Sci...362.1031F} and for the SFH from \cite{2014MadauDickinson}, respectively. Independent of the chosen model parameters, the integrated fluxes are reduced compared to the cases with the SFH from \citet{2014ApJ...790..115M}. The flux values of $\Phi_{>17.3}$ for the case of the SFH from \citet{2014MadauDickinson} lie about one third below the ones when taking the best-fit SFH from \citet{2014ApJ...790..115M} (roughly corresponding to the $-$1$\sigma$ lower-limit case of \citet{2014ApJ...790..115M}), whereas there is still significant overlap between the flux values for the SFH of the \citet{2018Sci...362.1031F} and our fiducial flux values (also note the large uncertainty ranges). At the same time, the values of the slope parameter $E_0$ are increased (i.e. the spectral tails are lifted) when taking the SFH of the \citet{2018Sci...362.1031F} or the one from \citet{2014MadauDickinson} (see Table~\ref{tab:exp_fit} and Figure~\ref{fig:dsnb_SFR}). Apparently, the large degeneracy between the parameters entering the flux calculations impedes both precise predictions and the exclusion of models.

\subsection{Comparison with Previous Works}\label{subsec:comparison}

Finally, we compare our DSNB flux predictions with the results of other recent works. For instance, \citet{2017JCAP...11..031P} found a \nuebar-flux above $\unit{11}{MeV}$ in the range of $\unit{(1.4-3.7)}{cm^{-2}s^{-1}}$, with their highest-flux model being a factor of $\sim$3 below the SK limit of \citet{2012PhRvD..85e2007B}. In contrast, our fiducial model yields flux values of $\unit{4.6^{+3.9}_{-1.6}}{cm^{-2}s^{-1}}$ ($\unit{3.9^{+3.3}_{-1.3}}{cm^{-2}s^{-1}}$) above $\unit{11}{MeV}$ in the case we consider neutrino oscillations for NH (IH) \citep[to follow][]{2017JCAP...11..031P}, reaching very close to the SK bound (see Section~\ref{subsec:dsnb_SK_limit}). Likewise, the recent study by \citet{2018JCAP...05..066M} suggested a clearly lower DSNB flux compared to our work (see their figures~3 and 10). These differences between our DSNB estimates and the previous results can be understood by the large variations of the neutrino outputs between the different core-collapse events in our sets of SN and BH-formation models, as shown in Figure~\ref{fig:neutrino_outcomesystematics}. While progenitors at the low end of the considered ZAMS-mass range radiate $\Enutot\,\mathord{\simeq}\,\unit{2\,\mathord{\times}\,10^{53}}{erg}$, the emission increases to values of $\unit{(3\,\mathord{-}\,4)\,\mathord{\times}\,10^{53}}{erg}$ for progenitors above about (11--12)\,\msun. On the other hand, \citet{2017JCAP...11..031P} and \citet{2018JCAP...05..066M} applied the low-energy neutrino signals ($\Enutot\,\mathord{\simeq}\,\unit{2\,\mathord{\times}\,10^{53}}{erg}$) of the s11.2c and z9.6co models considered by them for the entire mass interval between $\sim$8\,\msun\ and $\sim$15\,\msun, which receives a high weight by the IMF in the integration over all core-collapse events. Moreover, both studies make use of failed-SN models which form BHs relatively quickly (within $\lesssim$2\,s after bounce) and therefore radiate less energy ($\lesssim$3.7\,$\times$\,$10^{53}$\,erg) than most of our failed explosions. Each of these two aspects accounts for a reduction of the integral flux by several ten percent compared to our work.

\citet{2018MNRAS.475.1363H} for the first time employed a larger set of neutrino signals in their DSNB study, including seven models of BH-forming, failed explosions. However, the total neutrino energies \Enutot\ radiated from their failed SNe are in general below $\sim$$\unit{3.5\,\mathord{\times}\,10^{53}}{erg}$ (see their figure~5). In contrast, we find total neutrino energies in the cases of failed explosions of up to $\unit{5.2\,\mathord{\times}\,10^{53}}{erg}$ for a NS mass limit of $\MBH\,\mathord{=}\,\unit{2.3}{\msun}$ and of up to $\unit{6.7\,\mathord{\times}\,10^{53}}{erg}$ when using our fiducial value, $\MBH\,\mathord{=}\,\unit{2.7}{\msun}$ (see Figures~\ref{fig:neutrino_outcomesystematics} and \ref{fig:appendix_E_tot_fSNe}), enhancing the integral flux by some ten percent compared to \citet{2018MNRAS.475.1363H}. Accordingly, our study suggests that in particular the inclusion of slowly-accreting progenitors that lead to late BH formation (not considered in previous works) is responsible for a significant contribution to the DSNB.\\\\\\

\section{Summary of Uncertainties} \label{sec:summary_uncertainties}

After having discussed numerous dependencies of the DSNB, we summarize our main results with their corresponding uncertainties in this section. Again, these uncertainties are considered in reference to our fiducial DSNB spectrum, which is based on the Z9.6\,\&\,W18 neutrino engine with 26.9\% BH-formation cases, a baryonic NS mass limit of 2.7\,\msun, a value of $\alphaBH\,\mathord{=}\,2.0$ for the instantaneous neutrino-emission spectrum of failed SNe, no additional contribution from low-mass NS-formation events (i.e., $\chi\,\mathord{=}\,0$; Equation~\eqref{eq:chi}), only single-star progenitors (i.e., no hydrogen-stripped helium stars), no neutrino flavor oscillations, and the best-fit SFH of \citet{2014ApJ...790..115M}. The corresponding DSNB uncertainties can be grouped into the following four categories:

{\bf(1)} \underline{Stellar-diversity uncertainties} (see Sections~\ref{subsec:dsnb_parameter_study}, \ref{subsec:dsnb_LM}, \ref{subsec:dsnb_binary}; Fig\-ures~\ref{fig:dsnb_parameter_study}, \ref{fig:dsnb_He}): These include the still un\-de\-ter\-mined fraction of BH-forming stellar core-collapse events; a possible, still poorly understood contribution from low-mass NS-formation events (AIC, MIC, or ultrastripped SNe); and the relative fraction of helium stars, which serve as a proxy for SN progenitors that have stripped their hydrogen envelopes as a consequence of binary interaction at the end of core-hydrogen burning.

{\bf(2)} \underline{Microphysical uncertainties} (see Sections~\ref{subsec:dsnb_parameter_study}, \ref{subsec:flavor_conversions}; Fig\-ures~\ref{fig:dsnb_parameter_study}, \ref{fig:dsnb_uncert}): These concern, on the one hand, the still in\-com\-plete\-ly known high-density EoS of NS matter with the corresponding NS-mass limit, and, on the other hand, possible effects of neutrino flavor conversions.

{\bf(3)} \underline{Modeling uncertainties} (see Sections~\ref{subsec:dsnb_parameter_study}, \ref{subsec:remaining_uncertainties};
Fig\-ures~\ref{fig:dsnb_parameter_study}, \ref{fig:dsnb_uncert}): These are connected to our numerical de\-scrip\-tion of the neutrino emission from successful and failed SNe. Here we subsume approximations of the spectral-shape parameter (\alphaBH) for the instantaneous neutrino-emission spectrum, of the total neutrino energy loss from NS- and BH-formation events, and of the mean energy of the time-integrated \nuebar\ spectrum.

{\bf(4)} \underline{Astrophysical uncertainties} (see Section~\ref{sec:fiducial_model}; Fig\-ure~\ref{fig:dsnb_SFR}): These refer to the still insufficiently con\-strained cosmic SFH, for which we tested different rep\-re\-sen\-ta\-tions.

\begin{figure*}
	\includegraphics[width=\textwidth]{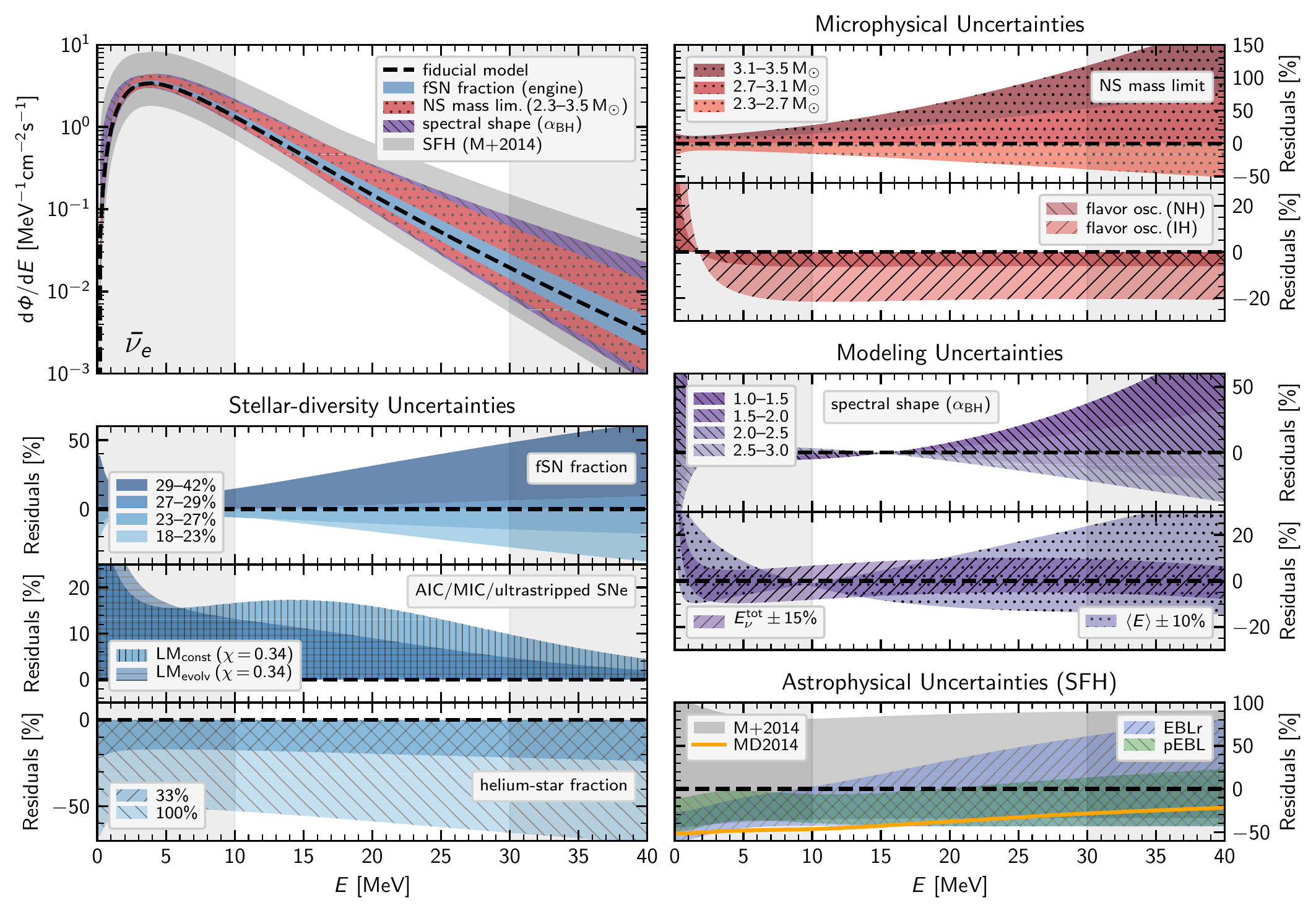}
	\caption{
	Overview of DSNB uncertainties. The upper left panel shows the \nuebar-flux spectrum, $\diff\Phi/\diff E$, of our fiducial DSNB model (dashed line) together with its major uncertainties stacked on top of each other (shaded/hatched bands): the failed-SN (fSN) fraction (17.8\% to 41.7\% of core-collapse progenitors depending on the strength of the neutrino engine); the NS baryonic mass limit (2.3\,\msun\ to 3.5\,\msun); the instantaneous spectral-shape parameter for the emission from failed SNe ($1.0\,\mathord{\leqslant}\,\alphaBH\,\mathord{\leqslant}\,3.0$); and the uncertainty connected to the cosmic SFH \citep[$\pm1\sigma$ limits of][]{2014ApJ...790..115M}. The resulting ``total'' uncertainty band is the same as in Figure~\ref{fig:dsnb_extremes}. The lower left panel and the right panels show the residuals of our DSNB models where only one parameter is changed relative to the fiducial model, while all other parameters are kept at their default values, grouped by stellar-diversity uncertainties, microphysical uncertainties, modeling uncertainties, and astrophysical uncertainties (see Section~\ref{sec:summary_uncertainties} for more details). LM$_\mathrm{const}$ and LM$_\mathrm{evolv}$ denote cases where the rate densities of low-mass NS-formation events (AIC, MIC, ultrastripped SNe) are constant or evolve with redshift, respectively (both for a value of $\chi\,\mathord{=}\,0.34$, which corresponds to a relative abundance of low-mass NS-formation events of 34\% compared to ``conventional'' core-collapse SNe plus fSNe; Equation~\eqref{eq:chi}). In each panel, gray-shaded vertical bands frame the approximate detection window.\\
	\label{fig:summary_uncertainties}}
\end{figure*}

The upper left panel of Figure~\ref{fig:summary_uncertainties} shows our fiducial DSNB \nuebar-flux spectrum with its main uncertainties (failed-SN fraction, NS baryonic mass limit, and spectral shape of the neutrino emission from failed SNe in terms of \alphaBH), stacked on top of each other. The uncertainty of the SFH is additionally applied to the upper and lower limits of the uncertainty range. The impact of the different uncertainties according to the four categories listed above is illustrated by their corresponding residuals relative to the fiducial spectrum in the four additional panels of Figure~\ref{fig:summary_uncertainties}.

Concerning stellar-diversity uncertainties, a large failed-SN fraction can enhance the DSNB spectrum by up to $\sim$50\%, whereas a considerable fraction of helium stars can shift the spectrum in the opposite direction by about the same margin. Among the microphysical uncertainties, the NS baryonic mass limit has the major impact, but an assumed value of 3.5\,\msun\ appears to be on the extreme side in view of current gravitational-wave and kilonova constraints, which seem to point to a mass limit around 2.7\,\msun\ \citep[e.g.,][]{2017ApJ...850L..19M}, which we applied for our fiducial spectrum. Future gravitational-wave and kilonova measurements as well as astrophysical observations by NICER \citep{2019ApJ...887L..24M} are likely to constrain this mass limit with increasingly better precision. Among the modeling uncertainties, which are specific to our approach based on large sets of core-collapse simulations with approximate neutrino treatment, the spectral-shape parameter \alphaBH\ has the dominant influence (up to $\sim$35\% enhancement of the DSNB \nuebar\ spectrum at a neutrino energy of 30\,MeV seem possible). However, this uncertainty as well as the (subdominant) ones connected to the total gravitational binding-energy release and the mean energy of the radiated neutrinos will also be reduced once the NS EoS is better determined and neutrino-signal predictions from detailed transport calculations for large sets of NS- and BH-formation events become available.

Finally, the SFH can make changes of the DSNB \nuebar\ spectrum by up to a factor of two and is certainly a much desirable aspect for further improvements through astronomical observations. If this can be achieved, DSNB measurements will provide an interesting handle to deduce information on the stellar core-collapse diversity, whose effects were the main focus of our work. Conversely, if theoretical and observational advances lead to a better understanding of the population of core-collapse progenitors and their final destinies (i.e., their fates as successful or failed SNe), the forthcoming detection of the DSNB will be able to yield valuable constraints on the SFH, complementing information from surveys for astronomical transients such as LSST \citep{2002SPIE.4836...10T}, which may not be able to reveal the rate of intrinsically faint stellar-death events.

\section{Conclusions} \label{sec:conclusions}

In this work we aimed at performing a comprehensive investigation of current astrophysical uncertainties in the predictions of the DSNB flux spectrum. Our study was based on large sets of single-star models \citep{2016ApJ...821...38S} and helium-star models \citep{2020ApJ...890...51E} for successful and failed SNe. The helium-star progenitors from \citet{2019ApJ...878...49W} were considered as a proxy of massive stars who evolved to the onset of stellar core collapse after stripping their hydrogen envelopes at the end of core-hydrogen burning through binary interaction, e.g., by common-envelope evolution or Roche-lobe overflow \citep[see][]{2012Sci...337..444S}. The progenitor sets contained between 100 and 200 stellar models with ZAMS masses between $<$9\,M$_\odot$ and 120\,M$_\odot$. These models were exploded (or failed to explode) in spherically symmetric simulations with the \textsc{Prometheus-HotB} code, employing a parametrized neutrino engine that was calibrated to reproduce the basic properties of the well-studied SNe of SN~1987A and the Crab Nebula.

Our stellar core-collapse models provided the total energy output in neutrinos from NS- and BH-formation events as well as the time-dependent mean energies of the radiated neutrinos, specifically of \nuebar. Since the treatment of the PNS cooling and its neutrino emission in these large model sets was only approximate, we compared our estimates of the total neutrino energy loss with the gravitational binding energies of NSs (up to their mass limit) as given by the radius dependent fit formula of \citet{2001ApJ...550..426L}. We found good agreement for NS radii of 11--12\,km, which is the range favored by recent astrophysical observations and nuclear theory and experiments. Moreover, we used NS- and BH-formation simulations with the \textsc{Prometheus-Vertex} code (which employs a state-of-the-art treatment of neutrino transport based on a Boltzmann-moment-closure scheme and a mixing-length treatment of PNS convection) to calibrate degrees of freedom in our approximate neutrino signals, for example the shape of the time-dependent neutrino spectrum, which we characterized by the widely used $\alpha$-fit of \citet{2003ApJ...590..971K}. We note that our treatment of the neutrino emission by successful and failed SNe is not based on a detailed microphysical PNS model, but nevertheless our procedure of combining information from \textsc{Prometheus-HotB} simulations with neutrino data from \textsc{Prometheus-Vertex} models enables our study to capture the generic properties of neutrino signals radiated from NS- and BH-formation cases.

In the course of our investigation we varied the neutrino engine, whose power is connected to the properties of the progenitor model considered for SN~1987A, yielding different relative fractions of successful SN events in contrast to failed explosions with BH formation. Moreover, we explored the effects of alternative paths to NS formation besides the stellar core-collapse channel, which could be associated with the accretion-induced or merger-induced collapse (AIC or MIC) of white dwarfs or with ECSN and ultrastripped core-collapse progenitors in close binary systems. All of these cases would preferentially lead to the formation of rather low-mass NSs with little postshock accretion, for which reason we treat this component in analogy to the ECSNe \citep{2010PhRvL.104y1101H} that are included in our standard set of stellar core-collapse models. We also varied the still uncertain NS mass limit (above which a transiently stable, accreting PNS collapses to a BH) between the currently measured largest masses of galactic neutron stars (2.3\,M$_\odot$ baryonic and $\sim$2.0\,M$_\odot$ gravitational) and the maximum mass that can be stabilized by still viable microphysical EoSs (3.5\,M$_\odot$ baryonic and $\sim$2.75\,M$_\odot$ gravitational). Moreover, we varied the shape parameter, $\alphaBH$, of the time-dependent neutrino emission spectrum from failed explosions and considered, in a standard way, the effects of neutrino flavor oscillations.

Our fiducial case employs a neutrino engine that is fully compatible with observationally determined NS and BH masses as well as chemogalactic constraints on SN nucleosynthesis, a NS mass limit of 2.7\,M$_\odot$ baryonic and $\sim$2.25\,M$_\odot$ gravitational mass (compatible with recent limits from GW170817), and a best-fit $\alpha$-spectrum for the time-dependent neutrino emission. With the SFH adopted from \cite{2014ApJ...790..115M}, it yields a total DSNB \nuebar-flux of $28.8^{+24.6}_{-10.9}$\,cm$^{-2}$s$^{-1}$ with a contribution of $6.0^{+5.1}_{-2.1}$\,cm$^{-2}$s$^{-1}$ in the energy interval of [10,30]\,MeV, which is most favorable for measurements. Our best value of the predicted flux for \nuebar\ energies $>$\,17.3\,MeV is $1.3^{+1.1}_{-0.4}$\,cm$^{-2}$s$^{-1}$, which is slightly lower than the result of
1.6\,cm$^{-2}$s$^{-1}$
published by \citet[with an update at NNN05]{2003APh....18..307A} and about a factor of two below the current SK limit (see \citealt{2012PhRvD..85e2007B}; preliminary, updated value of 2.7\,cm$^{-2}$s$^{-1}$ at 90\% CL by the Super-Kamiokande Collaboration; El Hedri et al. 2020, poster at Neutrino 2020; Nakajima et al. 2020, talk at Neutrino 2020).

Because of the currently expected narrow mass range of ECSNe from single stars, these events yield a negligible contribution to the DSNB. Similarly, the tested alternative low-mass NS-formation channel via AIC and MIC events or SNe from ultrastripped progenitors can contribute on a significant level ($\geqslant$\,10\%) only in the case of an implausibly large constant event rate or in the case of an evolving rate on the level of the cosmic SN~Ia rate. But even then the enhancement of the DSNB spectrum would happen mainly at low neutrino energies $\lesssim$\,10\,MeV and thus outside of the most favorable energy window for detection.

Our study confirms previous results \citep[e.g.,][]{2009PhRvL.102w1101L, 2012PhRvD..85d3011K, 2015ApJ...804...75N, 2016ApJ...827...85H, 2018ApJ...869...31H}, which were based on the consideration of exemplary cases of BH formation, that an increased fraction of failed SNe flattens the exponential-like decline of the DSNB spectrum beyond its peak and lifts the high-energy tail of the spectrum. This effect can be observed both in our model sets with weaker neutrino engines, where a larger fraction of stars collapses to BHs, and, particularly strongly, in those model sets where we assumed a high value for the maximum NS mass. The rise of the high-energy spectrum is mainly connected to core-collapse events with a long delay time until BH formation, where the mass-accreting PNS radiates harder neutrino spectra and releases a considerably higher total binding energy. Correspondingly, the high-energy tail of the DSNB spectrum varies by a factor of 6.6 at 30\,MeV and the DSNB flux values above 17.3\,MeV, $\Phi_{>17.3}$, by a factor of 3.9 between the limits of 0.8\,cm$^{-2}$s$^{-1}$ and 3.1\,cm$^{-2}$s$^{-1}$ (for the models S19.8-BH2.3-$\alpha$2.0 compared to W20-BH3.5-$\alpha$2.0). A similar effect, though considerably weaker (about 14\% increase of $\Phi_{>17.3}$ relative to our fiducial case), can be witnessed when the radiated neutrino spectra from failed explosions are considered to be antipinched ($\alphaBH\,\mathord{=}\,1$) at all times instead of being Maxwell-Boltzmann like ($\alphaBH\,\mathord{=}\,2$).

A larger population of hydrogen-stripped binary progenitors of SNe can have a significant impact on the DSNB spectrum, because, compared to single stars, the ZAMS mass range of stars that experience stellar core collapse is shifted upwards by $\sim$5\,M$_\odot$ to the more IMF-suppressed high-mass regime (compare Figure~\ref{fig:neutrino_outcomesystematics_He} with Figure~\ref{fig:neutrino_outcomesystematics}). At the same time, a lower fraction of BH-formation events reduces the high-energy tail. Correspondingly, we found a reduction of the total DSNB \nuebar-flux by $\sim$18\% (53\%) and a reduction of $\Phi_{>17.3}$ by $\sim$20\% (60\%) if 33\% (100\%) of the core-collapse progenitors evolve as helium stars (see Figure~\ref{fig:dsnb_He}). Neutrino flavor oscillations have an effect that is, at most, of roughly comparable magnitude. A complete swap of $\bar\nu_e$ and $\nu_x$ (the most extreme case) reduces our predictions of the total DSNB \nuebar-flux again by $\sim$16\% and of $\Phi_{>17.3}$ by $\sim$21\% relative to our fiducial case.

A major uncertainty in all predictions of the DSNB, however, is the still insufficiently constrained stellar core-collapse rate. With a defined form for the stellar IMF this refers to uncertainties in the cosmic SFH, which render all estimates uncertain within a factor of roughly 3 (considering the $\pm$1\,$\sigma$ range of \citealt{2014ApJ...790..115M}). Rigorously constraining individual inputs of the DSNB by measurements is further hampered by the existing large degeneracies between different effects of relevance. Nevertheless, the most extreme cases included in our study, which combine a very large fraction of BH-forming core-collapse events (up to an IMF-weighted fraction of 42\%) and/or the highest considered value of the NS mass limit (3.5\,M$_\odot$ baryonic and $\sim$2.75\,M$_\odot$ gravitational mass), seem to be ruled out by the current SK limit already.

Some of the physical quantities entering the DSNB calculations can be expected to be better constrained in the not too distant future. An increasing number of gravitational-wave detections from binary-NS mergers \citep{2010CQGra..27q3001A} will yield more information on the maximum NS mass and NS radii, placing tighter constraints on the high-density EoS; a steadily improved statistics of binary BH mergers might lead to better constraints on BH formation events and progenitors (see, e.g., \citealt{2020ApJ...896...56W}); long-baseline neutrino oscillation experiments should be able to determine the neutrino mass hierarchy \citep[e.g.,][]{2013arXiv1307.7335L}; and upcoming wide-field surveys such as LSST \citep{2002SPIE.4836...10T} will measure the rate of \textit{visible} SNe (below $z\,\mathord{\sim}\,1$) to good accuracy.

Complementary to these perspectives, future observations of the DSNB will probe the \textit{entire} population of stellar core-collapse events with its full diversity, particularly including \textit{faint} and \textit{failed} explosions \citep[cf.][]{2010PhRvD..81h3001L}. This opens the chance to better constrain the cosmic core-collapse rate as well as the fraction of BH-forming, failed SNe \citep{2018JCAP...05..066M}. Moreover, the DSNB may even carry the imprint of new physics \citep[e.g.,][]{2004PhRvD..70a3001F, 2014JCAP...06..014F, 2018JCAP...06..019J, 2020arXiv200713748D, 2020arXiv201110933T}. These exciting prospects for both particle and astrophysics motivate ongoing efforts to steadily improve the theoretical predictions of the DSNB. The next upgrade in this direction should be fully self-consistent successful and failed SN simulations with a detailed modeling of the neutrino signal radiated by the forming compact remnant.\\

\textit{Note added:} When our paper was already in the production process, we got notice of a new arXiv posting by \citet{2020arXiv201208524H}, dealing with the impact of mass transfer and mergers during binary evolution on the DSNB spectrum. We agree that this progenitor component of core-collapse SNe, which we did not take into account in our study, can potentially increase the DSNB flux, partially compensating the reducing influence of stripped progenitors discussed in our work. However, the relevant effects of mass transfer and mergers depend on a variety of uncertain processes during stellar evolution and are hard to assess in quantitative detail.\\

Our results are made available for download upon request on the following website: \url{https://wwwmpa.mpa-garching.mpg.de/ccsnarchive/archive.html}

\acknowledgments
We thank the anonymous referee for useful comments, which helped improve our paper. We also thank G. Raffelt, J. Sawatzki, and L. Oberauer for stimulating discussions. We are grateful to G. Stockinger for improving the treatment of energy transfer by neutrino-nucleon scattering in the neutrino-transport solver of \textsc{Prometheus-HotB}, to A. Heger for providing his z9.6 progenitor model, and to R. Bollig for providing his s20.0 neutrino signals. Funding by the Deutsche Forschungsgemeinschaft (DFG, German Research Foundation) through Sonderforschungsbereich (Collaborative Research Center) SFB-1258 ``Neutrinos and Dark Matter in Astro- and Particle Physics (NDM)'' and under Germany's Excellence Strategy through Cluster of Excellence ORIGINS (EXC-2094)---390783311, and by the European Research Council through Grant ERC-AdG No.~341157-COCO2CASA is acknowledged.

\software{\textsc{Prometheus-HotB} \citep{1996A&A...306..167J, 2003A&A...408..621K, 2006A&A...457..963S, 2016ApJ...818..124E, 2020ApJ...890...51E}, NumPy \citep{NumPy}, Scipy \citep{SciPy}, IPython \citep{IPython}, Matplotlib \citep{Matplotlib}, Bibmanager \citep{bibmanager}.}\\

\clearpage

\section*{Appendix}
\renewcommand\thesection{\Alph{section}}
\setcounter{section}{0}

\renewcommand{\theHsection}{appendixsection.\Alph{section}}

\renewcommand\thefigure{\thesection\arabic{figure}}
\renewcommand\theequation{\thesection\arabic{equation}}
\renewcommand\thetable{\thesection\arabic{table}}


\section{Extrapolation of neutrino signals}\label{appendix:extrapolation}
\setcounter{figure}{0}
\setcounter{equation}{0}
\setcounter{table}{0}

\renewcommand\theHfigure{appendixA_figure.\arabic{figure}}
\renewcommand\theHequation{appendixA_equation.\arabic{equation}}
\renewcommand\theHtable{appendixA_table.\arabic{table}}

\newcommand{\Lcorezero}{\ensuremath{L_0^\mathrm{core}}}

\begin{figure*}[ht!]
	\includegraphics[width=\textwidth]{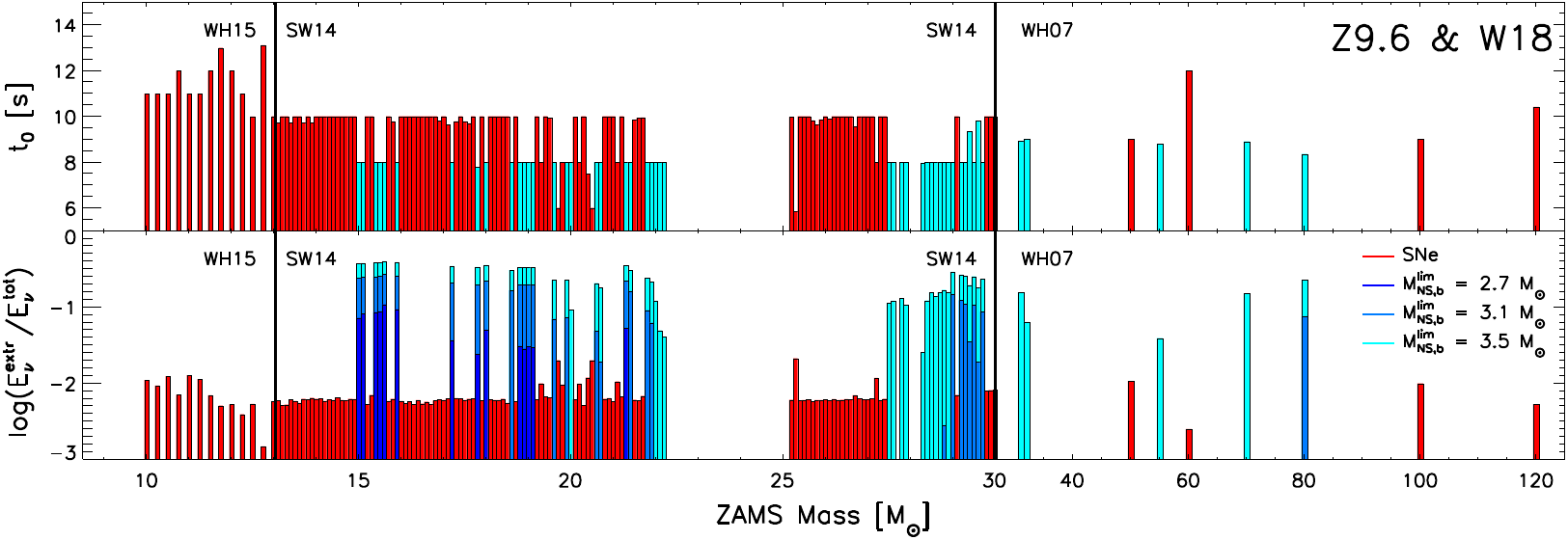}
	\caption{Systematics of our signal extrapolation over the range of progenitor models from the WH15, SW14, and WH07 sets for the Z9.6\,\&\,W18 engine (cf. Figure~\ref{fig:neutrino_outcomesystematics}). In the upper panel, the starting time of our extrapolation, $t_0$, is given. The lower panel shows the relative fraction of the total radiated neutrino energy arising from the extrapolation (note the logarithmic scale). Both quantities are plotted versus ZAMS mass. Red bars indicate successful SN explosions (including rare fallback SNe), while BH-forming, failed SNe are marked in dark blue, light blue, and cyan corresponding to a baryonic NS mass limit of $\unit{2.7}{\msun}$, $\unit{3.1}{\msun}$, and $\unit{3.5}{\msun}$, respectively (no extrapolation is needed for the case of $\unit{2.3}{\msun}$). The time $t_0$ is independent of the mass limit. Progenitors below a ZAMS mass of $\unit{10}{\msun}$ as well as fast-accreting BH cases do not require extrapolation.\\
		\label{fig:extrapolation}}
\end{figure*}

In our analysis as described in Sections~\ref{sec:simulation_setup} and \ref{sec:DSNB_formulation}, we employ the neutrino signals from successful SNe (including rare cases of fallback SNe) up to $\unit{15}{s}$ post bounce, at which time their luminosities have declined to a level that is not relevant for our purpose of estimating the DSNB; moreover at late times the NS temperature drops and therefore the mean spectral energies of the emitted neutrinos shift out of the DSNB detection window. In contrast, the signals from failed explosions have to be followed until the accreting NS reaches the mass limit for BH formation, \MBH, which may take tens of seconds in cases of low mass-accretion rates and high \MBH\ (see upper panel of Figure~\ref{fig:neutrino_outcomesystematics}). Not all of our successful or failed SN simulations could be carried out long enough because of rising computational costs or due to numerical problems emerging at late times (after several seconds). We thus extrapolate these neutrino signals after the computational end at post-bounce time $t_0$. In the upper panel of Figure~\ref{fig:extrapolation}, $t_0$ is plotted against ZAMS mass for our reference engine model Z9.6\,\&\,W18. Typically, our extrapolation starts at around 8--10\,s, whereas no extrapolation was needed for a few successful SNe near the low-mass end and for fast-accreting failed SNe with short-lived NSs (see top panel of Figure~\ref{fig:neutrino_outcomesystematics}). Even if the exact values of $t_0$ are slightly different for our other neutrino engines, the overall picture remains the same.

The cooling phases of our successful, NS-forming SNe can be described approximately by an exponential decline of the neutrino signal at sufficiently late times after shock revival, when the mass accretion onto the hot PNS has ceased and the diffusion of neutrinos from the core defines the emission \citep[][]{1986ApJ...307..178B, 1995A&A...296..145K, 1999ApJ...513..780P}. We thus extrapolate the signals of our successful SNe according to
\begin{equation}\label{eq:L_core}
L_\mathrm{core}(t) = \Lcorezero\,\E^{-(t-t_0)/\tau}
\end{equation}
for all neutrino species \nui, with $\Lcorezero\,\mathord{=}\,L_{\nui}(t_0)$ being the corresponding luminosity at the end our simulations at time $t_0$ and $\tau\,\mathord{=}\,\tau_{\nui}$ being a core-cooling timescale, which we obtain from least-squares fits over the last $\unit{2}{s}$ of the computed neutrino signals. Our values for $\tau$ typically range between $1$ and $\unit{4}{s}$, in agreement with the work by \citet{Huedepohl:2014} \citep[also see][table~1]{2016MNRAS.460..742M}. The lower panel of Figure~\ref{fig:extrapolation} shows the relative contributions to the total radiated neutrino energies from our extrapolations (in the time interval $t_0\,\mathord{\leqslant}\,t\,\mathord{\leqslant}\,\unit{15}{s}$ for the cases of successful explosions). They lie below $\sim$1--2\% for all successful SNe, which illustrates that a further extrapolation of the exponentially declining signals beyond 15\,s is not necessary. Similar results are obtained for all our engine models. The mean neutrino energies, $\langle E_{\nui}(t) \rangle$, are simply extrapolated by keeping them constant at their final values at $t_0$, which, because of the small contribution from the neutrino emission at late times ($t\,\mathord{>}\,t_0$) to the time-integrated signals, has no significant influence on our DSNB predictions and is therefore unproblematic.

In the cases of BH forming, failed SNe on the other hand, the continued infall of matter feeds an accretion luminosity in addition to the diffusive flux from the core \citep{1988ApJ...334..891B}. Therefore, we describe the total neutrino emission (of all species) as the sum of a core and an accretion component, $\Lnutot(t)\,\mathord{=}\,L_\mathrm{core}(t)\,\mathord{+}\,L_\mathrm{acc}(t)$. For the accretion luminosity, we follow the description by \citet{1988ApJ...334..891B},
\begin{equation}\label{eq:L_acc}
L_\mathrm{acc}(t) = \eta\,\frac{G M_\mathrm{NS,b}(t) \dot{M}_\mathrm{NS,b}(t)}{R_\mathrm{NS}(t)}~,
\end{equation}
with the gravitational constant $G$ and an adjustable efficiency parameter $\eta$ \citep[cf.][]{2009A&A...499....1F, Huedepohl:2014, 2014ApJ...788...82M}. For computational reasons, we take the late-time evolution of the progenitor-dependent (baryonic) PNS mass, $M_\mathrm{NS,b}(t)$, and accretion rate, $\dot{M}_\mathrm{NS,b}(t)$, from pure hydrodynamic simulations with the neutrino engine switched off and an open inner boundary of the computational grid placed in the supersonically infalling matter exterior to the stalled accretion shock. As we do not have model-based information on the time-dependent radius, $R_\mathrm{NS}(t)$, of the contracting PNS, we adopt equation~(9) of \citet{2016MNRAS.460..742M},
\begin{equation}\label{eq:R_NS}
R_\mathrm{NS}(t) = \left[\,R_1^{\,3}\left(\frac{\dot{M}_\mathrm{NS,b}(t)}{\msun\,\mathrm{s}^{-1}}\right)\left(\vphantom{\frac{\dot{M}}{M}}\frac{M_\mathrm{NS,b}(t)}{\msun}\right)^{\!\!-3}+R_2^{\,3}~\right]^{1/3}.
\end{equation}
We find that the late phases of those failed-SN simulations that are carried on beyond $\unit{10}{s}$ (21 cases in the N20 and 72 in the W20 sets, 17 of them beyond $\unit{20}{s}$) are reproduced by Equations~\eqref{eq:L_acc} and \eqref{eq:R_NS} with an accuracy of a few percent when we choose the parameter values $R_1\,\mathord{=}\,\unit{40}{km}$, $R_2\,\mathord{=}\,\unit{11}{km}$ and an accretion efficiency $\eta\,\mathord{=}\,0.51$.\footnote{The absolute values of $R_1$ and $R_2$ can be chosen somewhat arbitrarily since the adjustable parameter $\eta$ compensates for shifts of $L_\mathrm{acc}$ in Equation~\eqref{eq:L_acc}. For consistency with measured NS radii, we take $R_2\,\mathord{=}\,\unit{11}{km}$ (see footnote~\ref{fn:R_NS}). The resulting best-fit value of $R_1\,\mathord{=}\,\unit{40}{km}$ is much smaller than the $\unit{120}{km}$ in \citet{2016MNRAS.460..742M}, which reflects the moderate core contraction in our simulations.} Similar values for $\eta$ were found by \citet{2009A&A...499....1F}, \citet{Huedepohl:2014}, and \citet{2014ApJ...788...82M}. We apply this description of the accretion luminosity to all of our extrapolated failed-SN signals, independently of the engine model. For the core luminosity (of all neutrino species) in failed explosions, we employ Equation~\eqref{eq:L_core} with an initial value $\Lcorezero\,\mathord{=}\,\Lnutot(t_0)\,\mathord{-}\,L_\mathrm{acc}(t_0)$ and a core-cooling timescale $\tau\,\mathord{=}\,\tau_{\nux}$\,($\sim$\,1\,s) from a least-squares fit of the heavy-lepton neutrino signal between $\unit{3}{s}$ and $\unit{6}{s}$ after bounce in each model. During this phase, $L_{\nux}$ is dominated by its core component and can be well approximated by an exponential decline. We hence adopt this prescription also for the core luminosities of electron-type neutrinos, which are not as readily accessible \citep[cf.][]{Huedepohl:2014, 2014ApJ...788...82M}. In the extrapolation, the relative contributions of the different neutrino species to the total emission are kept constant at their final values obtained at the end of the simulations (i.e., $L_{\nui}(t)\,\mathord{=}\,f_{\nui} \Lnutot(t)$, with the factor $f_{\nui}\,\mathord{=}\,L_{\nui}(t_0)/\Lnutot(t_0)$ equally applied to core and accretion components).

As can be seen in the lower panel of Figure~\ref{fig:extrapolation}, our extrapolation accounts for up to $\sim$40\% of the total radiated neutrino energy for the case of a NS mass limit of $\unit{3.5}{\msun}$ in the most extreme conditions, while no extrapolation is required for a limiting NS mass of $\unit{2.3}{\msun}$. This is true for all of our engine models. The mean neutrino energies from slowly-accreting failed SNe, where the extrapolation has the biggest influence, flatten to rather constant values ($\sim$20\,MeV) at late times in simulations that could be carried on for more than $\sim$$\unit{10}{s}$. We thus extrapolate the mean neutrino energies in failed SNe by keeping them constant at their final values at $t_0$, in analogy to what we do in the cases of successful SNe. We tested other extrapolation schemes, but found that the time-integrated spectra are largely insensitive to the late-time description of the mean energies.

\section{Total energies of radiated neutrinos}\label{appendix:total_energies}
\setcounter{figure}{0}
\setcounter{equation}{0}
\setcounter{table}{0}

\renewcommand\theHfigure{appendixB_figure.\arabic{figure}}
\renewcommand\theHequation{appendixB_equation.\arabic{equation}}
\renewcommand\theHtable{appendixB_table.\arabic{table}}

\begin{figure*}
	\includegraphics[width=\textwidth]{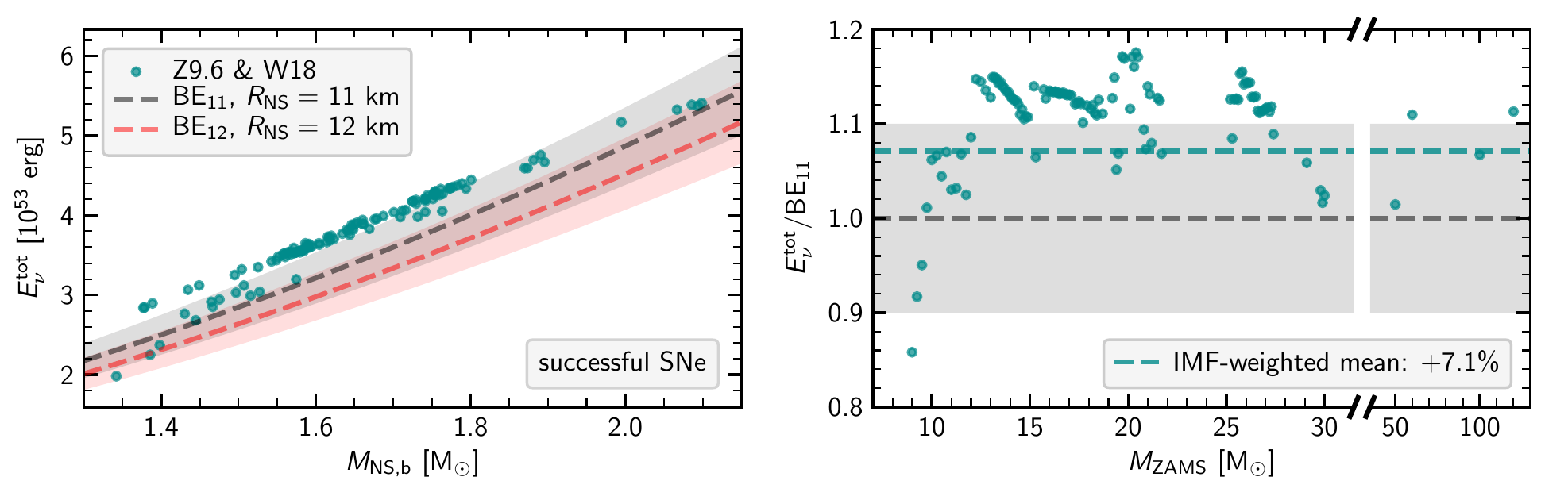}
	\caption{Comparison of the total neutrino energies, \Enutot, radiated by the successful explosions of our reference set (Z9.6\,\&\,W18) with the gravitational binding energies (BE) of the relic NSs as estimated with an analytic expression from \citet{2001ApJ...550..426L}. In the left panel, the relation between \Enutot\ and the baryonic NS mass, $M_\mathrm{NS,b}$, is shown (turquoise dots). The gray (red) dashed line indicates the NS's binding energy as a function of $M_\mathrm{NS,b}$, computed with Equation~\eqref{eq:LattimerPrakash}, assuming a NS radius of $\unit{11}{km}$ ($\unit{12}{km}$). The shaded bands correspond to deviations of $\pm$10\%. In the right panel, the ratio of the total radiated neutrino energy to BE is plotted versus ZAMS mass for a NS radius of $\unit{11}{km}$. The dashed turquoise line additionally indicates the IMF-weighted mean value, which deviates by +7.1\% from BE. Note the scale break at $M_\mathrm{ZAMS}\,\mathord{\sim}\,\unit{30}{\msun}$.
	\label{fig:appendix_E_tot_SNe}}
\end{figure*}

Both in successful and failed core-collapse SNe, the neutrino emission is fed by the release of gravitational binding energy (BE) from an assembling PNS, which either cools down to become a stable NS or further collapses to a BH. To assess the viability of our DSNB flux predictions, we compare the total radiated neutrino energy, \Enutot, obtained from our simulations with an analytic estimate of the binding energy. For this purpose, we adopt equation~(36) of \citet{2001ApJ...550..426L}, which connects the PNS's baryonic mass, $M_\mathrm{NS,b}$, with its gravitating mass, $M_\mathrm{NS,g}$, assuming a final (cold) NS radius $R_\mathrm{NS}$:
\begin{equation}\label{eq:LattimerPrakash}
\frac{\mathrm{BE}/c^2}{M_\mathrm{NS,g}} = \frac{0.6 \beta}{1 - 0.5 \beta}~~,
\end{equation}
with $\mathrm{BE}/c^2\,\mathord{\equiv}\,M_\mathrm{NS,b}\,\mathord{-}\,M_\mathrm{NS,g}$ and the dimensionless parameter $\beta\,\mathord{\equiv}\,GM_\mathrm{NS,g}/R_\mathrm{NS}c^2$.

In the left panel of Figure~\ref{fig:appendix_E_tot_SNe}, $E_\nu^\mathrm{tot}$ of our successful explosions in the Z9.6\,\&\,W18 set is plotted against the baryonic mass of the relic NS (turquoise dots). We compare these values with the corresponding gravitational binding energies BE$_{11}$ and BE$_{12}$ (gray and red dashed lines), computed with Equation~\eqref{eq:LattimerPrakash} for an assumed final NS radius of $\unit{11}{km}$ and $\unit{12}{km}$, respectively. The shaded bands indicate deviations of $\unit{\pm10}{\%}$ from the analytic relations. In the right panel, we also show the ratio of the total radiated neutrino energy to BE for the case of $R_\mathrm{NS}\,\mathord{=}\,\unit{11}{km}$, plotted against the zero-age main sequence mass $M_\mathrm{ZAMS}$ of the progenitors.

Our simulations feature good overall agreement with Equation~\eqref{eq:LattimerPrakash}, compatible with the PNS of a successful SN radiating essentially its entire gravitational binding energy in the form of neutrinos. Assuming a NS radius of $\unit{11}{km}$, 93\% of the successful explosions in our Z9.6\,\&\,W18 set deviate by less than 15\% from the analytic fit provided by \citet{2001ApJ...550..426L}. Most of our simulations overestimate the total radiated neutrino energy on the order of 10\%, but for the majority of low-mass progenitors the values of $E_\nu^\mathrm{tot}$ are close to or below BE$_{11}$, which leads to an IMF-weighted mean deviation of +7.1\%. If we assume $R_\mathrm{NS}\,\mathord{=}\,\unit{12}{km}$ $(\unit{13}{km})$ instead, the deviation increases to a value of +15.6\% (+24.1\%) above the analytic description. In Table~\ref{tab:IMF-weighted_deviations_from_LP2001}, we show the IMF-weighted mean deviations for all of our engine models.

\begin{deluxetable}{lCCC}[t!]
	\tablecaption{
	IMF-weighted deviations of \Enutot\ of successful SNe from the analytic description by \citet{2001ApJ...550..426L} for the NS gravitational binding energy (BE).
	\label{tab:IMF-weighted_deviations_from_LP2001}}
	\tablehead{
		\colhead{Engine Model} & \colhead{$R_\mathrm{NS}\,\mathord{=}\,\unit{11}{km}$} & \colhead{$R_\mathrm{NS}\,\mathord{=}\,\unit{12}{km}$} & \colhead{$R_\mathrm{NS}\,\mathord{=}\,\unit{13}{km}$}
	}
	\startdata
	Z9.6\,\&\,S19.8 & +11.8\% & +20.7\% & +29.6\% \\
	Z9.6\,\&\,N20 & +6.1\% & +14.6\% & +23.0\% \\
	Z9.6\,\&\,W18 & +7.1\% & +15.6\% & +24.1\% \\
	Z9.6\,\&\,W15 & +5.1\% & +13.5\% & +21.8\% \\
	Z9.6\,\&\,W20 & +7.0\% & +15.6\% & +24.1\%
	\enddata
	\tablecomments{For the computations of BE, Equation~\eqref{eq:LattimerPrakash} was used with final NS radii of $\unit{11}{km}$, $\unit{12}{km}$, or $\unit{13}{km}$.}
\end{deluxetable}

\begin{figure*}
	\includegraphics[width=\textwidth]{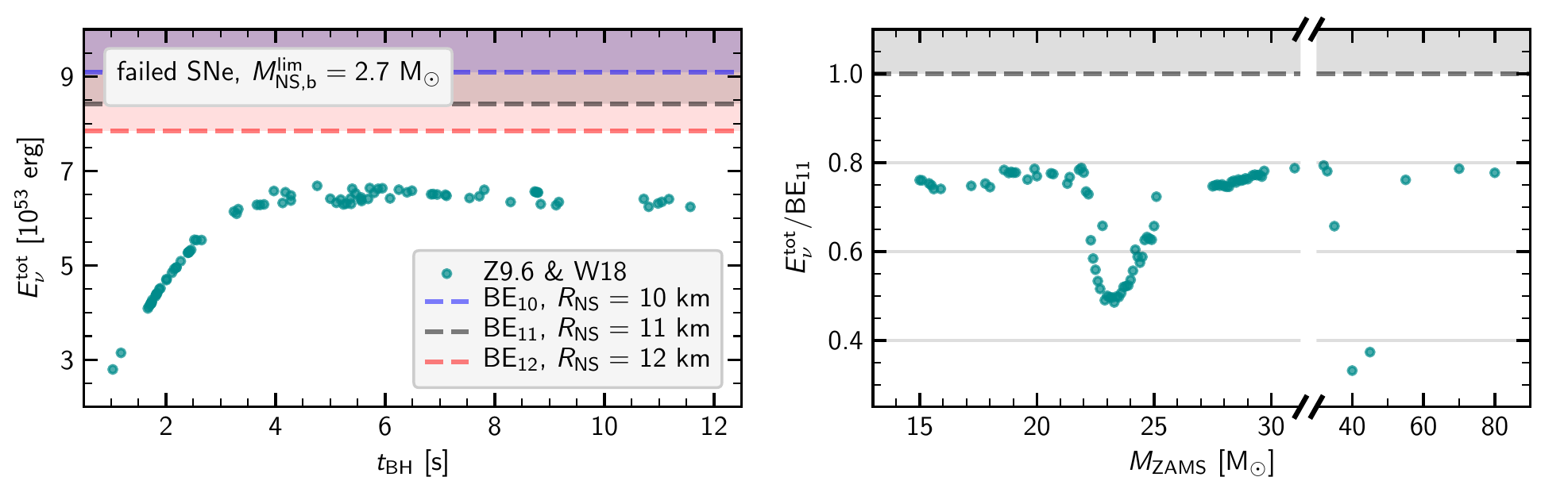}
	\caption{Comparison of the total neutrino energies, \Enutot, radiated by the failed explosions of our reference set (Z9.6\,\&\,W18, $\MBH\,\mathord{=}\,\unit{2.7}{\msun}$) with the maximally available reservoir of gravitational binding energy (BE) as given by the analytic fit formula of \citet{2001ApJ...550..426L}. The left panel shows \Enutot\ versus the time until BH formation (turquoise dots). The three dashed lines (in blue, gray, and red) indicate the total gravitational binding energies, according to Equation~\eqref{eq:LattimerPrakash}, of a NS with an assumed maximum baryonic mass of $\MBH\,\mathord{=}\,\unit{2.7}{\msun}$ and assumed radius of $\unit{10}{km}$, $\unit{11}{km}$, and $\unit{12}{km}$, respectively. In the right panel, the ratio of the radiated neutrino energy to the maximally available gravitational binding energy is plotted versus the progenitor's ZAMS mass for an assumed NS radius of $\unit{11}{km}$. Note the scale break at $M_\mathrm{ZAMS}\,\mathord{\sim}\,\unit{30}{\msun}$.\\
	\label{fig:appendix_E_tot_fSNe}}
\end{figure*}

\begin{deluxetable*}{lCCCC}[t!]
	\tablecaption{
	Maximally available gravitational binding energies (BE) for the cases of failed SNe.
	\label{tab:BE_fSNe}}
	\tablehead{
		\colhead{Baryonic NS Mass Limit} & \colhead{BE$_{10}$ [$10^{53}$ erg]} & \colhead{BE$_{11}$ [$10^{53}$ erg]} & \colhead{BE$_{12}$ [$10^{53}$ erg]} & \colhead{BE$_{13}$ [$10^{53}$ erg]}
	}
	\startdata
	$\MBH=\unit{2.3}{\msun}$ & 6.8~(77.8\%) & 6.3~(84.1\%) & 5.9~(90.5\%) & 5.5~(96.8\%) \\
	$\MBH=\unit{2.7}{\msun}$ & 9.1~(77.3\%) & 8.4~(83.4\%) & 7.8~(89.5\%) & 7.3~(95.6\%) \\
	$\MBH=\unit{3.1}{\msun}$ & 11.6~(76.0\%) & 10.8~(81.8\%) & 10.1~(87.6\%) & 9.4~(93.5\%) \\
	$\MBH=\unit{3.5}{\msun}$ & 14.4~(74.2\%) & 13.4~(79.7\%) & 12.5~(85.3\%) & 11.8~(90.8\%)
	\enddata
	\tablecomments{The values of BE are computed according to Equation~\eqref{eq:LattimerPrakash} for different values of the baryonic NS mass limit, \MBH, and for different NS radii ($\unit{10}{km}$, $\unit{11}{km}$, $\unit{12}{km}$, and $\unit{13}{km}$). In parentheses we give the largest value of the ratio \Enutot/BE for all of our failed explosion models and all neutrino engines. Note the slightly larger value of 83.4\% for \Enutot/BE$_{11}$ in the case of $\MBH\,\mathord{=}\,\unit{2.7}{\msun}$ compared to the $\sim$80\% in the right panel of Figure~\ref{fig:appendix_E_tot_fSNe}, which shows the case of the Z9.6\,\&\,W18 neutrino engine.}
\end{deluxetable*}

Compared to successful explosions, the total energy reservoir that could be released in neutrinos by BH-forming, failed SNe is generally higher if the PNS at the limiting mass remained stable until it has emitted its entire gravitational binding energy before it collapses to a BH (see Table~\ref{tab:BE_fSNe}). However, the binding energy of a maximum-mass NS constitutes just an upper limit for the radiated neutrino energy \Enutot, because BH formation typically occurs before the NS has cooled to a cold state, terminating the neutrino emission before the total gravitational energy is carried away by neutrinos. This can be seen in the left panel of Figure~\ref{fig:appendix_E_tot_fSNe}, where we plot \Enutot\ for the failed SNe of our reference set (Z9.6\,\&\,W18, $\MBH\,\mathord{=}\,\unit{2.7}{\msun}$) against the time until BH formation (turquoise dots). Only the slowly-accreting cases (with $t_\mathrm{BH}\,\mathord{\gtrsim}\,\unit{3}{s}$) come close to the maximally available binding energy according to Equation~\eqref{eq:LattimerPrakash}, which is indicated by a blue, gray, and red dashed line for NS radii of $\unit{10}{km}$, $\unit{11}{km}$, and $\unit{12}{km}$, respectively. In the right panel, we show the ratio of the radiated to the maximally available energy for an assumed NS radius of $\unit{11}{km}$ versus the ZAMS mass range of the corresponding progenitors.

For all of our simulations the neutrino emission from failed SNe lies well below the analytically computed energy limit. For our reference set shown in Figure~\ref{fig:appendix_E_tot_fSNe}, at most 80\% of BE$_{11}$ are radiated before a BH forms, while the progenitors at around 22--25\,\msun\ and $\sim$$\unit{40}{\msun}$, which exhibit very high mass-accretion rates (see footnote~\ref{fn:compactness} and upper panel of Figure~\ref{fig:neutrino_outcomesystematics}), feature considerably lower percentages ($\sim$30--60\%). The results for the other neutrino engines are very similar, because the emission from a failed SN is dominated by the progenitor-dependent accretion component rather than the PNS core emission. For larger NS radii applied in Equation~\eqref{eq:LattimerPrakash}, the ratio \Enutot/BE tends towards unity, as can be seen in Table~\ref{tab:BE_fSNe} (values in parentheses).\\\\

\section{Flavor rescaling}\label{appendix:rescaling}
\setcounter{figure}{0}
\setcounter{equation}{0}
\setcounter{table}{0}

\renewcommand\theHfigure{appendixC_figure.\arabic{figure}}
\renewcommand\theHequation{appendixC_equation.\arabic{equation}}
\renewcommand\theHtable{appendixC_table.\arabic{table}}

\begin{deluxetable*}{lCCCCCcC}
	\tablecaption{
	Flavor fractions and conversion factors.
	\label{tab:flavor_ratios}}
	\tablehead{
		\colhead{Model} & \colhead{$\tilde{\xi}_{\nuebar}$} & \colhead{$\tilde{\xi}_{\nue}$} & \colhead{$\tilde{\xi}_{\nux}$} & \colhead{$\lambdaE$} & \colhead{$\lambdaa$} & \colhead{Compact Remnant} & \colhead{$M_\mathrm{NS,b}$ [$\msun$]}
	}
	\startdata
	\textsc{Vertex}, s11.2co, LS220 & 0.166 & 0.194 & 0.160 & 0.990 & 0.808 & NS & 1.366 \\
	\textsc{Vertex}, z9.6co, LS220 & 0.155 & 0.173 & 0.168 & 0.992 & 0.810 & NS & 1.361 \\
	\textsc{Vertex}, z9.6co, SFHo & 0.157 & 0.176 & 0.167 & 0.990 & 0.790 & NS & 1.363 \\
	Average (``L'') & {\bf0.159} & {\bf0.181} & {\bf0.165} & {\bf0.991} & {\bf0.803} & $-$ & $-$ \\
	\hline
	\textsc{Vertex}, s20.0, SFHo & 0.172 & 0.176 & 0.163 & 0.965 & 0.813 & NS & 1.947 \\
	\textsc{Vertex}, s27.0co, LS220 & 0.172 & 0.181 & 0.162 & 0.957 & 0.807 & NS & 1.776 \\
	\textsc{Vertex}, s27.0co, SFHo & 0.170 & 0.179 & 0.163 & 0.973 & 0.810 & NS & 1.772 \\
	Average (``H'') & {\bf0.171} & {\bf0.179} & {\bf0.163} & {\bf0.965} & {\bf0.810} & $-$ & $-$ \\
	\hline
	\textsc{Vertex}, s40s7b2c, LS220 (``F'') & {\bf0.212} & {\bf0.257} & {\bf0.133} & {\bf1.068} & {\bf0.639} & BH (fast; $\unit{0.57}{s}$) & (2.320) \\
	\textsc{Vertex}, s40.0c, LS220 (``S'') & {\bf0.231} & {\bf0.251} & {\bf0.129} & {\bf0.940} & {\bf0.724} & BH (slow; $\unit{2.11}{s}$) & (2.279)
	\enddata
	\tablecomments{Relative fractions $\tilde{\xi}_{\nuebar}$, $\tilde{\xi}_{\nue}$, and $\tilde{\xi}_{\nux}$ of the total energy \Enutot\ radiated in the neutrino species \nuebar, \nue, and \nux\ (Equation~\eqref{eq:xi_tilde}), and conversion factors $\lambdaE\,\mathord{\equiv}\,(\langle E_{\nux} \rangle/\langle E_{\nuebar} \rangle)^{\scriptsize\textsc{PV}}$ (Equation~\eqref{eq:lambda_E}) and $\lambdaa\,\mathord{\equiv}\,(\alphax/\alphaea)^{\scriptsize\textsc{PV}}$ (Equation~\eqref{eq:lambda_a}), listed for eight models that were simulated with the 1D version of the \textsc{Prometheus-Vertex} code. The lines are grouped according to the masses of the compact remnants. For the models s40s7b2c and s40.0c, the values in parentheses indicate the post-bounce times of BH formation and the corresponding baryonic PNS masses at these times. The conversion factors applied to our \textsc{Prometheus-HotB} models (according to the four cases ``L'', ``H'', ``F'', and ``S'' for low-mass NSs, high-mass NSs, fast BH formation, or slow BH formation, respectively; see text for the details) are highlighted in boldface. Note that $\tilde{\xi}_{\nue}\,\mathord{+}\,\tilde{\xi}_{\nuebar}\,\mathord{+}\,4 \tilde{\xi}_{\nux}\,\mathord{=}\,1$.}
\end{deluxetable*}

The approximate treatment of the microphysics and the relatively modest contraction of our inner grid boundary in the considered core-collapse simulations result in underestimated luminosities of the heavy-lepton neutrinos, as mentioned in Sections~\ref{sec:simulation_setup} and \ref{sec:DSNB_formulation}. Consequently, we introduce a rescaling factor $\tilde{\xi}/\xi$ for the time-integrated neutrino spectra in Equation~\eqref{eq:time-integr_spec}, where
\begin{equation}
\xi = \xi_{\nui} = \Enuitot / \Enutot
\end{equation}
and
\begin{equation}
\tilde{\xi} = \tilde{\xi}_{\nui} = \left(E_{\nui}^\mathrm{tot} / E_{\nu}^\mathrm{tot}\right)^{\scriptsize\textsc{PV}}
\label{eq:xi_tilde}
\end{equation}
denote the relative fractions of the total neutrino energy \Enutot\ radiated in the species \nui\ before (i.e., as obtained from the \textsc{Prometheus-HotB} simulations) and after this readjustment, respectively. We perform the rescaling in terms of relative energy contributions to the total loss of gravitational binding energy by neutrinos (rather than in terms of relative neutrino numbers), because the energy obeys a conservation law in contrast to neutrino numbers. For setting the new weights $\tilde{\xi}_{\nu_i}$ we refer to six successful explosion models (artificially exploded in 1D) and two failed SNe that were computed with the \textsc{Prometheus-Vertex} (PV) code \citep{2002A&A...396..361R, 2006A&A...447.1049B} and are listed in Table~\ref{tab:flavor_ratios}: z9.6co and s27.0co, both simulated with the LS220 \citep{1991NuPhA.535..331L} as well as the SFHo EoS \citep{2013ApJ...774...17S} and discussed in detail in \citet{2016NCimR..39....1M}; the unpublished model s20.0 of a 20\,\msun\ progenitor of \citet{2007PhR...442..269W}, computed with the SFHo EoS in the same way as the four models mentioned before (Robert Bollig, 2018, private communication); and the three models s11.2co, s40.0c, and s40s7b2c from \citet{Huedepohl:2014}, all of them computed with the LS220 EoS. The suffix ``c'' of the model names  indicates the use of a mixing-length treatment for PNS convection, and the suffix ``o'' that mean-field potentials are taken into account in the charged-current neutrino-nucleon interactions (see \citealt{2016NCimR..39....1M} for details). The neutrino signals of all eight models can be found in the Garching Core-collapse Supernova Archive.\footnote{\url{https://wwwmpa.mpa-garching.mpg.de/ccsnarchive/archive.html} (access provided upon request)}

Although we constrain our analysis in most parts on the emitted \nuebar\ signals, we need information on the time-integrated spectra, $\diff N_{\nux}/\diff E$, also for heavy-lepton neutrinos in our discussion of flavor oscillation effects in Section~\ref{subsec:flavor_conversions}. Instead of taking the outcome of (too approximate) SN and BH-formation models, we directly employ the spectral shape from \citet{2003ApJ...590..971K},
\begin{eqnarray}
\frac{\diff N_{\nux}}{\diff E} =&& \frac{(\alphax + 1)^{(\alphax + 1)}}{\Gamma(\alphax + 1)} \frac{E_{\nux}^\mathrm{tot}}{\Emeanx^2} \left(\frac{E}{\Emeanx}\right)^{\alphax} \nonumber \\
&&\times\exp\left[-\frac{(\alphax + 1)E}{\Emeanx}\right]\:,
\label{eq:nux_spectrum}
\end{eqnarray}
with the total energy radiated in a single heavy-lepton neutrino species $E_{\nux}^\mathrm{tot}\,\mathord{=}\,\tilde{\xi}_{\nux} \Enutot$, the time-averaged mean neutrino energy $\langle E_{\nux} \rangle\,\mathord{=}\,\lambdaE \langle E_{\nuebar} \rangle$, and the spectral-shape parameter $\alphax\,\mathord{=}\,\lambdaa \alphaea$ of the time-integrated \nux\ spectrum.\footnote{The bar in the symbols $\overline{\alpha}_{\nu_{i}}$ indicates that the shape parameters refer to the \textit{time-integrated} spectra rather than the \textit{instantaneous} spectra (see Section~\ref{subsec:spectra} and Appendix~\ref{appendix:spectral_parameters}).} Here, $\langle E_{\nuebar} \rangle$ and \alphaea\ are computed from the time-integrated spectra of \nuebar\ obtained in our large set of core-collapse simulations. For the conversion factors
\begin{equation}
\lambdaE\,\mathord{\equiv}\,(\langle E_{\nux} \rangle/\langle E_{\nuebar} \rangle)^{\scriptsize\textsc{PV}}
\label{eq:lambda_E}
\end{equation}
and
\begin{equation}
\lambdaa\,\mathord{\equiv}\,(\alphax/\alphaea)^{\scriptsize\textsc{PV}},
\label{eq:lambda_a}
\end{equation}
we take the values from the \textsc{Prometheus-Vertex} models in Table~\ref{tab:flavor_ratios}. The shape parameters, $\overline{\alpha}\,\mathord{=}\,\overline{\alpha}_{\nui}$, and mean neutrino energies, $\langle E \rangle\,\mathord{=}\,\langle E_{\nui} \rangle$, of the time-integrated spectra, $\diff N/\diff E\,\mathord{=}\,\diff N_{\nui}/\diff E$, are computed as
\begin{equation}\label{eq:alpha}
\overline{\alpha} = \frac{2\langle E \rangle^2 - \langle E^2 \rangle}{\langle E^2 \rangle - \langle E \rangle^2}~~,
\end{equation}
with
\begin{equation}\label{eq:mean_energy}
\langle E \rangle = \frac{\int\diff E E (\diff N/\diff E)}{\int\diff E (\diff N/\diff E)}~~,
\end{equation}
\begin{equation}\label{eq:mean_squared_energy}
\langle E^2 \rangle = \frac{\int\diff E E^2 (\diff N/\diff E)}{\int\diff E (\diff N/\diff E)}~~.
\end{equation}

The neutrino spectra of our successful explosions which form NSs with baryonic masses of $M_\mathrm{NS,b}\,\mathord{\leqslant}\,\unit{1.6}{\msun}$ are rescaled by the average conversion factors of the s11.2co and the two z9.6co models (upper part of Table~\ref{tab:flavor_ratios}; case ``L''). For SNe with $M_\mathrm{NS,b}\,\mathord{>}\,\unit{1.6}{\msun}$, we apply the average values of the s20.0 and s27.0co models (middle part of Table~\ref{tab:flavor_ratios}; case ``H''). In cases of failed explosions with BH formation (lower part of Table~\ref{tab:flavor_ratios}), we distinguish between fast-accreting ($t_\mathrm{BH}\,\mathord{<}\,\unit{2}{s}$; ``F'') and slowly-accreting ($t_\mathrm{BH}\,\mathord{\geqslant}\,\unit{2}{s}$; ``S'') cases. The spectra of our fast-accreting models (progenitors with high core compactness; see footnote~\ref{fn:compactness}) are rescaled according to \textsc{Vertex} model s40s7b2c, which forms a BH after $\unit{0.57}{s}$. For our slowly accreting cases with long delays until BH formation, which correlate with higher maximum NS masses and with progenitors that have a relatively lower core compactness, we employ the rescaling factors of model s40.0c, where BH formation occurs at $t_\mathrm{BH}\,\mathord{=}\,\unit{2.11}{s}$. For completeness, we also give the baryonic PNS masses just before the PNSs collapse to BHs in Table~\ref{tab:flavor_ratios}. Note that approximate flavor equipartition ($\tilde{\xi}_{\nuebar}\,\mathord{\simeq}\,\tilde{\xi}_{\nue}\,\mathord{\simeq}\, \tilde{\xi}_{\nux}$) is realized for successful SNe, whereas \nuebar\ and \nue\ dominate over heavy-lepton neutrinos in cases of failed explosions. This can be understood by the continued accretion of infalling matter, which is accompanied by $e^{\pm}$ captures on free nucleons in the PNS's accretion mantle \citep{2012ARNPS..62..407J}, giving rise to an enhanced accretion luminosity of electron-flavor neutrinos and antineutrinos (see Equation~\eqref{eq:L_acc}).

\section{Spectral Shapes}\label{appendix:spectra}
\setcounter{figure}{0}
\setcounter{equation}{0}
\setcounter{table}{0}

\begin{figure*}[ht!]
	\includegraphics[width=\textwidth]{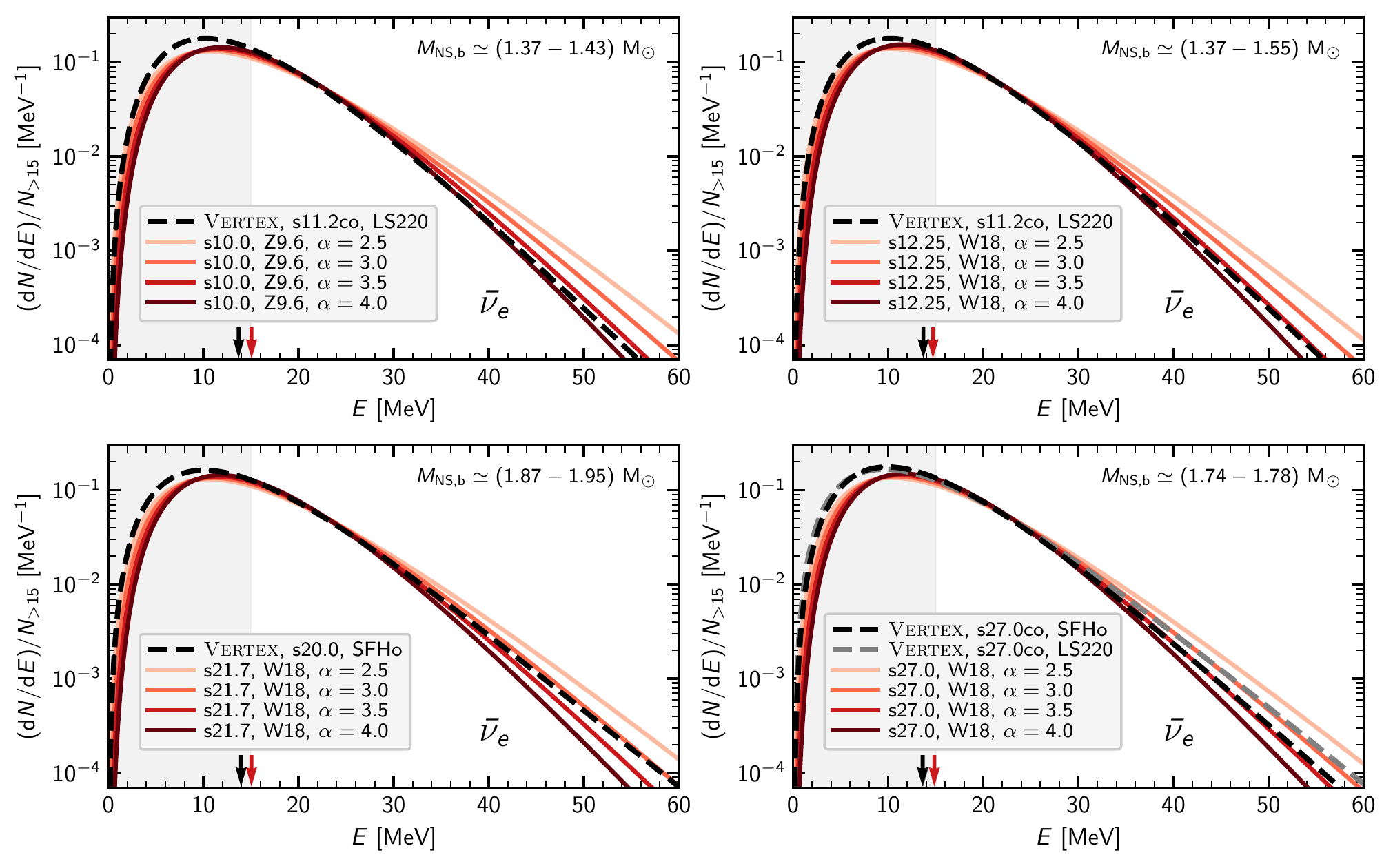}
	\caption{Time-integrated spectra, $\diff N/\diff E$, of electron antineutrinos (normalized by $N_{>15}\,\mathord{=}\,\int_{\unit{15}{MeV}}^{\infty}\diff E (\diff N/\diff E)$) from exemplary SN simulations of our Z9.6\,\&\,W18 set for different values of the instantaneous spectral-shape parameter $\alpha$ (red solid lines), compared to four SN models that were computed with the \textsc{Prometheus-Vertex} code (dashed lines). Arrows at the bottom of each panel mark the mean energies of the spectra (Equation~\eqref{eq:mean_energy}). The gray shaded vertical bands edge the most relevant energy region of $E\,\mathord{\gtrsim}\,\unit{15}{MeV}$ (see main text for details).\\
	\label{fig:appendix_spectra}}
\end{figure*}

\begin{figure*}[ht!]
	\includegraphics[width=\textwidth]{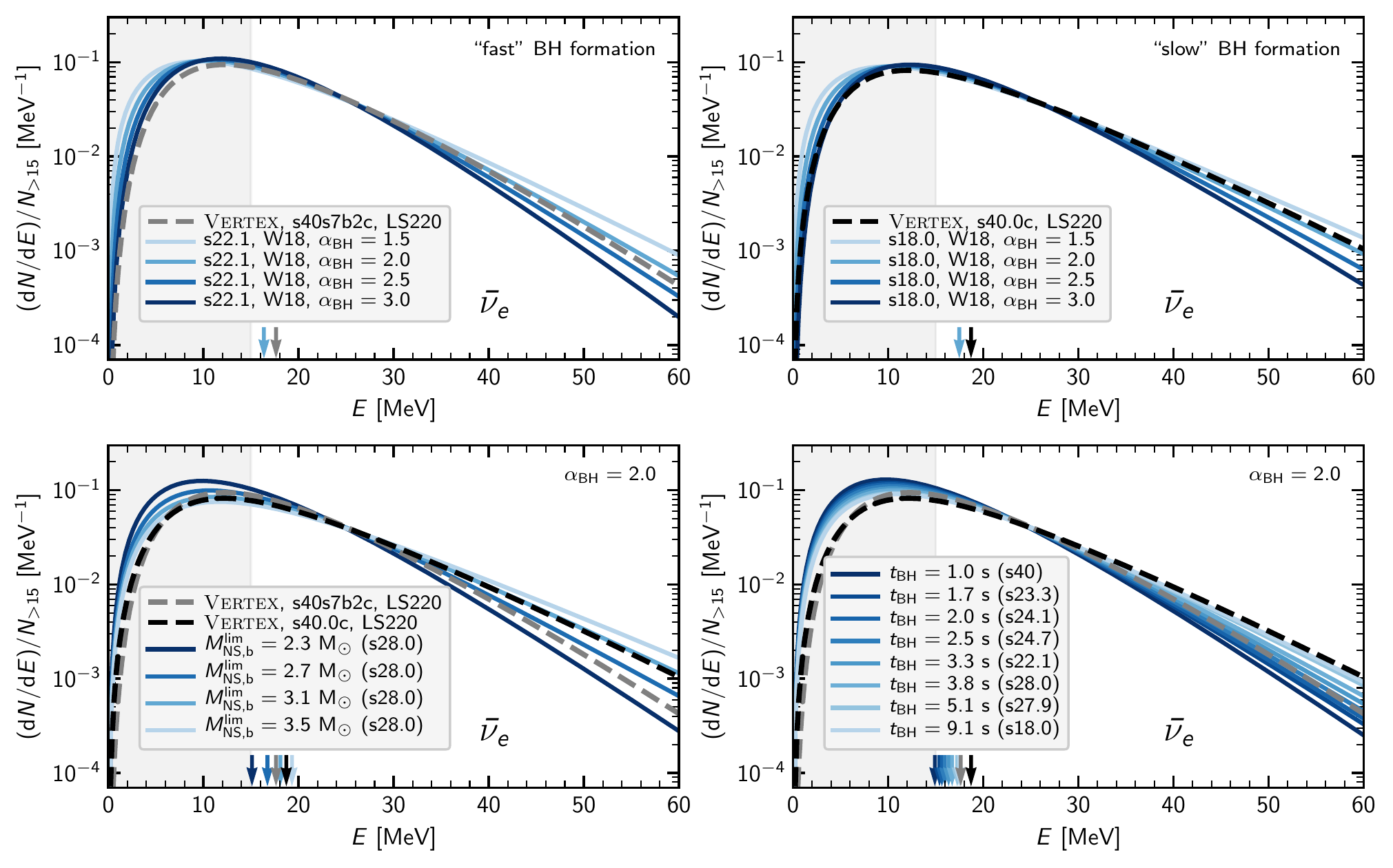}
	\caption{Time-integrated spectra, $\diff N/\diff E$, of electron antineutrinos (normalized by $N_{>15}\,\mathord{=}\,\int_{\unit{15}{MeV}}^{\infty}\diff E (\diff N/\diff E)$) from selected BH-formation simulations of our Z9.6\,\&\,W18 set (with $\MBH\,\mathord{=}\,\unit{2.7}{\msun}$; blue solid lines), compared to two BH-formation models that were computed with the \textsc{Prometheus-Vertex} code (dashed lines). In the upper panels the spectra from two exemplary progenitors (s22.1 and s18.0) for different values of the instantaneous spectral-shape parameter $\alphaBH$ are compared to the \textsc{Vertex} models s40s7b2c and s40.0c, which form BHs relatively ``fast'' (after 0.57\,s at $\MBH\,\mathord{=}\,\unit{2.320}{\msun}$) or ``slowly'' (after 2.11\,s at $\MBH\,\mathord{=}\,\unit{2.279}{\msun}$), respectively. The lower left panel shows the spectra from the exemplary s28.0 progenitor for different choices of the NS baryonic mass limit \MBH; the lower right panel the spectra for eight different progenitors with increasing accretion times until BH formation (between 1.0\,s and 9.1\,s for the shown case of $\MBH\,\mathord{=}\,\unit{2.7}{\msun}$). In both lower panels $\alphaBH\,\mathord{=}\,2.0$ is taken. Arrows at the bottom of each panel mark the mean energies of the spectra (Equation~\eqref{eq:mean_energy}). The gray shaded vertical bands edge the most relevant energy region of $E\,\mathord{\gtrsim}\,\unit{15}{MeV}$ (see main text for details).\\
	\label{fig:appendix_spectra_fSNe}}
\end{figure*}

\renewcommand\theHfigure{appendixD_figure.\arabic{figure}}
\renewcommand\theHequation{appendixD_equation.\arabic{equation}}
\renewcommand\theHtable{appendixD_table.\arabic{table}}

Our simplified approach does not provide information on the spectral shape of the neutrino emission. As described in Section~\ref{sec:DSNB_formulation}, we therefore assume a spectral-shape parameter $\alpha$, which is constant in time and for which we adopt different values in Section~\ref{sec:parameter_study}. Here, we examine how well our time-integrated spectra match the outcome of more sophisticated simulations with time-dependent $\alpha$. Moreover, the range of values for the instantaneous shape parameters used in our study shall be motivated in this context.

In Figure~\ref{fig:appendix_spectra}, we compare the time-integrated spectra, $\diff N/\diff E$, of electron antineutrinos, obtained from exemplary SN simulations of our Z9.6\,\&\,W18 set for different values of the instantaneous spectral-shape parameter $\alpha$ with the spectra from four SN models that were computed with the 1D version of the \textsc{Prometheus-Vertex} code (also see Appendix~\ref{appendix:rescaling} and Table~\ref{tab:flavor_ratios} there). We take models in the same range of ZAMS masses as the \textsc{Vertex} models to compare with, and with NS baryonic masses $M_\mathrm{NS,b}$ similar to those of the \textsc{Vertex} calculations. Because neutrinos emitted with energies less than $\sim$15\,MeV fall below the detection window of 10--30\,MeV for most SNe after accounting for the cosmological redshift, we restrict our comparison to the (most relevant) high-energy range of $E\,\mathord{\gtrsim}\,\unit{15}{MeV}$. Accordingly, we normalize the spectra by $N_{>15}\,\mathord{=}\,\int_{\unit{15}{MeV}}^{\infty}\diff E (\diff N/\diff E)$ for better comparability of their shapes.

We find good overall agreement of our models with the \textsc{Vertex} simulations at energies $E\,\mathord{\gtrsim}\,\unit{15}{MeV}$. There is a noticeable mismatch left of the spectral peak, which is connected to slightly higher mean neutrino energies (by $\sim$1\,MeV) compared to the reference models computed with \textsc{Vertex} (see arrows in Figure~\ref{fig:appendix_spectra} and the values of $\langle E_{\nuebar} \rangle$ in Table~\ref{tab:spectral_parameters}). This cannot be avoided with our chosen normalization, but it is of no relevance as pointed out above. The best fits are achieved when we take an instantaneous shape parameter of $\alpha\,\mathord{=}\,3.5$ for SNe with low-mass NSs ($M_\mathrm{NS,b}\,\mathord{\leqslant}\,\unit{1.6}{\msun}$; upper panels) and $\alpha\,\mathord{=}\,3.0$ for SNe with $M_\mathrm{NS,b}\,\mathord{>}\,\unit{1.6}{\msun}$ (lower panels). This parameter choice is largely insensitive to variations of the progenitors or of our engine model. We thus use these ``best-fit'' values of $\alpha$ for successful SNe in all of our DSNB calculations (see Section~\ref{sec:fiducial_model}).

For the case of BH-forming, failed SNe we provide an analogous comparison of our time-integrated spectra with two \textsc{Prometheus-Vertex} models in Figure~\ref{fig:appendix_spectra_fSNe}. The two upper panels show the spectra from the exemplary models s22.1 and s18.0 (with our standard value of the NS baryonic mass limit of $\MBH\,\mathord{=}\,\unit{2.7}{\msun}$) for different values of the instantaneous spectral-shape parameter ($\alpha\,\mathord{=}\,\alphaBH$). These two models are chosen such that the mean neutrino energies of their spectra (Equation~\eqref{eq:mean_energy}) are not too different from the ones of the two \textsc{Vertex} models to compare with (see Table~\ref{tab:spectral_parameters}). We find a best fit of the spectra for $\alphaBH\,\mathord{=}\,2.0$ in both cases (i.e., when the instantaneous spectra are Maxwell-Boltzmann like at all times of emission). This value is used as our fiducial case for failed SNe (see Section~\ref{sec:fiducial_model}).

However, as the neutrino emission from BH-formation events is strongly dependent on the maximum stable mass of cold NSs as well as on the progenitor-specific accretion rates, we also investigate how the spectral shapes change under variation of \MBH\ (lower left panel) or the chosen progenitor (lower right panel). When the NS mass limit is increased from 2.3\,\msun\ to 3.5\,\msun, the mean energies of the time-integrated spectra rise by roughly 4\,MeV from $\sim$15\,MeV to $\sim$19\,MeV for the exemplary case of model s28.0 (see arrows in the lower left panel of Figure~\ref{fig:appendix_spectra_fSNe}), which leads to a flattened spectral slope. The same trend, i.e. higher mean energies and thus flatter spectral slopes, can also be seen for spectra from progenitors with increasingly late BH formation (compare, e.g., case s18.0 with the rapid BH-formation case s40 in the lower right panel of Figure~\ref{fig:appendix_spectra_fSNe}). Because the comparison to only two \textsc{Vertex} reference models cannot be ultimately conclusive for the optimal choice of the instantaneous $\alphaBH$ values, we perform a set of DSNB calculations with varied choices of $\alphaBH$ between 1 and 3 for our BH-formation cases in Section~\ref{subsec:dsnb_parameter_study}. Doing this, we intend to test the uncertainties connected to the spectral variations of the neutrino emission from failed explosions in a systematic way.\\

\section{Spectral Parameters}\label{appendix:spectral_parameters}
\setcounter{figure}{0}
\setcounter{equation}{0}
\setcounter{table}{0}

\renewcommand\theHfigure{appendixE_figure.\arabic{figure}}
\renewcommand\theHequation{appendixE_equation.\arabic{equation}}
\renewcommand\theHtable{appendixE_table.\arabic{table}}

For the sake of completeness and as a community service for the use in future studies, we provide in Table~\ref{tab:spectral_parameters} the spectral parameters (i.e., total radiated neutrino energies, \Enuitot, mean neutrino energies, $\langle E_{\nui} \rangle$, and spectral-shape parameters, $\overline{\alpha}_{\nui}$) for the time-integrated neutrino emission of all neutrino species (\nuebar, \nue, and \nux) and the same \textsc{Prometheus-Vertex} reference models as listed in Table~\ref{tab:flavor_ratios} as well as for the 8.8\,\msun\ ECSN (model ``Sf'') from \citet{2010PhRvL.104y1101H}. Moreover, we also list the values for a selected set of exemplary \textsc{Prometheus-HotB} models (as shown in Figures~\ref{fig:appendix_spectra} and \ref{fig:appendix_spectra_fSNe}). Note that the values of $\overline{\alpha}_{\nui}$ (Equation~\eqref{eq:alpha}) are generally somewhat ($\sim$5--15\%) smaller than the $\alpha$ parameters of the instantaneous neutrino emission (Equation~\eqref{eq:spectral_shape}), i.e., the time-integrated spectra are slightly wider than the instantaneous ones.

We also point out that in the \textsc{Prometheus-Vertex} simulations the total energy released in neutrinos by the cooling PNSs is EoS dependent and for BH-forming cases, in particular, it depends on the accretion time until the NS collapses to a BH. For a comparison of the results obtained from our \textsc{Prometheus-HotB} models with generic values based on the (EoS-independent but NS-radius dependent) fit formula of \citet[see Equation~\eqref{eq:LattimerPrakash}]{2001ApJ...550..426L}, we refer the reader to Appendix~\ref{appendix:total_energies}. For a comparative discussion of the time-integrated neutrino spectra of the \textsc{Prometheus-Vertex} simulations and our best-fit spectra for the \textsc{Prometheus-HotB} models (obtained by suitable choices of the values of the instantaneous shape parameter $\alpha$), we refer the reader to Appendix~\ref{appendix:spectra}.

\begin{deluxetable*}
	{lCCCCCCCCCcC}
	\tablecaption{Spectral parameters for selected \textsc{Prometheus-Vertex} and \textsc{Prometheus-HotB} models.\label{tab:spectral_parameters}}
	\tablehead{
		\colhead{Model} & \colhead{$E_{\nuebar}^\mathrm{tot}$} & \colhead{$E_{\nue}^\mathrm{tot}$} & \colhead{$E_{\nux}^\mathrm{tot}$} & \colhead{$\langle E_{\nuebar} \rangle$} & \colhead{$\langle E_{\nue} \rangle$} & \colhead{$\langle E_{\nux} \rangle$} & \colhead{$\overline{\alpha}_{\nuebar}$} & \colhead{$\overline{\alpha}_{\nue}$} & \colhead{$\overline{\alpha}_{\nux}$} & \colhead{Remnant} & \colhead{$M_\mathrm{NS,b}$} \\
		\cline{2-4} \cline{5-7} \cline{12-12}
		\colhead{} & \multicolumn{3}{c}{[$10^{52}$\,erg]} & \multicolumn{3}{c}{[MeV]} & \multicolumn{3}{c}{} & \colhead{} & \colhead{[\msun]}
	}
	\startdata
	\textsc{Vertex}, 8.8\,\msun\ ECSN (``Sf'') & 2.67 & 3.20 & 2.62 & 11.6 & 9.5 & 11.5 & 2.49 & 3.06 & 2.10 & NS & 1.366 \\
	\textsc{Vertex}, z9.6co, LS220 & 2.93 & 3.28 & 3.17 & 12.4 & 9.7 & 12.4 & 2.51 & 2.82 & 2.03 & NS & 1.361 \\
	\textsc{Vertex}, z9.6co, SFHo & 3.13 & 3.49 & 3.31 & 12.1 & 9.6 & 12.0 & 2.83 & 3.03 & 2.24 & NS & 1.363 \\
	\textsc{Vertex}, s11.2co, LS220 & 3.09 & 3.56 & 3.02 & 13.7 & 10.6 & 13.6 & 2.90 & 2.76 & 2.34 & NS & 1.366 \\
	\textsc{Vertex}, s27.0co, LS220 & 5.72 & 5.99 & 5.37 & 13.7 & 10.9 & 13.1 & 2.25 & 2.15 & 1.82 & NS & 1.776 \\
	\textsc{Vertex}, s27.0co, SFHo & 5.91 & 6.24 & 5.68 & 13.6 & 10.9 & 13.2 & 2.61 & 2.50 & 2.11 & NS & 1.772 \\
	\textsc{Vertex}, s20.0, SFHo & 7.36 & 7.53 & 6.96 & 14.0 & 11.3 & 13.5 & 2.48 & 2.31 & 2.02 & NS & 1.947 \\
	\textsc{Vertex}, s40s7b2c, LS220 & 4.49 & 5.44 & 2.81 & 17.6 & 14.4 & 18.8 & 2.52 & 2.08 & 1.61 & BH (0.57\,s) & (2.320) \\
	\textsc{Vertex}, s40.0c, LS220 & 8.62 & 9.38 & 4.83 & 18.7 & 15.7 & 17.6 & 1.95 & 1.58 & 1.41 & BH (2.11\,s) & (2.279) \\
	\hline
	s10.0, Z9.6, $\alpha\,\mathord{=}\,3.5$, ``L'' & 4.41 & 5.01 & 4.56 & 15.1 & 11.6 & 14.9 & 3.32$^{~3.78}_{~2.85}$ & 3.32$^{~3.78}_{~2.85}$ & 2.67$^{~3.04}_{~2.29}$ & NS & 1.430 \\
	s12.25, W18, $\alpha\,\mathord{=}\,3.5$, ``L'' & 5.55 & 6.30 & 5.74 & 14.7 & 11.5 & 14.6 & 3.29$^{~3.75}_{~2.83}$ & 3.32$^{~3.79}_{~2.86}$ & 2.64$^{~3.01}_{~2.27}$ & NS & 1.551 \\
	s27.0, W18, $\alpha\,\mathord{=}\,3.0$, ``H'' & 7.21 & 7.51 & 6.84 & 14.9 & 11.7 & 14.3 & 2.83$^{~3.28}_{~2.36}$ & 2.84$^{~3.31}_{~2.38}$ & 2.29$^{~2.66}_{~1.91}$ & NS & 1.742 \\
	s21.7, W18, $\alpha\,\mathord{=}\,3.0$, ``H'' & 7.87 & 8.20 & 7.46 & 15.0 & 12.1 & 14.5 & 2.82$^{~3.28}_{~2.36}$ & 2.82$^{~3.28}_{~2.36}$ & 2.29$^{~2.66}_{~1.91}$ & NS & 1.870 \\
	s40, W18, $\alpha_\mathrm{BH}\,\mathord{=}\,2.0$, ``F'' & 5.93 & 7.19 & 3.71 & 14.9 & 12.7 & 15.9 & 1.90$^{~2.37}_{~1.43}$ & 1.79$^{~2.23}_{~1.35}$ & 1.21$^{~1.51}_{~0.91}$ & BH (1.03\,s) & (2.7) \\
	s23.3, W18, $\alpha_\mathrm{BH}\,\mathord{=}\,2.0$, ``F'' & 8.67 & 10.51 & 5.42 & 15.4 & 13.2 & 16.4 & 1.87$^{~2.33}_{~1.41}$ & 1.76$^{~2.19}_{~1.32}$ & 1.19$^{~1.49}_{~0.90}$ & BH (1.67\,s) & (2.7) \\
	s24.1, W18, $\alpha_\mathrm{BH}\,\mathord{=}\,2.0$, ``S'' & 10.83 & 11.78 & 6.07 & 15.6 & 13.5 & 14.7 & 1.85$^{~2.31}_{~1.39}$ & 1.74$^{~2.16}_{~1.31}$ & 1.34$^{~1.67}_{~1.01}$ & BH (2.01\,s) & (2.7) \\
	s24.7, W18, $\alpha_\mathrm{BH}\,\mathord{=}\,2.0$, ``S'' & 12.30 & 13.39 & 6.90 & 16.0 & 13.8 & 15.0 & 1.82$^{~2.27}_{~1.37}$ & 1.70$^{~2.11}_{~1.28}$ & 1.32$^{~1.64}_{~0.99}$ & BH (2.46\,s) & (2.7) \\
	s22.1, W18, $\alpha_\mathrm{BH}\,\mathord{=}\,2.0$, ``S'' & 14.30 & 15.56 & 8.01 & 16.4 & 14.1 & 15.4 & 1.80$^{~2.24}_{~1.36}$ & 1.68$^{~2.09}_{~1.27}$ & 1.30$^{~1.62}_{~0.98}$ & BH (3.32\,s) & (2.7) \\
	s28.0, W18, $\alpha_\mathrm{BH}\,\mathord{=}\,2.0$, ``S'' & 14.53 & 15.81 & 8.14 & 16.7 & 14.6 & 15.7 & 1.76$^{~2.18}_{~1.33}$ & 1.62$^{~2.00}_{~1.22}$ & 1.27$^{~1.58}_{~0.96}$ & BH (3.79\,s) & (2.7) \\
	~~~~------------''------------ & 8.72 & 9.49 & 4.89 & 15.1 & 12.8 & 14.2 & 1.89$^{~2.36}_{~1.42}$ & 1.79$^{~2.23}_{~1.35}$ & 1.37$^{~1.71}_{~1.03}$ & BH (2.04\,s) & (2.3) \\
	~~~~------------''------------ & 18.46 & 20.09 & 10.35 & 18.1 & 16.1 & 17.0 & 1.63$^{~2.01}_{~1.24}$ & 1.45$^{~1.79}_{~1.10}$ & 1.18$^{~1.45}_{~0.89}$ & BH (5.64\,s) & (3.1) \\
	~~~~------------''------------ & 22.06 & 24.01 & 12.37 & 19.3 & 17.4 & 18.1 & 1.53$^{~1.89}_{~1.17}$ & 1.34$^{~1.65}_{~1.02}$ & 1.11$^{~1.36}_{~0.85}$ & BH (7.51\,s) & (3.5) \\
	s27.9, W18, $\alpha_\mathrm{BH}\,\mathord{=}\,2.0$, ``S'' & 14.61 & 15.90 & 8.19 & 17.3 & 15.1 & 16.2 & 1.69$^{~2.09}_{~1.28}$ & 1.51$^{~1.87}_{~1.14}$ & 1.22$^{~1.51}_{~0.92}$ & BH (5.11\,s) & (2.7) \\
	s18.0, W18, $\alpha_\mathrm{BH}\,\mathord{=}\,2.0$, ``S'' & 14.49 & 15.77 & 8.12 & 17.5 & 15.4 & 16.4 & 1.67$^{~2.07}_{~1.26}$ & 1.46$^{~1.80}_{~1.10}$ & 1.21$^{~1.50}_{~0.91}$ & BH (9.12\,s) & (2.7)
	\enddata
	\tablecomments{Total energies, \Enuitot\, radiated in the neutrino species $\nui\,\mathord{=}\,(\nuebar,\,\nue,\,\nux)$, mean energies, $\langle E_{\nui} \rangle$, and shape parameters, $\overline{\alpha}_{\nui}$, of the time-integrated \nui\ spectra according to Equations~\eqref{eq:mean_energy} and \eqref{eq:alpha}, respectively, listed for the same \textsc{Prometheus-Vertex} models as employed in Appendices~\ref{appendix:rescaling} and \ref{appendix:spectra} as well as for the 8.8\,\msun\ ECSN (model ``Sf'') from \citet{2010PhRvL.104y1101H}, and for the same subset of \textsc{Prometheus-HotB} models shown in Figures~\ref{fig:appendix_spectra} and \ref{fig:appendix_spectra_fSNe}. Note that $\Enutot\,\mathord{=}\,E_{\nuebar}^\mathrm{tot}\,\mathord{+}\,\,E_{\nue}^\mathrm{tot}\,\mathord{+}\,4\,E_{\nux}^\mathrm{tot}$. For our \textsc{Prometheus-HotB} models, we take $E_{\nui}^\mathrm{tot}\,\mathord{=}\,\tilde{\xi}_{\nui} \Enutot$, $\langle E_{\nux} \rangle\,\mathord{=}\,\lambdaE\,\langle E_{\nuebar} \rangle$ and $\alphax\,\mathord{=}\,\lambdaa\,\alphaea$ with the conversion factors ($\tilde{\xi}_{\nui}$, \lambdaE, \lambdaa; Equations~\eqref{eq:xi_tilde}, \eqref{eq:lambda_E}, \eqref{eq:lambda_a}) according to the four cases ``L'', ``H'', ``F'', and ``S'' in Table~\ref{tab:flavor_ratios} (see Appendix~\ref{appendix:rescaling}). The values of $\overline{\alpha}_{\nui}$ are given for our best-fit choices of the \textit{instantaneous} spectral-shape parameter (``$\alpha_\mathrm{best}$''; i.e., $\alpha\,\mathord{=}\,3.5$ for SNe with $M_\mathrm{NS,b}\,\mathord{\leqslant}\,\unit{1.6}{\msun}$, $\alpha\,\mathord{=}\,3.0$ for SNe with $M_\mathrm{NS,b}\,\mathord{>}\,\unit{1.6}{\msun}$, and $\alpha_\mathrm{BH}\,\mathord{=}\,2.0$ for failed SNe; see Appendix~\ref{appendix:spectra}), as well as for the choices of $\alpha_\mathrm{best}\,\mathord{+}\,0.5$ (in superscript) and of $\alpha_\mathrm{best}\,\mathord{-}\,0.5$ (in subscript). The NS-formation cases are sorted according to the baryonic masses of the remnant NSs ($M_\mathrm{NS,b}$; last column), the failed-SN cases according to the times of BH formation ($t_\mathrm{BH}$; second to last column); the baryonic PNS masses at these times are listed in the last column (values in parentheses).}
\end{deluxetable*}

\clearpage
\bibliographystyle{aasjournal}
\bibliography{bibliography}


\end{document}